\documentclass[fleqn,usenatbib]{mnras}
\usepackage{newtxtext,newtxmath}

\usepackage[T1]{fontenc}
\usepackage{ae,aecompl}

\usepackage{graphicx}	
\usepackage{amsmath}	
\usepackage{amssymb}	

\newcommand{\dm}{$\Delta m_{15} (B)$}

\def\lsim{\hbox{\rlap{\raise 0.425ex\hbox{$<$}}\lower 0.65ex\hbox{$\sim$}}}
\def\gsim{\hbox{\rlap{\raise 0.425ex\hbox{$>$}}\lower 0.65ex\hbox{$\sim$}}}

\def\code#1{\texttt{#1}}

\title[SN Ia Diversity with Kaepora]{Investigating the Diversity of Type I\lowercase{a} Supernova Spectra with the Open-Source Relational Database Kaepora}

\author[M. R. Siebert et al.]{
M. R. Siebert,$^{1}$\thanks{E-mail: msiebert@ucsc.edu}
R. J. Foley,$^{1}$
D. O. Jones,$^{1}$
R. Angulo,$^{1}$
K. Davis,$^{1}$
A. Duarte,$^{1}$
\newauthor
E. Strasburger,$^{1}$
M. Conlon,$^{2}$
N. Kazmi,$^{3}$
R. Nishimoto,$^{4}$
M. Schubert,$^{3}$
L. Sun,$^{3}$
\newauthor
and R. Tippens$^{5}$
\\
$^{1}$Department of Astronomy and Astrophysics, University of California, Santa Cruz, CA 95064\\
$^{2}$Department of Information Technology, University of Wisconsin-Madison, Madison, WI 53706\\
$^{3}$Department of Physics, University of Illinois at Urbana-Champaign, Urbana, IL 61801\\
$^{4}$Department of Computer Science, University of Illinois at Chicago, Chicago, IL 60607\\
$^{5}$Department of Astronomy, The University of Texas at Austin, Austin, TX 78712\\
}

\date{Accepted XXX. Received YYY; in original form ZZZ}

\pubyear{2019}

\begin{document}
\label{firstpage}
\pagerange{\pageref{firstpage}--\pageref{lastpage}}
\maketitle

\begin{abstract}
We present a public, open-source relational database (we name \code{kaepora}) containing a sample of 4975 spectra of 777 Type Ia supernovae (SNe~Ia). Since we draw from many sources, we significantly improve the spectra by inspecting these data for quality, removing galactic emission lines and cosmic rays, generating variance spectra, and correcting for the reddening caused by both MW and host-galaxy dust. With our database, we organize this homogenized dataset by 56 unique categories of SN-specific and spectrum-specific metadata. With \code{kaepora}, we produce composite spectra of subpopulations of SNe~Ia and examine how spectral features correlate with various SN properties. These composite spectra reproduce known correlations with phase, light-curve shape, and host-galaxy morphology. With our large dataset, we are also able to generate fine-grained composite spectra simultaneously over both phase and light-curve shape. The color evolution of our composite spectra is consistent with other SN Ia template spectra, and the spectral properties of our composite spectra are in rough agreement with these template spectra with some subtle differences. We investigate the spectral differences of SNe Ia that occur in galaxies with varying morphologies. Controlling for light-curve shape, which is highly correlated with host-galaxy morphology, we find that SNe Ia residing in late-type and early-type galaxies have similar spectral properties at multiple epochs. However for SNe Ia in these different environments, their spectra appear to have Ca II near-infrared triplet features that have slightly different strengths. Although this is apparent in the composite spectra and there is some difference in the populations as seen by individual spectra, this difference is not large enough to indicate differences in the underlying populations. All individual spectra and metadata are available in our open-source database \code{kaepora} along with the tools developed for this investigation to facilitate future investigations of SN Ia properties.
\end{abstract}


\section{Introduction}\label{sec:intro}

Type Ia (SNe~Ia) are excellent cosmological probes and an important tool for understanding dark energy. In general, the accepted models for these phenomena involve the thermonuclear disruption of a carbon-oxygen white dwarf. Despite their reputation as ``standard candles", the observational properties of these explosions are intrinsically heterogeneous \citep{hatano00,CfA}. SN Ia luminosities can vary by an order of magnitude \citep{pskovskii11}. The ``standardizability'' of these events only became evident after observing some basic correlations between these optical properties. Mainly, SNe~Ia  exhibit an empirical relationship between peak luminosity and light-curve width \citep{pskovskii11,phillips93, riess96}. This relationship has been attributed to the varying amount of $^{56}$Ni produced in these explosions \citep{arnett82,nugent,stritzinger06}.

The normalization of SN Ia luminosities via this relationship allows for precise distance measurements on extragalactic scales. Measurements of SNe Ia were used to show that the expansion of our universe is currently accelerating \citep{Riess98,Perlmutter99}. Today, after making corrections for light-curve shape and color, there remains an intrinsic scatter that may belie unaccounted SN physics \citep{fk11,scolnic18,jones18}. This scatter will soon become a dominant source of systemic uncertainty for SN~Ia distance measurements \citep{jones18}. 

Although valuable, these basic correlations do not adequately explain the observed spectral diversity. For example, SNe~Ia with the same light-curve shape and color exhibit a wide range of photospheric velocities at maximum light \citep{hatano00}. Spectra provide an abundance of information about SN~ejecta (e.g., nucleosynthesis, ionization states, kinematics, etc.) necessary for constructing a complete picture of these phenomena. 

Despite their proven cosmological utility, open questions remain regarding the physical processes that can produce the observed range of diversity. Detailed studies of individual events have been essential for improving our understanding of how spectral features change with phase and luminosity. The well-sampled spectroscopic observations of SN 2011fe solidified our understanding of the optical properties of normal SNe Ia from very early to very late epochs \citep{nugent11, pereira13, shappee13, mazzali14}. Observations of the spectroscopically distinct SNe 1991T \citep{phillips92, filippenko92b} and 1991bg \citep{filippenko92a,leibundgut93} highlight how the temperature of the SN ejecta can impact spectral features. SN 1991T exhibited a large intrinsic luminosity and showed very little Ca H\&K and Si II $\lambda 5972$ absorption. In addition to its low intrinsic luminosity, SN 1991bg showed strong Ti II absorption near $4300$~\AA\ at maximum light. These effects have been attributed to a sequence in the photospheric temperature where peculiar events like SNe 1991T and 1991bg lie in the high- and low-temperature tails of this relationship, respectively \citep{nugent}. This sequence could be described by the tight correlation between light-curve shape and the ratio of the depth Si II $\lambda 5972$ and $\lambda 6355$ ($\mathcal{R}$(Si II)). While very useful, detailed analyses of a few individual events cannot fully represent the observed spectroscopic diversity. 

\citet{branch87} first showed that ejecta velocities of SNe~Ia are heterogeneous. Studies of larger samples have revealed that the nature of the diversity of SNe~Ia properties is indeed multidimensional \citep{hatano00,li01,benetti05,fsk11, CfA}. For normal SNe~Ia, the velocities derived from Si II $\lambda 6355$ do not correlate with the decline-rate parameter \dm \space \citep{hatano00}. \citet{fk11} showed that high-velocity (HV) SNe~Ia have redder intrinsic colors. Furthermore, subgroups of SNe~Ia exhibit a large range of velocity gradients for the Si II $\lambda 6355$ feature. \citet{benetti05} showed that for normal SNe~Ia, this velocity gradient does not correlate with \dm. Additional observations have also revealed the presence of high-velocity features (HVF) in numerous events. These typically appear as double-troughed absorption profiles in Si II $\lambda 6355$, Ca II H\&K, or the  Ca II near-infrared triplet \citep{gerardy04,mazzali05a,mazzali05b,foley12,childress13,silverman15}. HVF strength was shown to correlate with the maximum-light velocity of Si II $\lambda 6355$ \citep{childress14}. Filling this photometric and spectroscopic parameter space has proven difficult even with large individual SN data sets.

Studies of the mean spectroscopic properties of SNe~Ia have been particularly useful for cosmology. Distance estimators require spectral models so that photometric measurements can be calibrated. Spectral models are also useful for estimating bolometric luminosities and thus $^{56}$Ni yields \citep{howell09}. \citet{nugent02} first developed a spectral template for the purpose of representing the temporal evolution of normal SNe~Ia. Similarly, \citet{hsiao07} constructed a mean spectral template time series using ${\sim} 600$ spectra of $\sim$100 normal SNe~Ia. Sparse sampling of the observed spectroscopic diversity has limited most data-driven spectral templates to single-parameter models (typically phase). The Nugent and Hsiao template spectra are only representative of normal SNe~Ia and do not vary with \dm. To best represent the heterogeneous nature of SNe~Ia, all available spectroscopic data and metadata should be compiled.  

Large efforts have been made to improve the accessibility of spectroscopic SN data. Online databases such as SUSPECT \citep{richardson01}, WISEREP \citep{yaron12}, and the Open Supernova Catalog \citep{guillochon17} have all provided access to large samples of spectra covering decades of observations. However, since the origins of these data are heterogeneous, studying the bulk properties of these data has been challenging. These spectra have been reduced using different techniques and come from a variety of telescopes and instruments. Thus, the available spectra vary greatly in many basic properties (e.g., wavelength coverage, resolution, signal-to-noise ratio, reduction quality, existence of telluric correction, variance spectra, etc.) Furthermore, online databases do not often link these samples to their respective calibrated light-curves and photometric metadata. While some groups have released large homogenized spectroscopic samples \citep{CfA,bsnip,CSP}, no efforts have been made to make use of all available SN~Ia spectra at once. 

In order to properly combine these data, we have developed a framework to homogenize these spectra. By taking steps like inspecting individual spectra for quality, removing unwanted features, generating variance spectra, correcting for Milky-Way (MW) extinction, and interpolating to constant resolution, we are able to provide a fully homogenized sample containing 4975 spectra of 777 SNe Ia and their associated event specific and spectrum specific metadata. One way we aim to improve the efficiency of SN Ia  analyses is via leveraging the power of relational queries. A relational database model simplifies data retrieval and provides a structure that is easily scalable in order to accommodate the rapidly increasing number of SN observations. In this work, we present a comprehensive relational database for observations of SNe~Ia. We generate composite spectra from subsets of the data in order to validate the data itself and our processing techniques. The database used in this work and all composite spectra presented below can be found online\footnote{\href{https://msiebert1.github.io/kaepora/}{https://msiebert1.github.io/kaepora/}}. The source code developed here is open and freely available for use\footnote{\href{https://github.com/msiebert1/kaepora}{https://github.com/msiebert1/kaepora}}. Installation instructions and example code are also available online\footnote{\href{https://kaepora.readthedocs.io/en/latest/index.html}{https://kaepora.readthedocs.io/en/latest/index.html}}. Here we outline how to query the database for subsamples of individual spectra, and produce composite spectra if desired.

In Section 2, we overview the demographics of the dataset. In Section 3, we outline the reprocessing steps that we execute to homogenize all spectroscopic data and the methods we use for spectroscopic analyses. In Section 4 we present our composite spectra generated from various subsets of these data, compare these composites spectra with other template spectra, and use these composite spectra to investigate SN~Ia spectral variations related to their host-galaxy environments. Finally in Section 5, we summarize our results and discuss the future potential for uses of our relational database. 

\begin{figure*}
\begin{center}
\includegraphics[angle=0,width=3.2in]{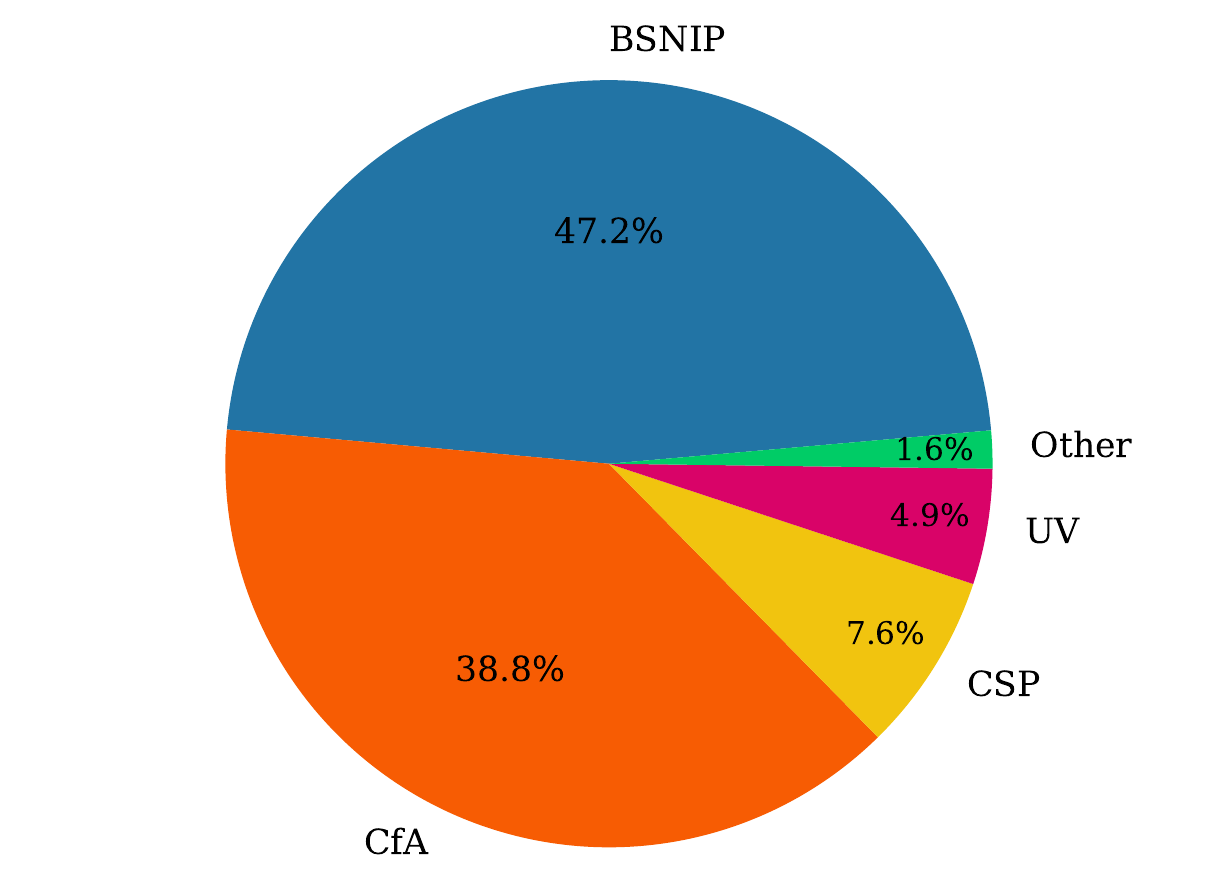}
\includegraphics[angle=0,width=3.2in]{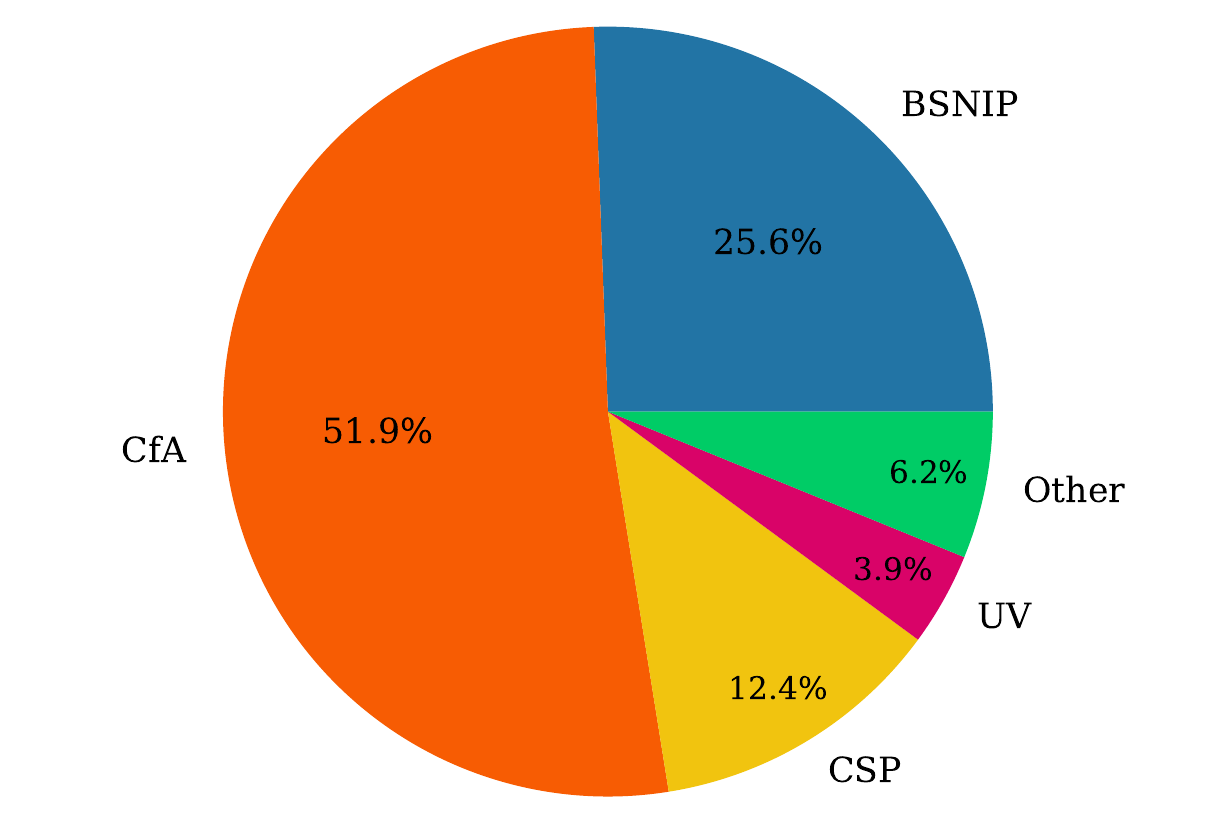}
\caption{(\textit{left}): Contributions of each major data source to the total number of SNe in the database. (\textit{right}): Contributions of each major data source to the total number of spectra in the database. The BSNIP sample provides the largest number of individual SNe, while the CfA sample tends to provide the largest number of spectra per SN.}\label{sources}
\end{center}
\end{figure*}

\begin{figure}
    \includegraphics[width=3.2in]{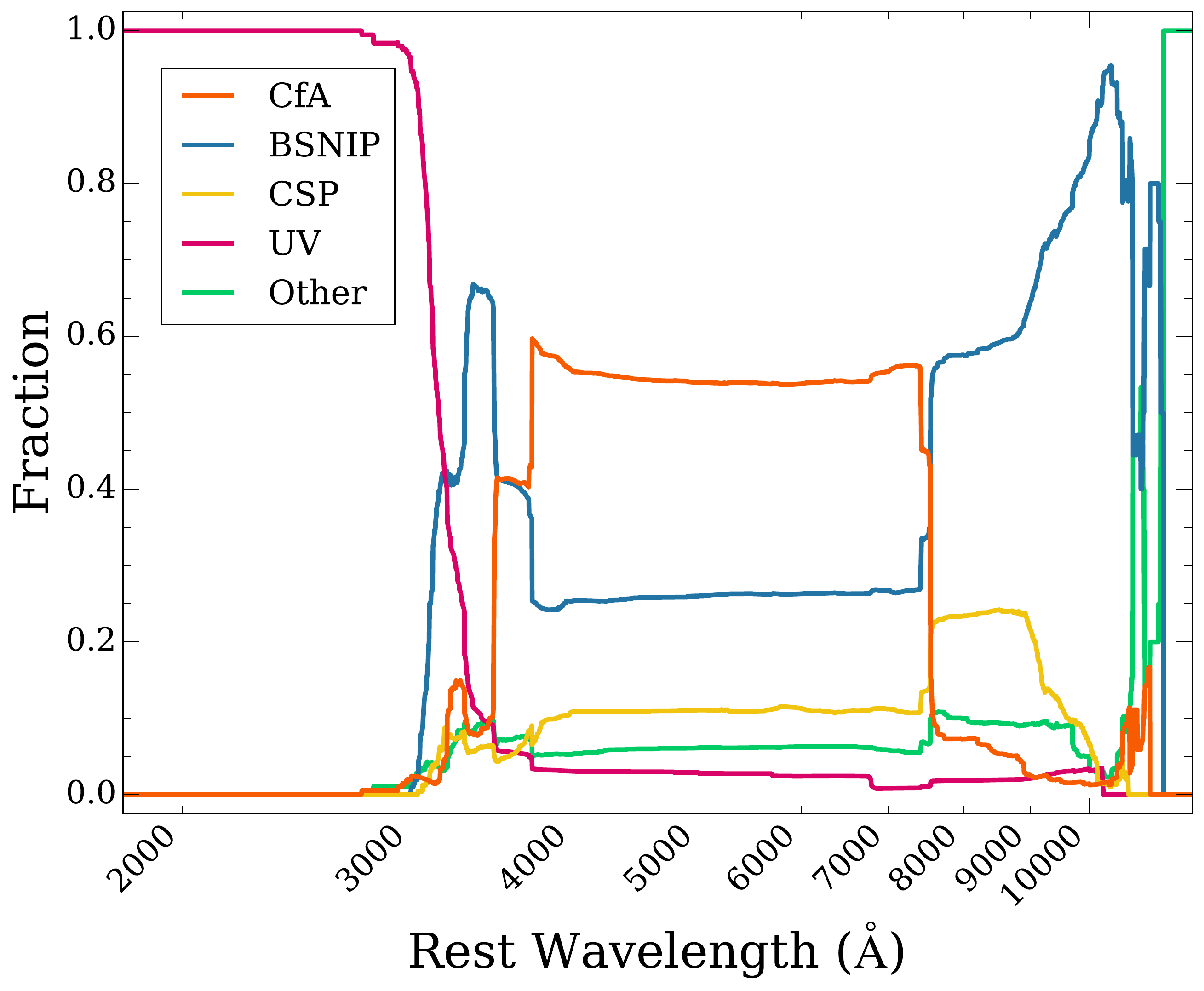}
  \caption{Fraction of all spectra coming from the major sources as a function of wavelength. The CfA sample is the dominant source of spectra in the range of ${\sim } 3500-7500$~\AA. The BSNIP sample is incredibly important for reaching wavelengths ${>}7500$~\AA, and \textit{Swift} and \textit{HST} spectra provide the valuable data ${<}3000$~\AA.}\label{fig:sources_wav}
\end{figure}

\section{Sample} \label{sec:samp}
Our database currently contains 4975 spectra of 777 SNe~Ia. The full sample is comprised of spectra from heterogeneous data sources. The majority of our data are sourced from the Center for Astrophysics (CfA) Supernova Program \citep{CfA}, the Berkeley SN Ia Program (BSNIP; \citealt{bsnip}), and the Carnegie Supernova Project (CSP; \citealt{CSP}). These sources provide 2584, 1273, and 616 spectra respectively. We also include 195 \textit{Swift} and \textit{HST} UV spectra, and 307 spectra from various other sources detailed in Table \ref{tab:1}. The relative contributions to the total number of objects and spectra from each source are shown in Figure \ref{sources}. While BSNIP and the CfA have observed a comparable number of SNe, the CfA has acquired more spectra per object. Figure \ref{fig:sources_wav} shows the relative contributions of these sources as a function of wavelength. The two largest samples, CfA and BSNIP, are complementary. The CfA data are very important for representing SNe Ia at a range of epochs, and the BSNIP spectra are vital for their larger wavelength range (extending from ${\sim} 3000 - 10000$~\AA). These data are linked to a variety of event-specific and spectrum-specific metadata. We also include light-curves from the Open Supernova Catalog \citep{OSC} that overlap with our sample of SNe~Ia. 

Individual spectra have been inspected by eye for quality. Those that are dominated by noise or contain large wavelength gaps are discarded. Spectra with considerable telluric contamination have been identified and flagged. Details for individual spectra are provided in Table \ref{tab:1}. Associated metadata for each SN contained within the database are described in Table \ref{tab:2}.

\begin{table*}
\centering
\begin{tabular}{lccrcc}
\hline
 SN Name   &     MJD &   Phase &   S/N & Wavelength Range   &   Reference \\
\hline
 1980N     & 44584.0 &    -1.8 &   1.9 & 1886 - 3247        &           1 \\
 1980N     & 44586.0 &     0.2 &   1.8 & 1875 - 3261        &           1 \\
 1980N     & 44589.0 &     3.2 &   1.2 & 1883 - 3271        &           1 \\
 1980N     & 44590.0 &     4.2 &   1.8 & 1883 - 3239        &           1 \\
 1980N     & 44596.0 &    10.1 &   1.5 & 1899 - 3202        &           1 \\
 1980N     & 44597.0 &    11.1 &   1.4 & 1923 - 3237        &           1 \\
 1980N     & 44620.0 &    34.0 &   0.6 & 1899 - 3240        &           1 \\
 1981B     & 44672.0 &     1.0 &   3.8 & 1910 - 3213        &           1 \\
 1981B     & 44673.0 &     2.0 &   2.2 & 1891 - 3261        &           1 \\
 1981B     & 44674.0 &     2.9 &   2.5 & 1862 - 3298        &           1 \\
 1986G     & 46556.0 &    10.5 &   0.7 & 1995 - 3348        &           1 \\
 1986G     & 46558.0 &    12.5 &   1.8 & 1939 - 3257        &           1 \\
 1986G     & 46562.0 &    16.5 &   1.4 & 1923 - 3286        &           1 \\
 1986G     & 46565.0 &    19.5 &   1.9 & 1873 - 3326        &           1 \\
 1986G     & 46569.0 &    23.5 &   1.5 & 1907 - 3270        &           1 \\
 1989A     & 47643.0 &    83.8 &  11.6 & 3450 - 9000        &           2 \\
 1989B     & 47572.0 &     7.5 & 211.6 & 3450 - 8450        &           3 \\
 1989B     & 47578.0 &    13.5 &  12.1 & 3450 - 7000        &           3 \\
 1989B     & 47643.0 &    78.3 &  56.3 & 3300 - 9050        &           3 \\
 1989B     & 47717.0 &   152.1 &  14.1 & 3900 - 6226        &           3 \\
 1989B     & 47558.0 &    -6.5 &   2.6 & 1226 - 3348        &           1 \\
 1989M     & 47716.0 &     2.5 & 221.1 & 3080 - 10300       &           2 \\
\hline
\end{tabular}
\caption{Full spectral sample. \textbf{References}: (1) \citet{2008ApJ...686..117F}; (2) \citet{2012MNRAS.425.1789S}; (3) \citet{1994AJ....108.2233W}; (4) \citet{1992AJ....104.1543F}; (5) \citet{1993Natur.365..728R}; (6) \citet{2001AJ....121.1648M}; (7) \citet{1992ApJ...384L..15F}; (8) Unknown; (9) \citet{2012AJ....143..126B}; (10) \citet{1997ASIC..486.....R}; (11) \citet{1996MNRAS.278..111P}; (12) \citet{2001MNRAS.321..254S}; (13) \citet{1999AJ....117.2709L}; (14) \citet{2001ApJ...546..734L}; (15) \citet{2001ApJ...549L.215C}; (16) \citet{2001PASP..113..308M}; (17) \citet{2003AJ....125.1087G}; (18) \citet{2001PASP..113.1178L}; (19) \citet{2003ApJ...595..779V}; (20) \citet{2011AJ....142...74K}; (21) \citet{2005ApJ...632..450L}; (22) \citet{2004MNRAS.348..261B}; (23) \citet{2008MNRAS.384..107E}; (24) \citet{2003PASP..115..453L}; (25) \citet{2006AJ....132..189J}; (26) \citet{2008MNRAS.388..971P}; (27) \citet{2005A&A...436.1021}; (28) \citet{2006MNRAS.369.1880E}; (29) \citet{2007A&A...469..645S}; (30) \citet{2005A&A...429..667A}; (31) \citet{2013ApJ...773...53F}; (32) \citet{2007AIPC..937..311L}; (33) \citet{2009ApJ...697..380W}; (34) \citet{2007A&A...471..527G}; (35) \citet{2006PASP..118..722C}; (36) \citet{2007PASP..119..360P}; (37) \citet{2010ApJ...708.1748F}; (38) \citet{2007ApJ...654L..53T}; (39) \citet{2007ApJ...669L..17H}; (40) \citet{2011A&A...526A..28O}; (41) \citet{2008ApJ...675..626W}; (42) \citet{2009PASJ...61..713Y}; (43) \citet{2007ApJ...671L..25S}; (44) \citet{2010PASP..122....1Z}; (45) \citet{2011MNRAS.410..585S}; (46) \citet{2018MNRAS.479..517P}; (47) \citet{2014ApJ...786..134M}; (48) \citet{2009AJ....138..376F}; (49) \citet{2013ApJ...769L...1F}; (50) \citet{2013A&A...554A..27P}; (51) \citet{2014MNRAS.439.1959M}; (52) \citet{2015MNRAS.453.3300A}; (53) \citet{2015MNRAS.452.4307P}; (54) \citet{2016MNRAS.461.1308F} (This table is available in its entirety in a machine-readable form in the online journal. A portion is shown here for guidance regarding its form and content.)}
\label{tab:1}
\end{table*}
\begin{table}
\centering
\begin{tabular}{ccc}
\hline
 Property                               &   Number of SNe \\
\hline
 $t_{max}$ (B)                          &             444   \\
 Redshift                               &             773  \\
 $\Delta m_{15} (B)$                    &             274 \\
 $\Delta m_{15} (B)$ from fit parameter &              97\\
 $A_V$                                  &             314\\
 Host Morphology                        &             525\\
 $M_B$                                  &             120\\
 $B-V$                                  &             120\\
 Velocity at Max $m_B$                  &             290\\
 Carbon Presence                        &             193\\
 Hubble Residual                        &             149\\
\hline
\end{tabular}
\caption{The main queryable spectral and photometric properties of the SNe included in the database. We also include but do not list additional metadata from \citet{hicken09} and \citet{betoule14}.}
\label{tab:2}
\end{table}

\subsection{Nominal Sample}

We explore the diversity of spectral properties of SNe~Ia for a large subsample of these data. The main criterion for selecting this subsample from the total SN sample was whether there was a calibrated light-curve covering the time of maximum light. With this, each spectrum allows for a measurement of the phase relative to $B$-band maximum light for each spectrum in this subsample. We also do not consider SNe classified as Type Iax as they are proposed to originate from a different progenitor scenario (103 spectra of 13 SNe~Iax); \citep{Iax}. Since the reddening caused by host-galaxy dust causes the relative strengths of spectral features and overall continuum to change, we restrict our sample again to SNe~Ia where we can correct for this effect using MLCS reddening parameters \citep{jha07}. These requirements limit us to  3485 spectra of 305 SNe~Ia for our analysis. From here on this sample will be referred to as our ``nominal'' sample. 
\begin{figure}
    \includegraphics[width=3.2in]{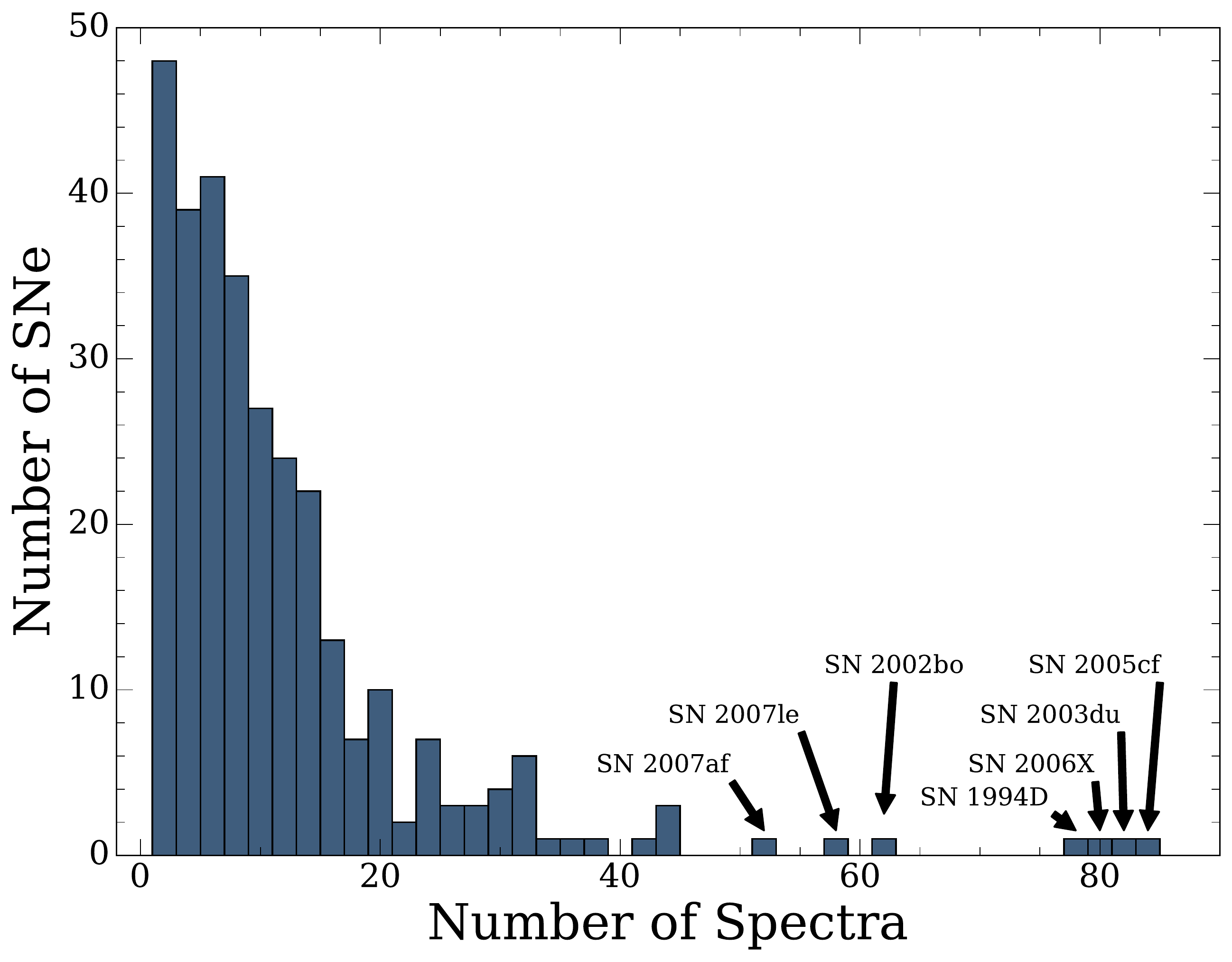}
  \caption{Histogram of the number of individual spectra per SN in the nominal sample.}\label{fig:dem1}
\end{figure}
\begin{figure}
    \includegraphics[width=3.2in]{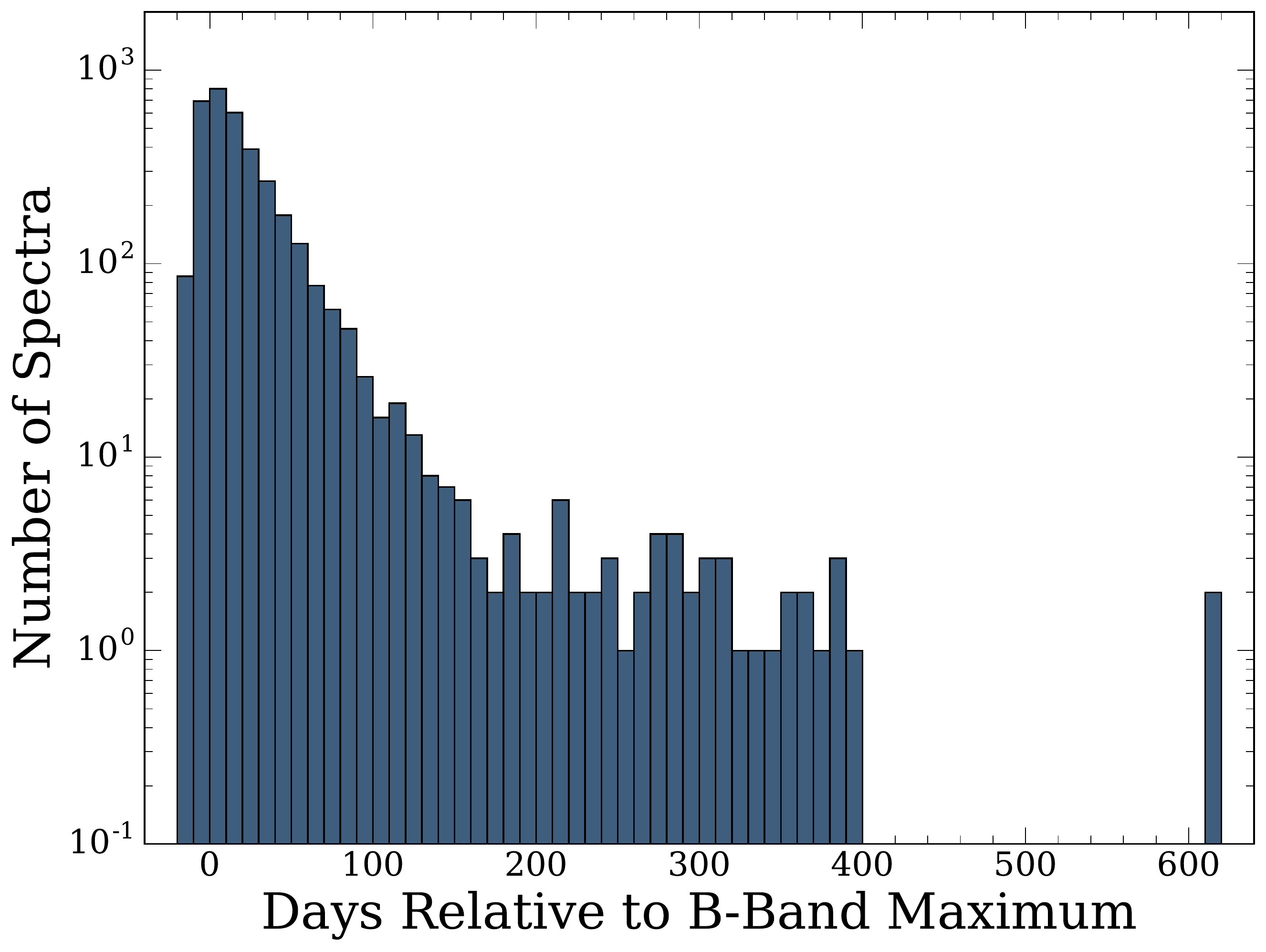}
  \caption{Histogram of the number of individual SN spectra at each epoch in the nominal sample.}\label{fig:dem2}
\end{figure}
\begin{figure}
    \includegraphics[width=3.2in]{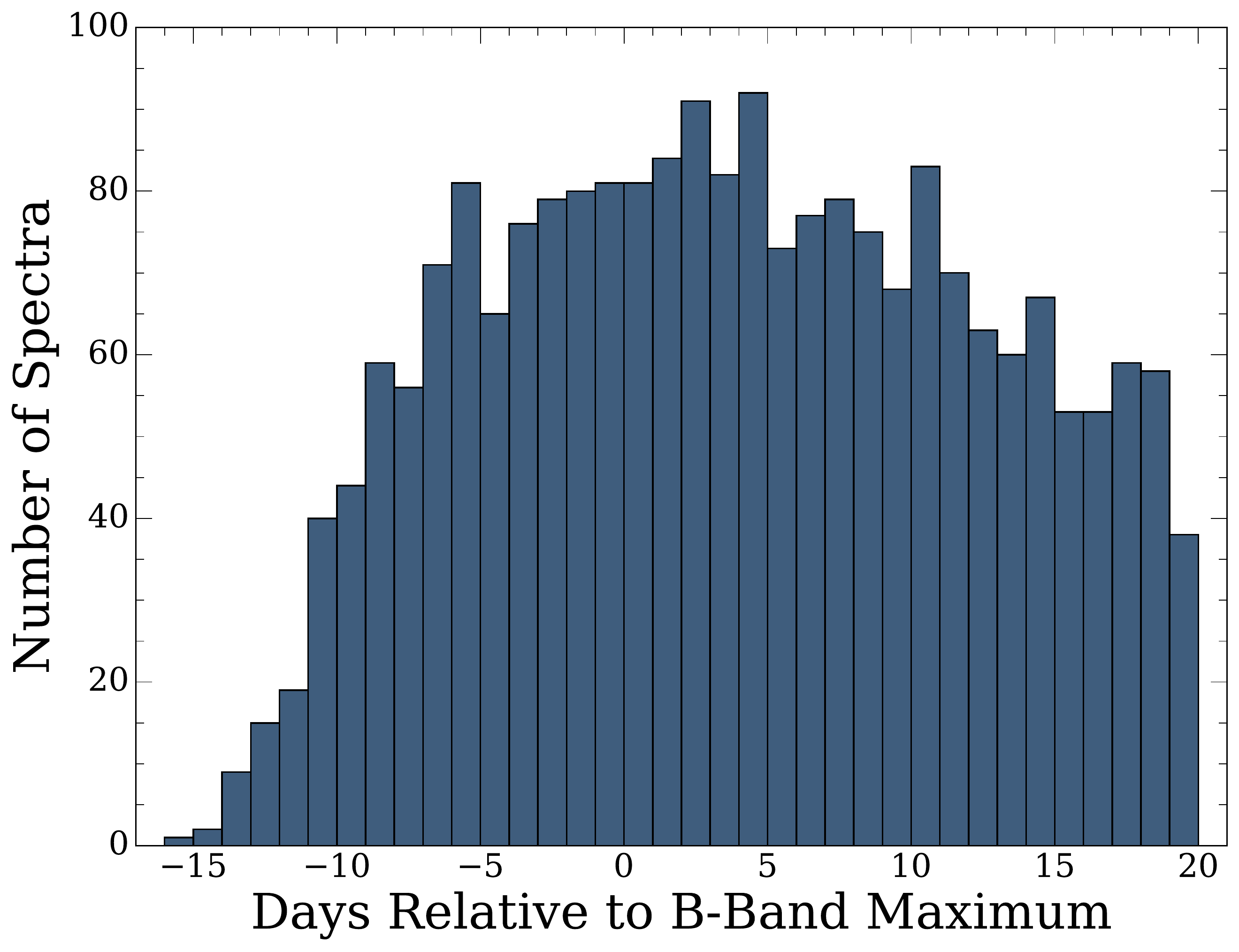}
  \caption{Histogram of the number of individual SN spectra at each epoch
within 20 days of maximum brightness in the nominal sample.}\label{fig:dem3}
\end{figure}

\begin{figure}
    \includegraphics[width=3.2in]{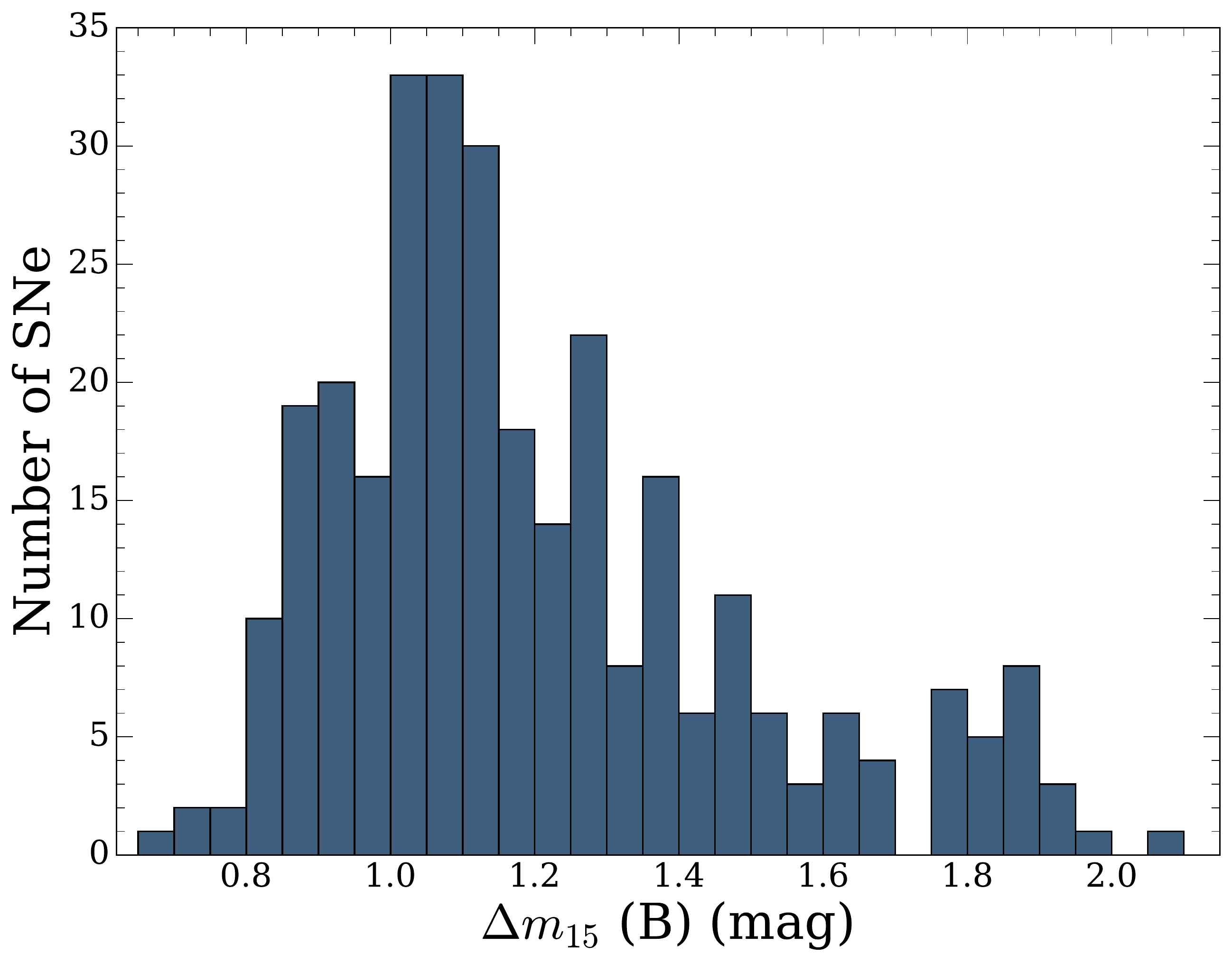}
  \caption{Histogram of the number of SNe~Ia per \dm\ bin. This distribution is skewed towards slower-declining events and peaks at \dm\ $ = 1.12$ mag (typical for normal SNe~Ia.)}\label{fig:dem4}
\end{figure}
\begin{figure}
    \includegraphics[width=3.2in]{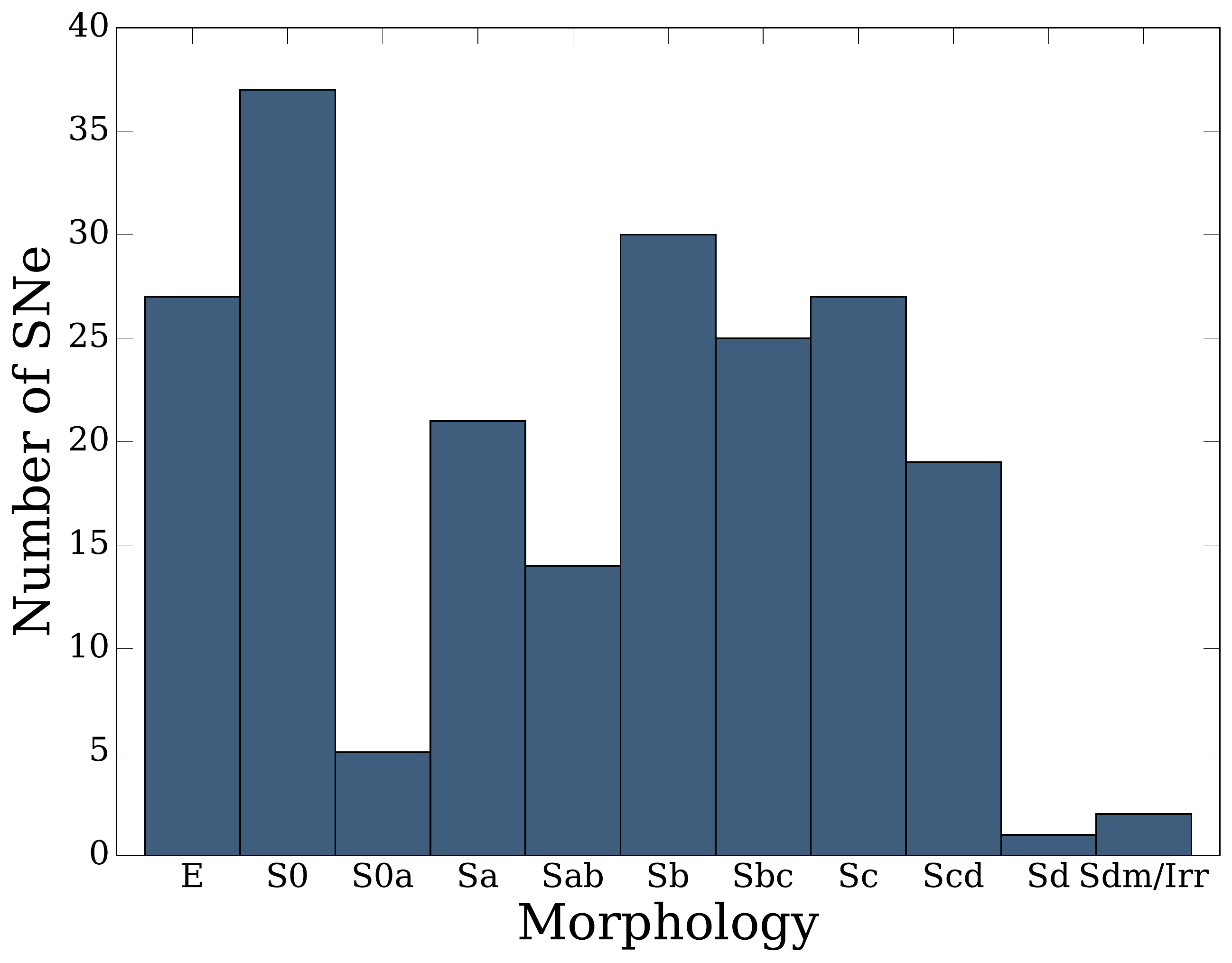}
  \caption{Histogram of the number of SNe~Ia per bin in host galaxy morphology.}\label{fig:dem5}
\end{figure}
\begin{figure}
    \includegraphics[width=3.2in]{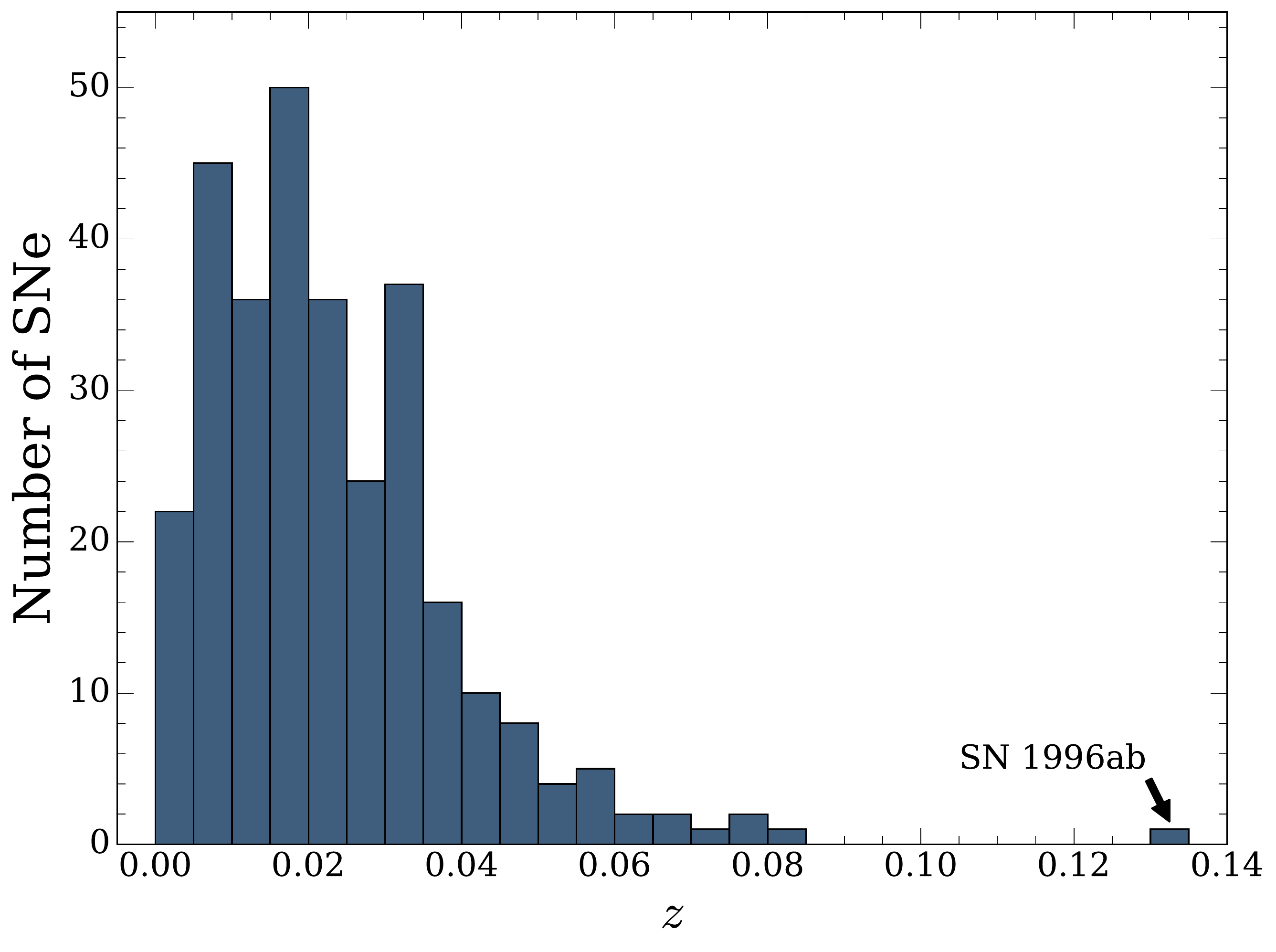}
  \caption{Histogram of the number of SNe~Ia per redshift bin.}\label{fig:dem6}
\end{figure}
\begin{figure}
    \includegraphics[width=3.2in]{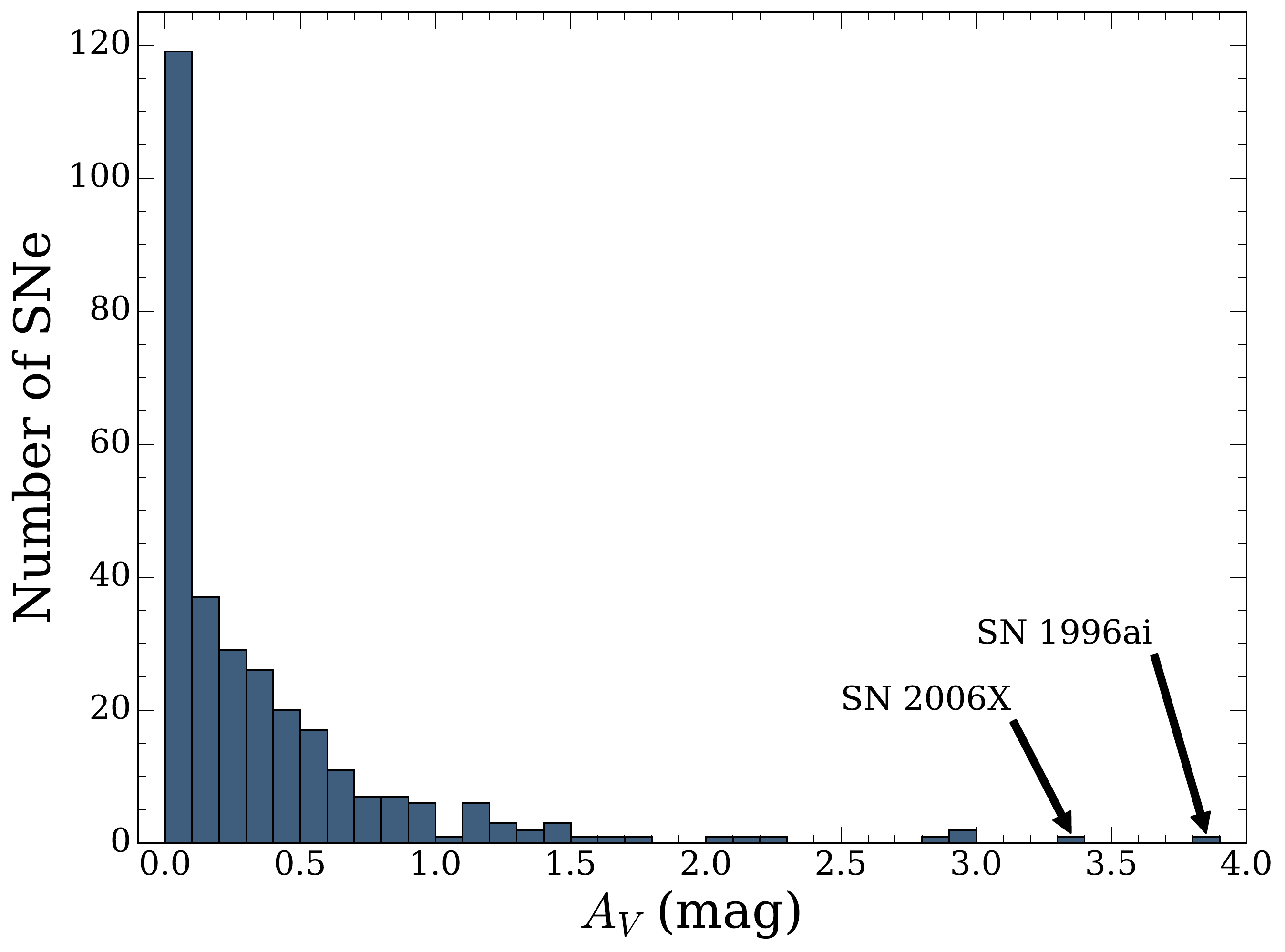}
  \caption{Histogram of the number of SNe~Ia per host-galaxy extinction bin.}\label{fig:dem7}
\end{figure}
\begin{figure*}
\begin{center}
\includegraphics[angle=0,width=6.8in]{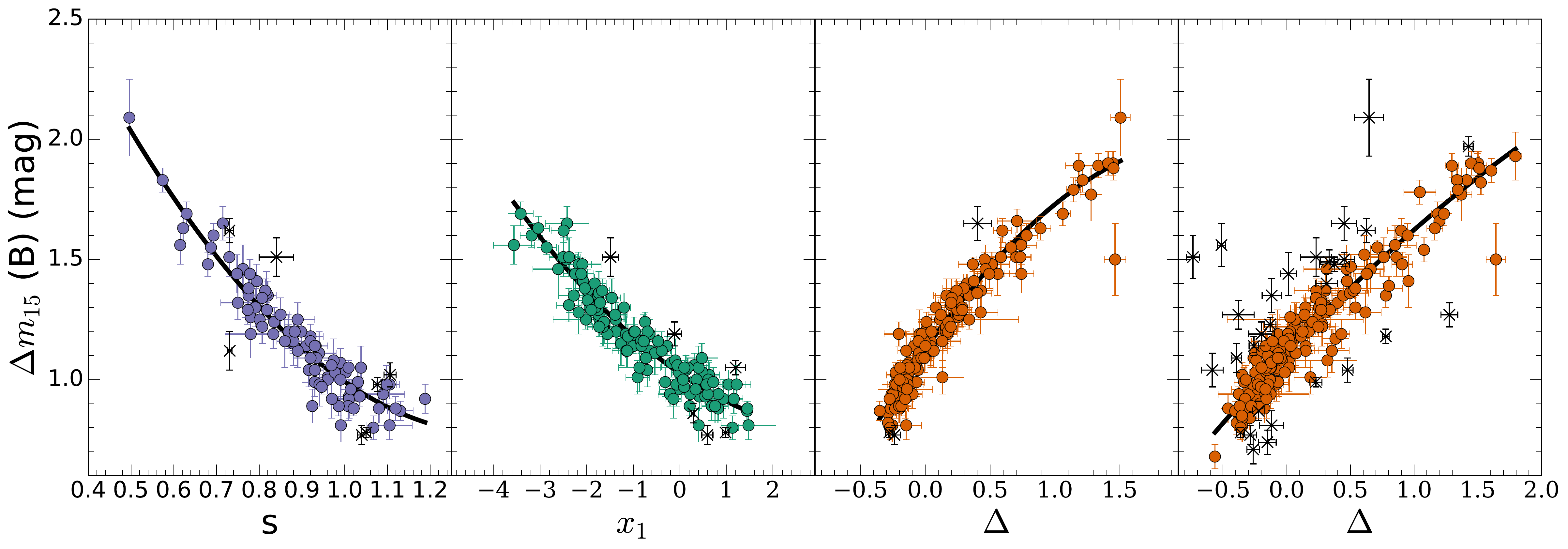}
\caption{Power-law relationships between \dm\ and the light-curve shape parameters $s$ (SALT; \textit{first}), $x_1$ (SALT2; \textit{second}), $\Delta$ (MLCS using $R_V=3.1$; \textit{third}), and $\Delta$ (MLCS using $R_V=2.5$; \textit{fourth}). We use these relationships to determine an estimate of \dm\ when there is no explicit measurement available. In all panels, points marked with a black ``x'' are $3$-$\sigma$ outliers and are not included when fitting the relation.}\label{fig:dm15_stretch}
\end{center}
\end{figure*}

Some of the characteristics of the nominal sample are shown in Figures \ref{fig:dem1}  $-$ \ref{fig:dem7}. All SNe in the sample are at low-redshift (median $z=0.02$), vary greatly in their light-curve shapes ($0.68 \leq$ \dm\ $\leq 2.09$ mag), and occur in a diversity of host environments. For SNe~Ia without an explicit measurement of \dm, we estimate \dm\ based on its power-law relationships with SALT, SALT2 and MLCS31 shape parameters (Figure \ref{fig:dm15_stretch}). $s$, $x_1$, and $\Delta$ are taken from tables 1, 2, and 3 of \citealt{hicken09}. We also include $\Delta$ estimates from another set of MLCS fits using $R_V = 2.5$ \citep{jha07}. For SNe that have been fit by more than one of these methods, we determine \dm\ from the power-law relationship with the lowest on outlier fraction and RMS uncertainty. Using a consistent sample of 85 SNe among fitting methods, we estimate the outlier fractions of SALT, SALT2, MLCS31, and MLCS25 to be 0.082, 0.035, 0.024, and 0.071 respectively, and the RMS uncertainties to be 0.071, 0.066, 0.070, and 0.074 respectively. Thus, we choose to give MLCS31 highest priority, followed by SALT2, SALT, and then MLCS25 for determining \dm\ from a given light-curve shape parameter.

\section{Data Homogenization}
The heterogeneous nature of spectral observations and reduction techniques has been a significant barrier to studying the bulk properties of SN~Ia spectra across multiple sources. We have developed a framework to process SN spectra in order to facilitate analyses of large data sets. Prior to storing a spectrum in the database, several steps are taken to ensure homogeneity. These processing steps are applied to every spectrum in the database in addition to the nominal sample. 

\subsection{Variance Spectrum Generation}\label{sec:vargen}
\begin{figure}
\includegraphics[angle=0,width=3.2in]{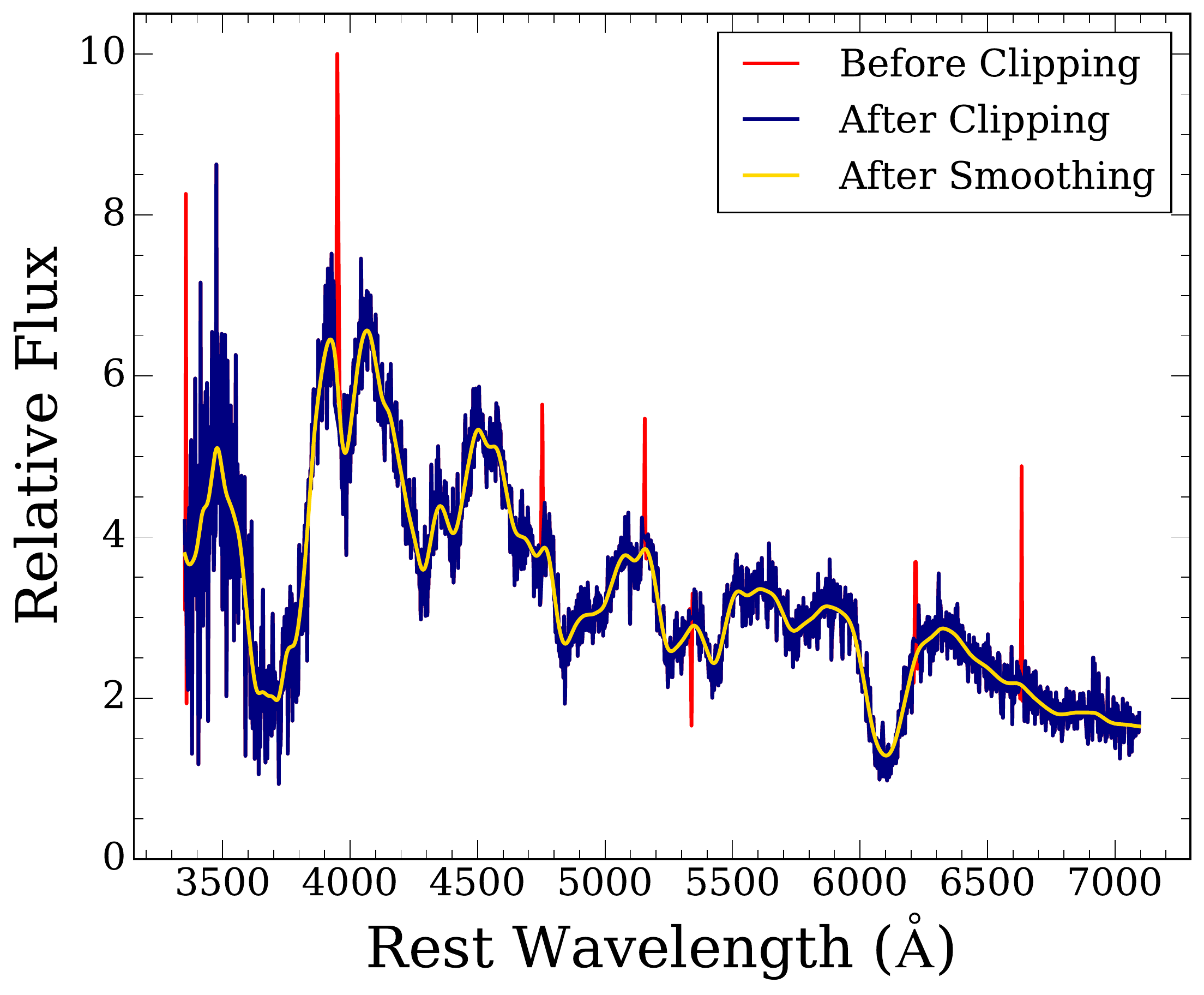}
\caption{Spectra SN~2005lz before (red curve) and after (blue curve) removing unwanted features using the method described in section \ref{sec:clip}. The smoothed spectrum (with $d\lambda / \lambda = 0.004$; yellow curve) is used to estimate a preliminary uncertainty in order to determine the clipping threshold.}\label{fig:rep1}
\end{figure}

\begin{figure}
\includegraphics[angle=0,width=3.2in]{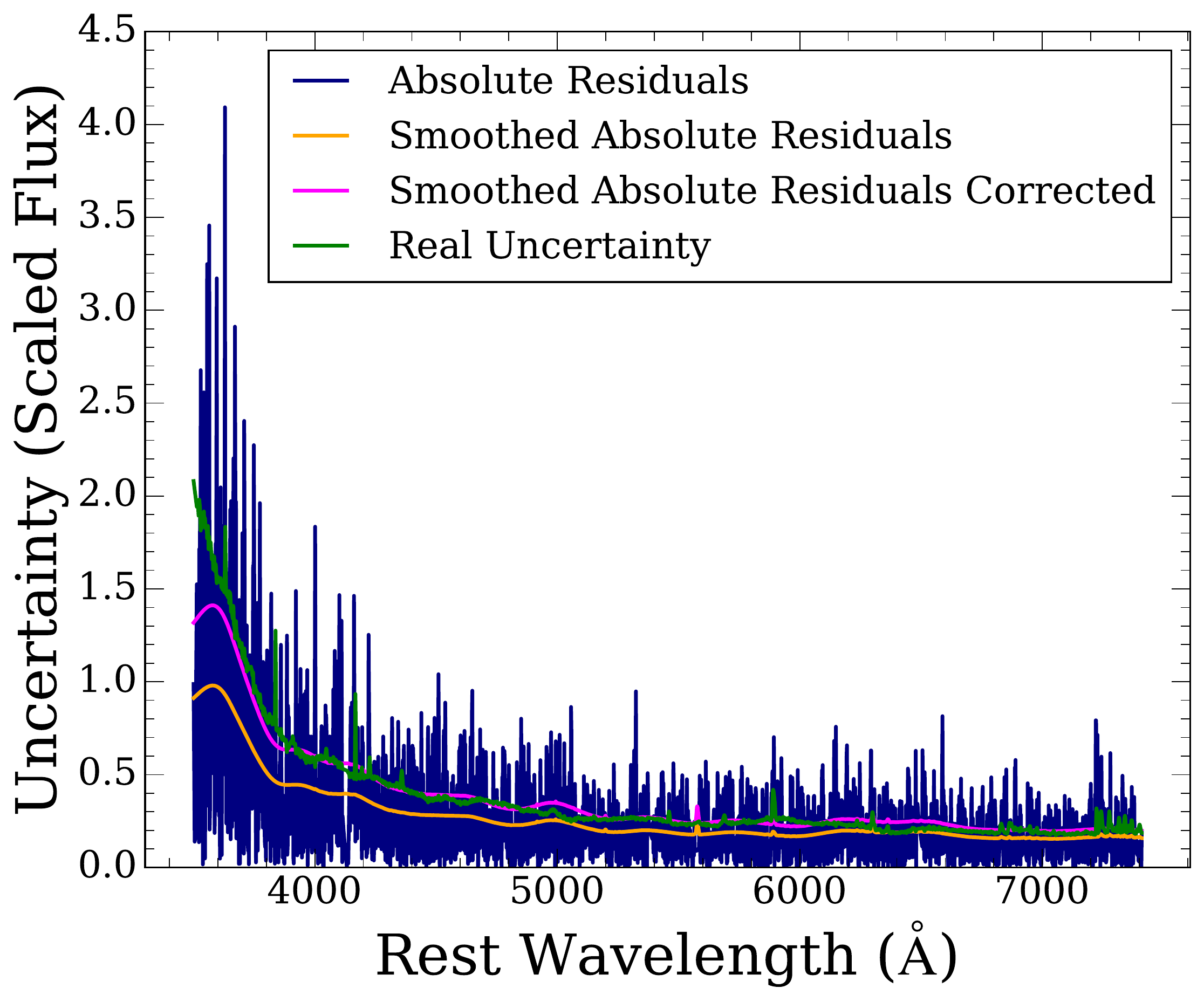}
\caption{Uncertainty spectra corresponding to the spectrum of SN~2005lz presented in Figure \ref{fig:rep1}. The blue curve is the absolute residual spectrum from the difference between the clipped spectrum and the smoothed spectrum, the green curve is the real uncertainty spectrum from the CfA archive, the yellow curve is the smoothed absolute residual spectrum, and the pink curve is the same spectrum after applying a small scaling in order to better match the green curve.}\label{fig:rep2}
\end{figure}

For many spectra, we do not have a corresponding variance spectrum. We illustrate the procedure we use to generate an approximate variance spectrum in Figures \ref{fig:rep1} and \ref{fig:rep2}. First, we create a smoothed spectrum using the inverse-variance Gaussian smoothing algorithm from \citealt{blondin06}. In this work, \citealt{blondin06} define a ``smoothing factor" ($d\lambda / \lambda$) that accounts for the intrinsic broadening of spectral features due to the large expansion velocities. Since our spectra vary greatly in S/N, one value of $d\lambda / \lambda$ is not sufficient to adequately smooth every spectrum. Using a sample of spectra with varying S/N, we empirically determined the best value of $d\lambda / \lambda$ for each spectrum. We fit a linear function of S/N to these data:
\begin{equation}
d\lambda / \lambda =
\begin{cases} 
      0.0045 & S/N < 2.5 \\
      4.61\cdot 10^{-3} - 4.52\cdot 10^{-5}*S/N & 2.5 \leq S/N \leq 80 \\
      0.001 & S/N > 80 
\end{cases}
\label{eq:smoothing}
\nonumber
\end{equation}
We apply constant values of $d\lambda / \lambda$ at the extremes of our S/N range to prevent over-/under-smoothing of the data. The yellow curve in Figure \ref{fig:rep1} shows a smoothed spectrum of SN~2005lz using this method. We then subtract the smoothed spectrum from the original spectrum (blue curve), take the absolute value, and smooth again using $d\lambda / \lambda = 0.015$ (orange curve in Figure \ref{fig:rep2}). We account for possible uncertainty from sky emission by scaling a template sky emission spectrum to to our smoothed absolute residual spectrum, multiplying by a constant factor, and adding this to our smoothed absolute residual spectrum. We chose a value for the constant factor that provides a reasonable match to the uncertainty due to sky emission seen in the CfA spectra. 

We compare the generated approximate uncertainty spectra to the associated uncertainty spectra of the CfA sample. We find that we consistently underestimate the uncertainty of the CfA sample at bluer wavelengths (see Figure \ref{fig:rep2}). To quantify this effect, we divide every CfA uncertainty spectrum by the uncertainty spectrum that we generate from the flux. We then fit a linear function of wavelength to each of these ratio spectra. For spectra without archival uncertainty spectra, we apply the mean fit as a wavelength-dependent scaling to each uncertainty spectrum in order to account for this discrepancy. After making this scaling correction, we obtain an uncertainty spectrum (pink curve) that closely matches the CfA uncertainty spectrum (green curve), especially at bluer wavelengths. 

\subsection{Removal of Residual Sky Lines, Cosmic Rays, and Galaxy Emission Lines} \label{sec:clip}
We attempt to remove any residual night sky lines, cosmic ray hits, and host-galaxy emission lines that are still present.  We generate a smoothed absolute residual spectrum via the same method in Section \ref{sec:vargen}. We interpolate the flux and the variance over regions where the residual spectrum exceeds this smoothed residual spectrum by ${>}5\sigma$, where $\sigma$ is the local scatter obtained from the variance spectrum. We also increase the threshold for clipping of negative residuals to $10\sigma$ in regions where absorption lines from Ca H\&K, Na I D, \ion{K}{1} $\lambda 7687$, and the 5780~\AA\ diffuse interstellar band may exist. The effects of this method are shown in Figure \ref{fig:rep1}. The narrow emission features (red) have been removed resulting in the blue curve.  

\subsection{Reddening Corrections, Deredshifting, and Re-binning}
We correct each flux and variance spectrum for MW-reddening using the Schlafly and Finkbeiner reddening map \citep{S&F}, a \citealt{Fitzpatrick} (F99) reddening law, and $R_V = 3.1$. We then deredshift the spectrum, rebin at equally spaced wavelength values ($\Delta \lambda = 2$~\AA) using a flux-conserving interpolation algorithm, and store the re-binned spectrum in the database. Since interpolation introduces unwanted correlation in our variance spectra, we scale our original variance spectrum to the rebinned variance spectrum. When available, we use MLCS reddening parameters to correct for the distortion introduced by host-galaxy reddening. Since there is a larger degree of uncertainty associated with this step, the non-corrected spectra are stored in the database and we perform this correction during our analysis.

\section{Generating Composite Spectra}
The large sample necessitates a reliable way to visualize and validate the data and our processing techniques. Composite spectra provide a reasonable solution to this problem. They attempt to provide the ``most representative'' average of a given subset of data. We use this tool to reproduce the known spectral trends in SNe~Ia in order to verify that our sample is of high quality. By maximizing S/N and reducing the impact of outliers, composite spectra make it easy to analyze the differences between complex spectral feature shapes.  This is an advantage over measurements of derived quantities from individual spectra like velocities, EWs, and line ratios, which significantly reduce the available information and may only apply to a subset of the data (e.g., not all spectra have obvious carbon absorption).

We show that composite spectra in combination with relational queries with \code{kaepora} provide for a fast means of validating known relationships and testing different hypotheses. We always find it useful to follow up potential trends seen in composite spectra with an analysis of the individual underlying spectra.

Here we describe our method for generating composite spectra from subsets of our nominal sample. The basic algorithm is adapted from the methods used in \citealt{foley08} and \citealt{foley12}. Reference these works for a more detailed description of the potential systematic errors introduced using these techniques. We require $A_V < 2$ for each SN included in the composite spectra generated for this work.

\subsection{Algorithm} \label{sec:compalg}
When combining spectra from a subsample of SNe, we make an effort to ensure that the selected sample of SNe are representative of their respective populations. Multiple spectra from a single SN may satisfy a given query and in this case, we first create individual inverse-variance weighted composite spectra for each of these SNe using the methods described below. If a spectrum does not overlap with any other spectra from the same SN we attempt to add it in during the next step (if there is still no overlap, the spectrum is ignored). Once the set of spectra (now including combined spectra) is selected, we scale the flux of all spectra to the flux of the spectrum with the largest wavelength range (the preliminary template spectrum) by minimizing the sum over wavelength of the absolute differences between each spectrum and this template spectrum. A new template spectrum is then created by averaging these spectra weighted by their inverse variance (or taking the median). Spectra in the original sample that did not have overlap with the largest wavelength spectrum may potentially overlap with this new template spectrum. This process is repeated until all possible spectra are included in the composite spectrum. The $50$~\AA\ at the beginning and end of each spectrum are ignored when constructing composite spectra in order to suppress any systematic errors that frequently occur in these regions. 

It is often the case that one or a few high-S/N spectra will strongly influence the overall continuum and spectral features of the composite spectrum. This is a larger problem for smaller sample sizes where the median may be a more appropriate representation of the underlying spectra. Highly unequal weight distributions also tend to cause discontinuities in the composite spectrum at wavelengths where highly weighted spectra end. These issues can be resolved by assigning equal weights to each spectrum at the expense of S/N of the composite spectrum. We re-weight using Gini coefficients to mitigate this issue. This solution is a middle-ground approach that avoids discontinuities while attempting to still maximize S/N. 

In this case, the Gini coefficient is used to measure how much the distribution of weight deviates from complete equality. It ranges from 0 to 1, where 0 means complete equality and 1 means complete inequality. Therefore in the context of combining spectra, a Gini coefficient of 0 indicates equal weight given to each spectrum (i.e., the average) and a gini coefficient of 1 indicates that a single spectrum carries all of the weight. Equation \ref{eq:gini} defines the Gini coefficient as half of the relative mean absolute difference. For our purposes, $w_i$ represents the total weight (sum of inverse variance) given to an individual spectrum in some wavelength range, and $N$ is the number of spectra contributing to the composite spectrum in that same wavelength range. 
\begin{equation}
G = \frac{\sum\limits_{i=1}^N \sum\limits_{j=1}^N |w_i - w_j|}{2N \sum\limits_{i=1}^N w_i}
\label{eq:gini}
\end{equation}
We separate the composite spectrum into $1000$~\AA\ regions and measure $G$. If $G > 0.6$ in any of these wavelength bins, we subsequently de-weight the individual spectrum that carries the most weight across its whole wavelength range by a constant factor. This process is iterated until $G < 0.6$ in all wavelength bins. We chose a value of $G < 0.6$ as our threshold because it tends to provide a reasonable compromise between the average and the inverse-variance weighted average. This method allows us to achieve better S/N in our composite spectrum than the average without a strong bias towards the highest S/N spectra. This tends to result in a smooth composite spectrum that is more representative of the underlying data.

\begin{figure}
\includegraphics[angle=0,width=3.2in]{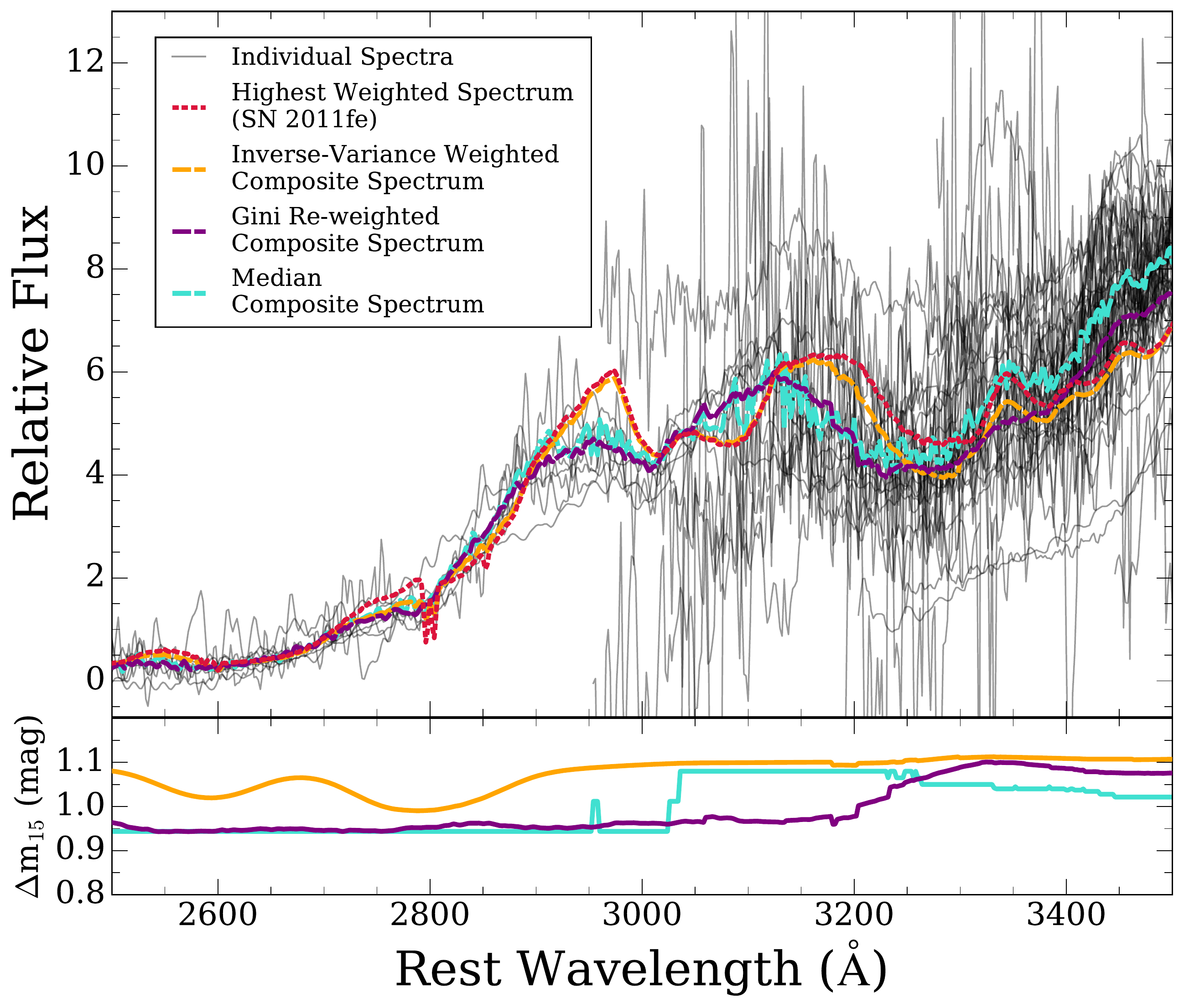}
\caption{(\textit{top panel}): Composite spectra generated from a subsample (black curves) using different methods. The red dashed curve is a high S/N spectrum of SN 2011fe. This spectrum carries the most weight in this wavelength range. The cyan curve is the median composite spectrum, the yellow curve is the inverse-variance weighted composite spectrum, and the purple curve is the Gini-weighted composite spectrum. The Gini-weighted composite spectrum is less influenced by SN 2011fe than the inverse-variance weighted composite spectrum while maintaining a larger S/N that the median composite spectrum. (\textit{bottom panel}): The average value of \dm\ as a function of wavelength for each composite spectrum.}\label{fig:gini}
\end{figure}

We demonstrate the effects using different methods to construct composite spectra in Figure \ref{fig:gini}. Here we have generated composite spectra from a sample of 180 spectra of 82 total SNe. These spectra have phase between $-4$ and $0$ days and \dm\ $<1.25$ mag. The black curves are individual spectra (or combined spectra from the same SN) of varying S/N, the red dashed curve is a high-S/N spectrum SN 2011fe, and the other colored curves are composite spectra generate from these individual spectra using three different methods; medianed (cyan), inverse-variance weighted (yellow), and ``Gini-weighted" (purple). As expected, the inverse-variance weighted composite spectrum (yellow curve) is strongly influenced by the high S/N spectrum of SN 2011fe. We find that the median composite spectrum works well in this wavelength range and the Gini-weighted composite spectrum is consistent with the median, but with higher S/N. We also see that as $N$ increases, the properties of the Gini-weighted composite spectrum approach the inverse-variance weighted composite spectrum. The average gini coefficients of the inverse-weighted composite spectrum and the Gini-weighted composite spectrum in this wavelength range are 0.86 and 0.57 respectively. The effective value of \dm\ for our Gini-weighted composite spectrum is consistent with the median at wavelengths ${<}2950$~\AA\ and then smoothly approaches the inverse-variance weighted value as $N$ increases. This is expected since SN 2011fe carries less overall weight as more spectra contribute to each weighted composite spectrum. Since we have demonstrated that Gini-weighted composite spectra are more reliable for sample sizes that vary with wavelength, the composite spectra presented in this work are constructed using this method unless specified otherwise. 

\subsection{Uncertainty of Composite Spectra} \label{sec:var}
To understand the subtle differences between composite spectra we need an estimate of the uncertainty at each wavelength. We implement a bootstrap resampling with replacement algorithm in order to estimate the variation about the average spectrum. 

Although not completely independent, the bootstrapping uncertainty is different from the root-mean-square error (RMSE) in several ways. The bootstrapping method assumes that the given sample probes the diversity of the underlying population. This diversity is manifested in the bootstrapping uncertainty by taking random resamples of the original set of spectra. This means that composite spectra with high S/N can still have large bootstrapping uncertainty if the underlying population is diverse. Therefore, the bootstrapping uncertainty provides information about the reproducibility of the average spectrum. The RMSE is typically smaller than the bootstrapping uncertainty and less affected by the diversity of the underlying population. At any given wavelength, the bootstrapping uncertainty may be estimated from the full distribution (and therefore asymmetric about the average spectrum) while the RMSE is defined as being symmetric about the average spectrum. 

The uncertainty that we observe is a combination of several effects. The blue end of our composite spectra are primarily dominated by the intrinsic variation of the sample which increases towards the UV \citep{foley08}, Poisson noise because we have fewer spectra contributing to the average, and the typically low detector responses in this region. We correct the continuum for host-galaxy extinction using an estimate of $A_V$ from an MLCS2k2 fit \citep{jha07}. Since we normalize our spectra by minimizing absolute differences in the overlap region with the template spectrum (see Section \ref{sec:compalg}), we effectively normalize near $5000\ $ \AA\ on average. Therefore, spectra with anomalous continua can cause more scatter farther away from this central wavelength. Since host-galaxy extinction corrections modify the continua of our spectra, we expect that the uncertainty at the red end of a composite spectrum is dominated by imperfect corrections for reddening.

\subsection{Example Composite Spectrum}
\begin{figure*}
\begin{center}
\includegraphics[width=6.1in]{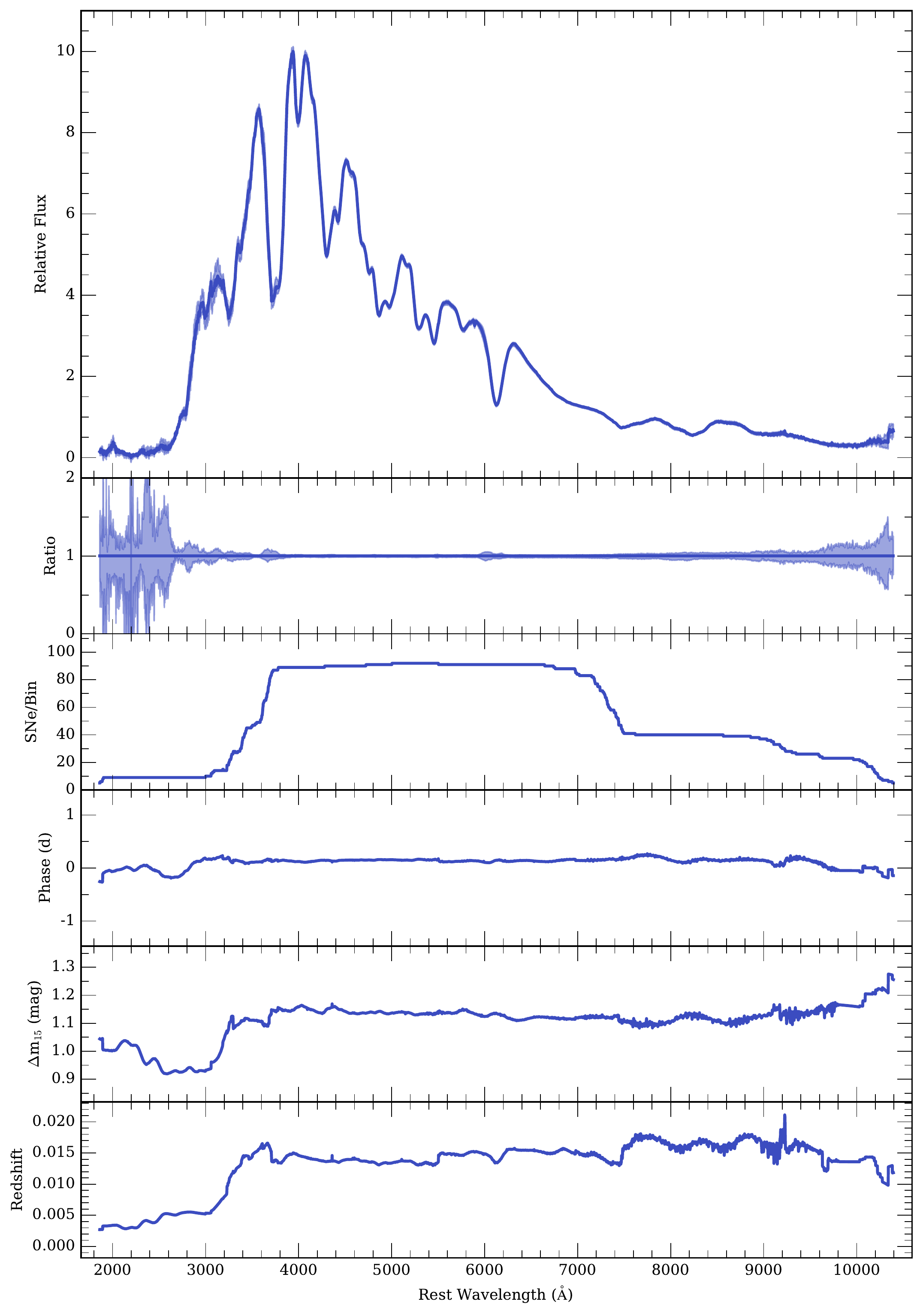}
\caption{(\textit{first panel}): Maximum-light composite spectrum created from our nominal sample consisting of spectra from 145 spectra of 96 SNe ($-1 < \tau < +1$ days, and \dm\ $< 1.8$ mag). The light-blue region is the $1 \sigma$ bootstrap sampling uncertainty. (\textit{second panel}): The light-blue region is the ratio of the $1 \sigma$ bootstrap sampling uncertainty relative to the composite spectrum. (\textit{third panel}): The number of individual spectra contributing to each wavelength bin. For this composite spectrum, a minimum of 5 and a maximum of 96 spectra contribute at any given wavelength. (\textit{fourth panel}): The average phase relative to maximum brightness as a function of wavelength for this composite spectrum. (\textit{fifth panel}): The average value of \dm\ as a function of wavelength for this composite spectrum. (\textit{sixth panel}): The average redshift of the composite spectrum as a function of wavelength.}\label{fig:max}
\end{center}
\end{figure*}

In Figure \ref{fig:max} we present a Gini-weighted composite spectrum constructed from 142 maximum-light spectra ($-1 < \tau < +1$ days, and \dm\ $< 1.8$ mag) of 96 SNe Ia. This composite spectrum has an average phase of $0.11$ days, an average \dm\ $=1.11$ mag, and an average $z=0.013$. In the first panel, the blue curve is the relative flux and the blue-shaded region is the $1 \sigma$ bootstrap resampling region. Our composite spectrum covers a wavelength range where we require at least 5 contributing SNe. We exclude the fastest declining events (\dm\ $< 1.8$ mag) because there continua are often far redder than the rest of the sample. Overall, the flux and the average properties of our maximum-light composite spectrum are very similar to those of a normal SN Ia. 

\begin{figure}
\includegraphics[width=3.2in]{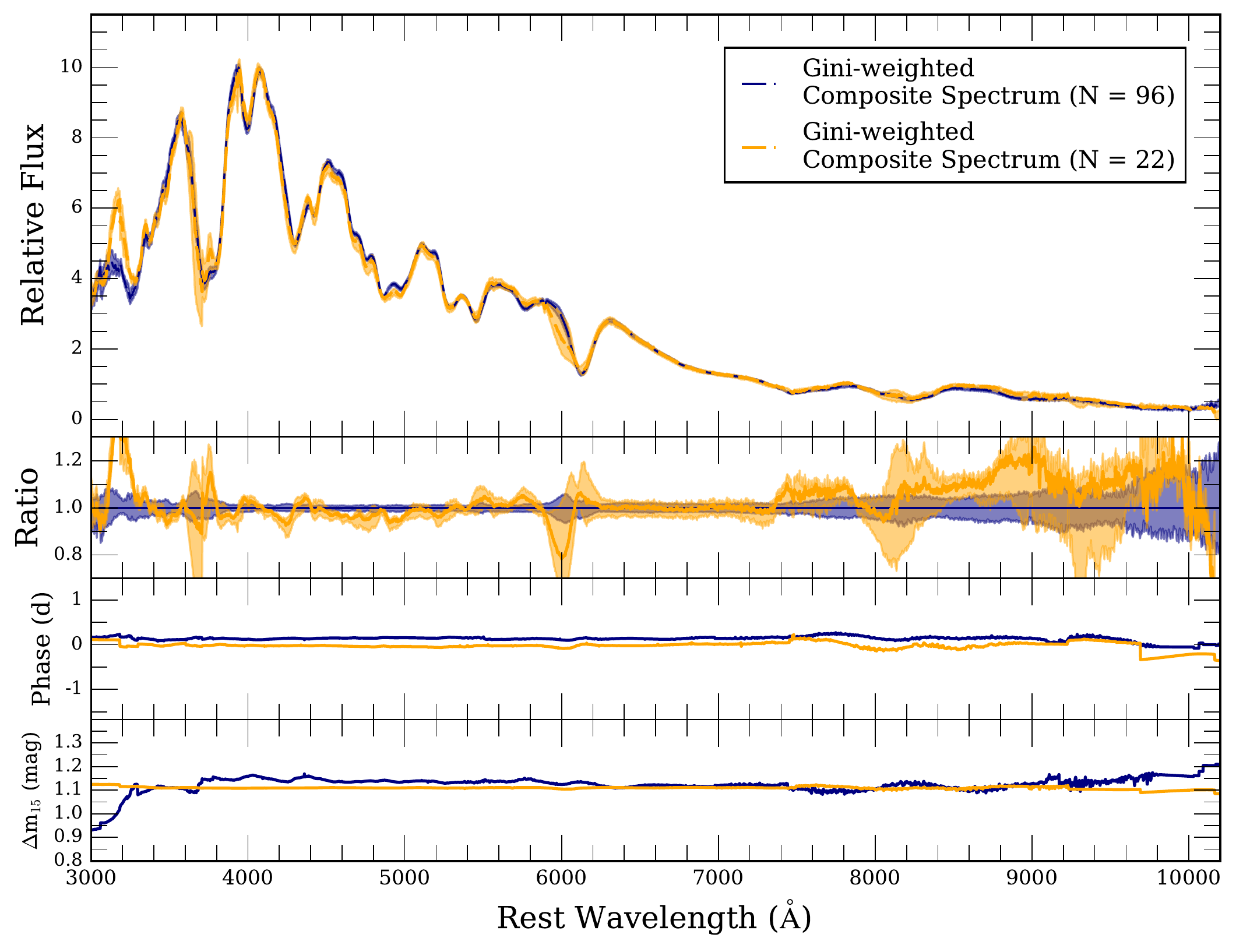}	
\caption{(\textit{first panel}): The maximum-light composite spectrum from Figure \ref{fig:max} (blue curve) compared to a maximum-light composite spectrum constructed from a subsample of 15 SNe~Ia with $-1 < \tau < +1$ days and $1.05 < $\dm\ $< 1.15$ mag (yellow curve). The shaded regions are the $1 \sigma$ bootstrap sampling uncertainties of the respective composite spectra. (\textit{second panel}): The ratio of the given composite spectra and $1\sigma$ bootstrap sampling  regions to the larger sample size composite spectrum. (\textit{third panel}): The average phase relative to maximum brightness as a function of wavelength for a given composite spectrum. (\textit{fourth panel}): The average value of \dm\ as a function of wavelength for a given composite spectrum.\label{fig:samp_size}}
\end{figure}

\begin{figure}
\includegraphics[width=3.2in]{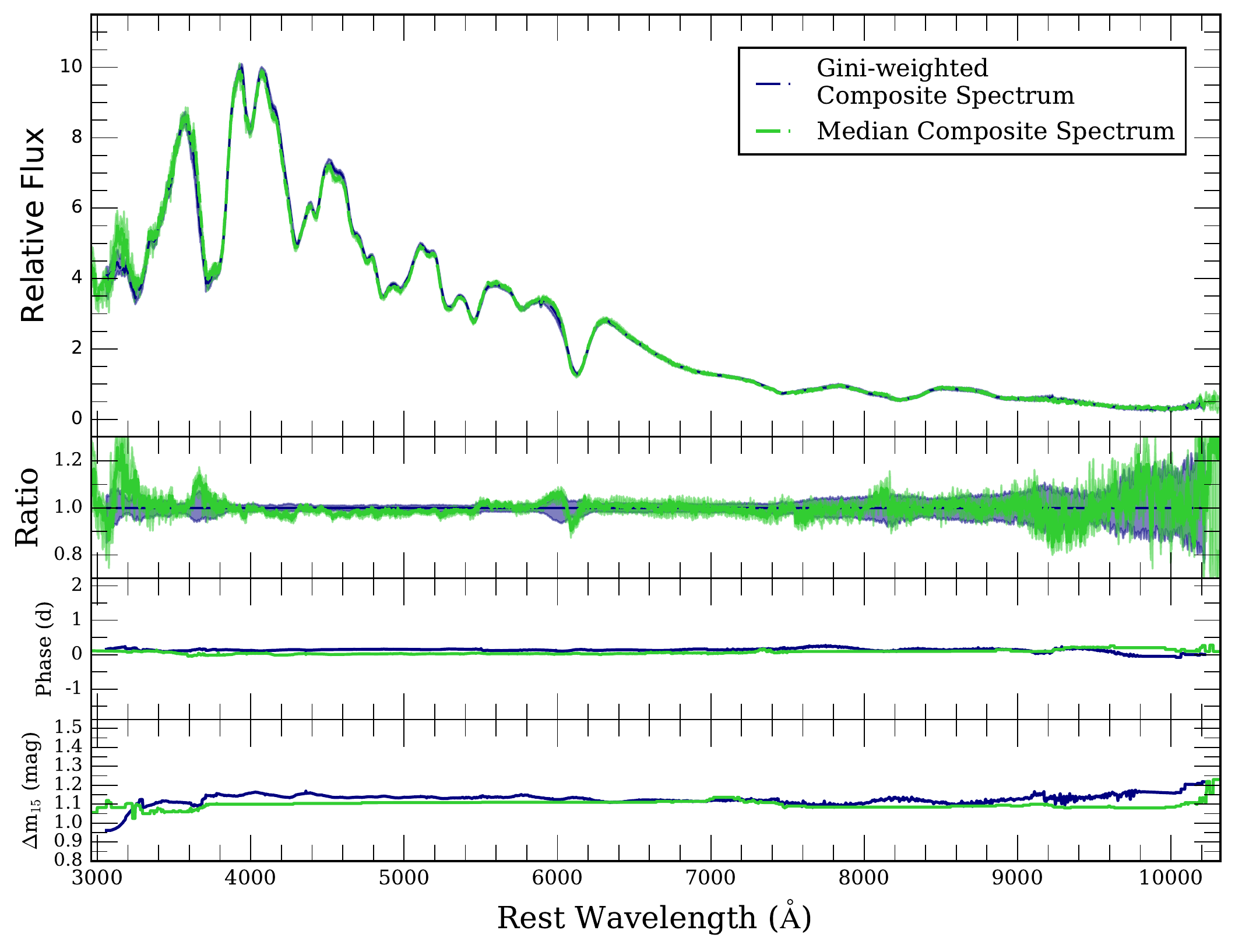}	
\caption{A comparison between Gini-weighted composite spectra and median composite spectra constructed from the subsample of 96 SNe~Ia with $-1 < \tau < +1$ days and \dm\ $<1.8$ mag (blue and green curves respectively). The panel format is the same as Figure \ref{fig:samp_size}. \label{fig:med_large}}
\end{figure}

The second panel contains the $1 \sigma$ bootstrap contours divided by the relative flux of the composite spectrum. This fractional uncertainty is largest below $3200$~\AA. The uncertainty also gradually increases redward of ${\sim} 6000$~\AA. The fractional uncertainty is smallest between these regions because this is both the region where most spectra overlap, and the region where most spectra are effectively normalized. There is a slight correlation of the fractional uncertainty with the Si II and Ca II absorption features. This correlation is likely caused by the variability of line velocities in the underlying sample.

Since the wavelength ranges of individual spectra can vary greatly, the number of of contributing events varies with wavelength. The third panel of the figure shows the number of SNe contributing to the composite spectrum as a function of wavelength. Since the CfA could obtain spectra frequently, it is more likely that to have spectra within a small phase window than other surveys. Therefore, a large portion of this subsample are spectra obtained with the FAST spectrograph. This instrument has a wavelength coverage of ${\sim} 3700 - 7600$~\AA\ \citep{fabricant98}. Thus the number of spectra contributing to the composite spectrum reaches its maximum in this wavelength range. 

 Additionally, since the variance of individual spectra vary with wavelength, the average properties of composite spectra can change with wavelength. Panels $4-6$ of Figure \ref{fig:max} show the variation of phase, \dm, and redshift with wavelength. There is a large shift in the average properties from the UV to the optical part of our composite spectrum. This change is due to the fact that UV spectra tend to be from intrinsically bright and nearby events. Overall, there is not much variation in these average properties in the optical, but there does seem to be some slight correlation with the stronger spectral features. It is also important to note our maximum-light composite spectrum is subject to Malmquist bias. However, analyses of subsamples with comparable \dm\ should not be as heavily impacted by this effect.

We find that composite spectra with similar average properties tend to look remarkably similar regardless of sample size. In Figure \ref{fig:samp_size} we compare our maximum-light composite spectrum  ($N$ = 96; blue curve) to a maximum-light composite spectrum constructed from a smaller sample of SNe ($N$ = 22) with the same average phase and \dm\ ($-1 < \tau < +1$ days and $1.05 < $\dm\ $< 1.15$ mag). This composite spectrum has an average phase of $0.13$ days, an average \dm\ $=1.09$ mag, and an average $z=0.015$. These properties are very similar to those of the composite spectrum constructed with a larger \dm\ window. Most of the differences are captured by the $1 \sigma$ bootstrap uncertainty regions and the largest differences occur in regions where the average light curve shapes differ between composite spectra (${\sim}3200$~\AA).

We also compare these Gini-weighted maximum-light composite spectra to  median composite spectra generated from the same subsamples (Figure \ref{fig:med_large}). Generally, the Gini-weighted composite spectrum is very similar to the corresponding median composite spectrum at all wavelengths and has higher S/N. The small differences tend occur at the large spectral features. The most discrepant regions are the Si II and Ca II absorption features, however the differences are consistent with the $1 \sigma$ bootstrap uncertainty region of the Gini-weighted composite spectrum and vice versa.

\section{Results and Analysis}
In this section, we use \code{kaepora} to generate subsets of our nominal sample with desirable average properties. We present composite spectra generated using our Gini-weighting method from these subsets that vary primarily with phase and light-curve shape. We demonstrate that our composite spectra reproduce known correlations between spectral and photometric properties. We obtain composite spectra with desired properties by making cuts on the certain parameter ranges. We often refer to this process as ``controlling for" a specific parameter. Finally, we investigate the spectral variation of SNe~Ia residing in different host-galaxy environments.

We compare our results to four sets of SN Ia template spectra and their respective color curves. The template spectra come from \citet{nugent02}, \citet{stern04}, \citet{hsiao07}, \citet{guy07}, and \cite{foley08} and will be referred to as the Nugent/Nugent-91T/Nugent-91bg, Hsiao, SALT2, and Foley template spectra respectively. These template spectra were developed with different scientific goals in mind. It is also important to note that the spectral samples used in constructing these spectral templates will inevitably overlap with some of the data in our nominal sample. It is difficult to know exactly which data are shared between template spectra, but we still expect to find similar results because of this overlap.

The Nugent and Hsiao template spectra were created for the purpose of determining $K$-corrections for samples of SNe~Ia. The Nugent template spectra are solely dependent on phase and use only local SNe~Ia. The Nugent template spectra are heavily influenced by SNe 1989B, 1992A, and 1994D. One of which has strong dust reddening (SN 1989B; \citealt{wells94}) and another of which has anomalous luminosity and colors (SN 1994D; \citealt{richmond95}; \citealt{patat96}). We include the Nugent template spectra in our comparisons but acknowledge that significant discrepancies may be caused by the weight of these atypical SNe. 

The Hsiao template spectra improve on the method for constructing spectral templates and provide a phase dependent spectral sequence for normal SNe~Ia.  The number of spectra contributing to these epochs ranges from 16 spectra at 85.6 days to 71 spectra at 0 days. Additionally, the spectral library used to construct these template spectra included a sample of high-$z$ SNe Ia. The total sample included 576 spectra of 99 SNe at $z<0.2$ and 32 spectra of 32 SNe at $z>0.2$ \citep{hsiao07}. While small compared to the total sample, the high-$z$ sample is restricted in phase and light-curve shape greatly influences the UV region of the Hsiao template spectra. The colors of the Hsiao template spectra are set to match the B-band template light curve from \citet{goldhaber01} and the UVR-band template light curves from \citet{knop03}.

SALT2 was developed to improve distance estimates of SNe Ia. This light-curve fitter models the evolution of the mean spectral energy distribution (SED) of SNe Ia along with its variation with light-curve shape and color \citep{guy07}. We are interested in improving our understanding of this multi-dimensional spectral surface with our composite spectra. The SALT2 model also makes use of high-$z$ SN data in order to constrain the UV continuum. In the sections below we compare the properties of our composite spectra to version 2.4 of the SALT2 spectral sequence model.

The Foley template spectra were created to investigate the potential evolution of SNe~Ia with redshift. The methods for constructing our composite spectra are similar to this work and make use of inverse-variance weighting to produce composite spectra. These composite spectra vary with phase, light-curve shape, and redshift. 

\subsection{Evolution of Phase-binned Composite Spectra}
\begin{figure*}
\begin{center}
\includegraphics[width=5.8in]{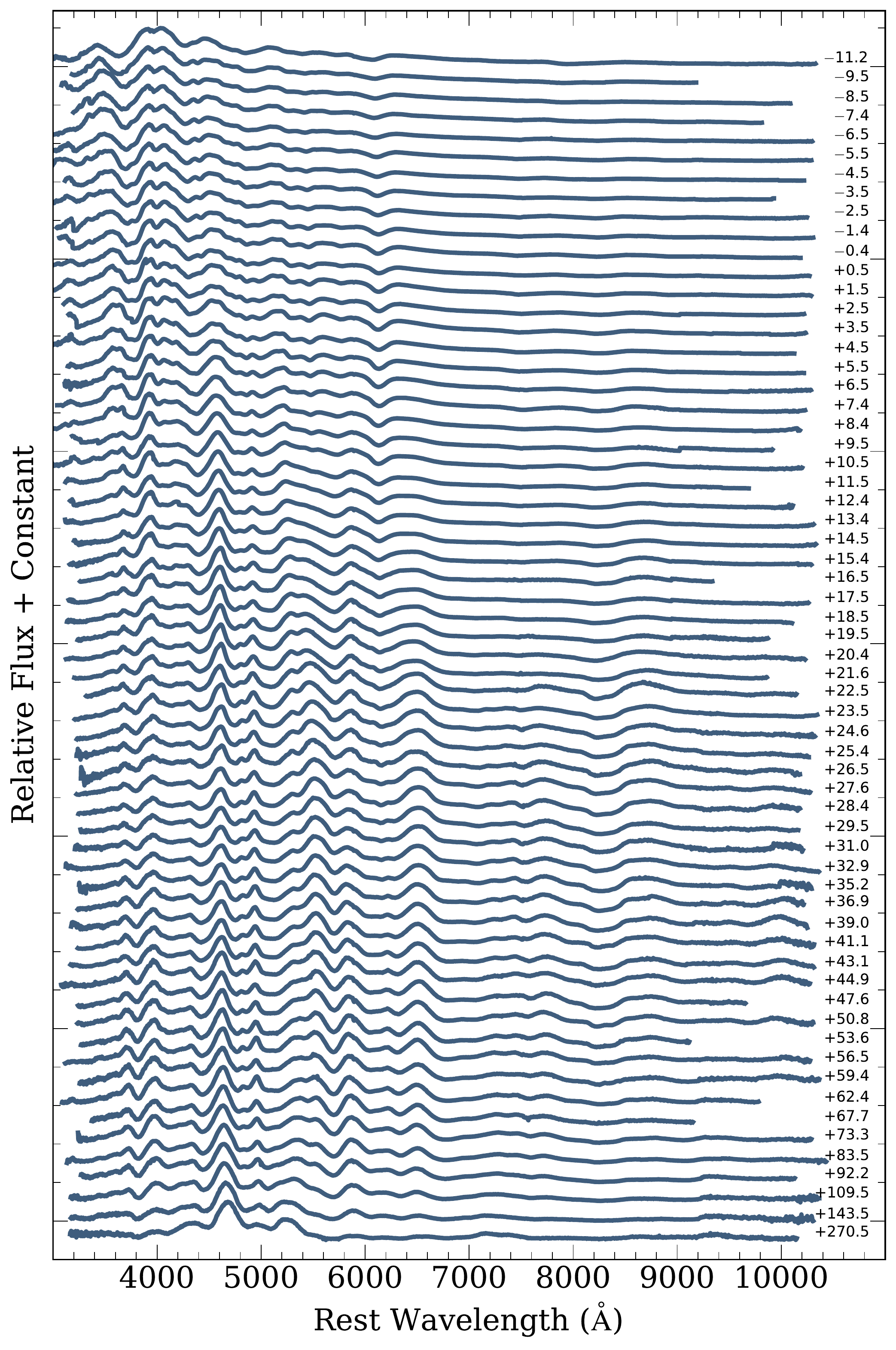}	
\caption{Composite spectra generated from the nominal sample showing the time evolution of spectral features of SNe~Ia. A time series is plotted with effective phases of $t = -11.3$ to $t = 271.2$ days relative to $B$ maximum brightness. Minimum bin sizes of 1 day are enforced near maximum brightness. Bin sizes are adjusted accordingly at later epochs to include a statistically significant ($N\geq 20$) sample for each composite spectrum.\label{fig:all_t}}
\end{center}
\end{figure*}

We first examine the spectral evolution of our composite spectra. In Figure \ref{fig:all_t}, we present 62 phase-binned composite spectra with phase ranging from $-11.2$ to $+270.5$ days. This set of composite spectra represents all of the spectra in our nominal sample. The time of maximum light of each SN is likely the largest source of error on phase (typical uncertainty $0.5$ days), therefore we do not assume that our composite spectra can differentiate spectral features with sub-day precision. We require a minimum bin size of 1 day and expand the size of the phase bin as needed as we probe later epochs. We require a minimum of 20 individual SNe per composite spectrum. We also require at least 5 spectra per wavelength bin and thus these composite spectra have differing wavelength ranges. For clarity, the $1 \sigma$ bootstrap uncertainty regions and other composite spectrum properties are are not displayed. However, each of these composite spectra can be decomposed in a similar way to our maximum-light composite spectrum in Figure \ref{fig:max}. 

The spectral features evolve smoothly with time and clearly show how the average SN Ia transitions from its photospheric phase, dominated by absorption from intermediate mass elements (IMEs), to the nebular phase, which is dominated by forbidden emission from iron-group elements (IGEs). There is a slight tendency to see small discontinuities near the edges of the composite spectra where $N$ is small and near ${\sim} 7600$~\AA\ (the red end of the CfA sample's wavelength coverage). The overall effect on the continuum is small and most often captured by the bootstrapping uncertainty region. Our composite spectra are comprised of spectra coming from a diversity of events, therefore, at any specific phase normal selection biases will apply. In particular, composite spectra for later phases are more likely to be influenced by intrinsically bright events.

\begin{figure*}
\begin{center}
\includegraphics[width=6.4in]{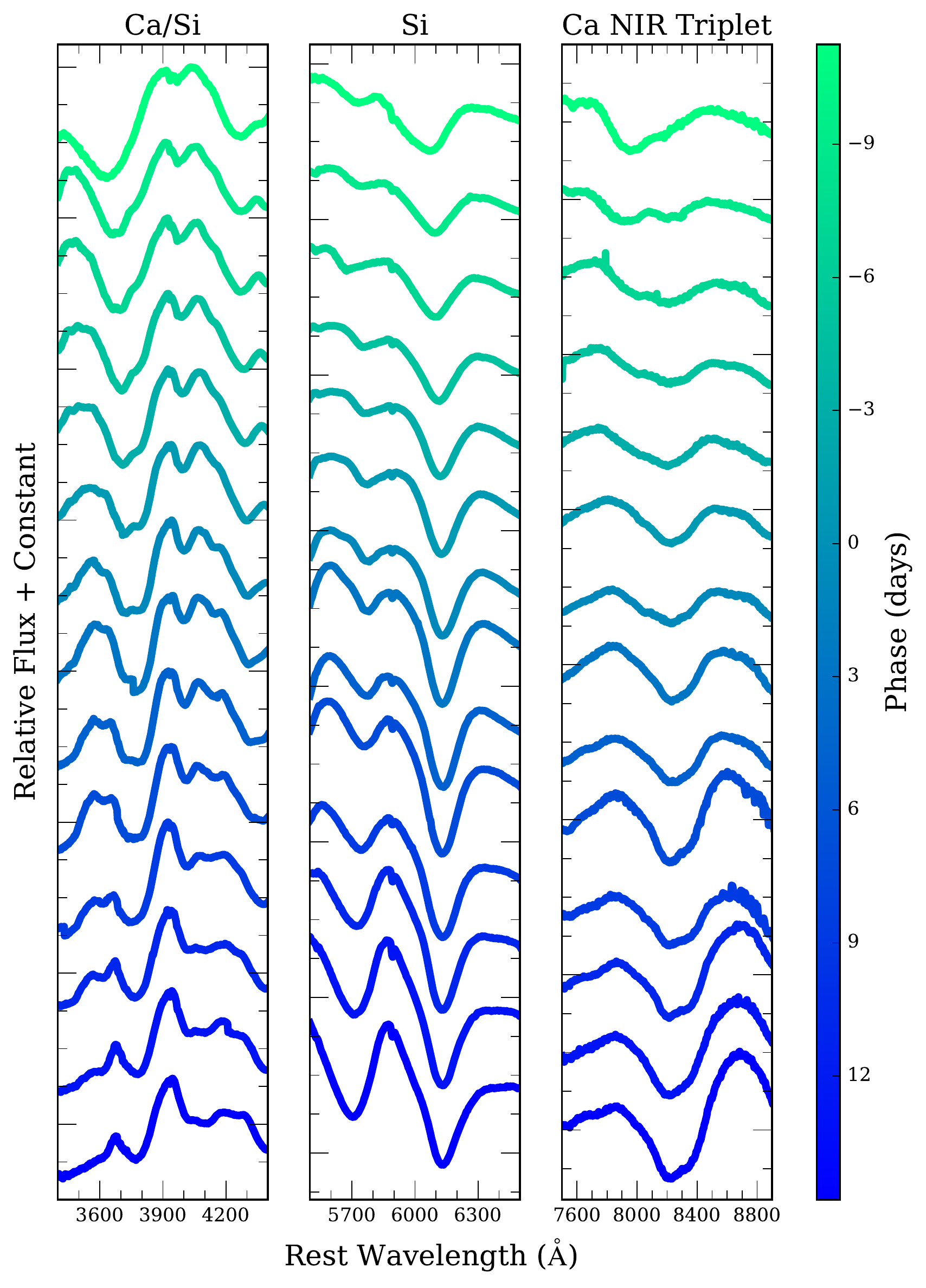}
\caption{Zoomed-in composite spectra within 16 days of maximum light created from the nominal sample. The spectral absorption features associated with Ca II H$\&$K, Si II, and the Ca II NIR triplet are shown in the left, middle and right panels respectively. Bin sizes of 2 days are used for each of these composite spectra.}\label{fig:zoom_in}
\end{center}
\end{figure*}

In Figure \ref{fig:zoom_in}, we present 16 phase-binned composite spectra ranging from $-11$ days relative to maximum light to $16$ days after maximum light using a bin size of 2 days. The Si II and Ca II absorption features evolve smoothly with time indicating a steady change in photospheric velocity. The minima of these features progress towards redder wavelengths indicating a smooth decrease in velocity with time.  

Beyond MW and host galaxy extinction corrections, we do not make any modifications to the continua of our composite spectra. The $B-V$ color evolution of these composite spectra is very similar to that of normal SNe~Ia. This similarity means that the relative flux calibration among different surveys is good enough to produce reliable mean continua. 

\begin{figure}
    \includegraphics[width=3.2in]{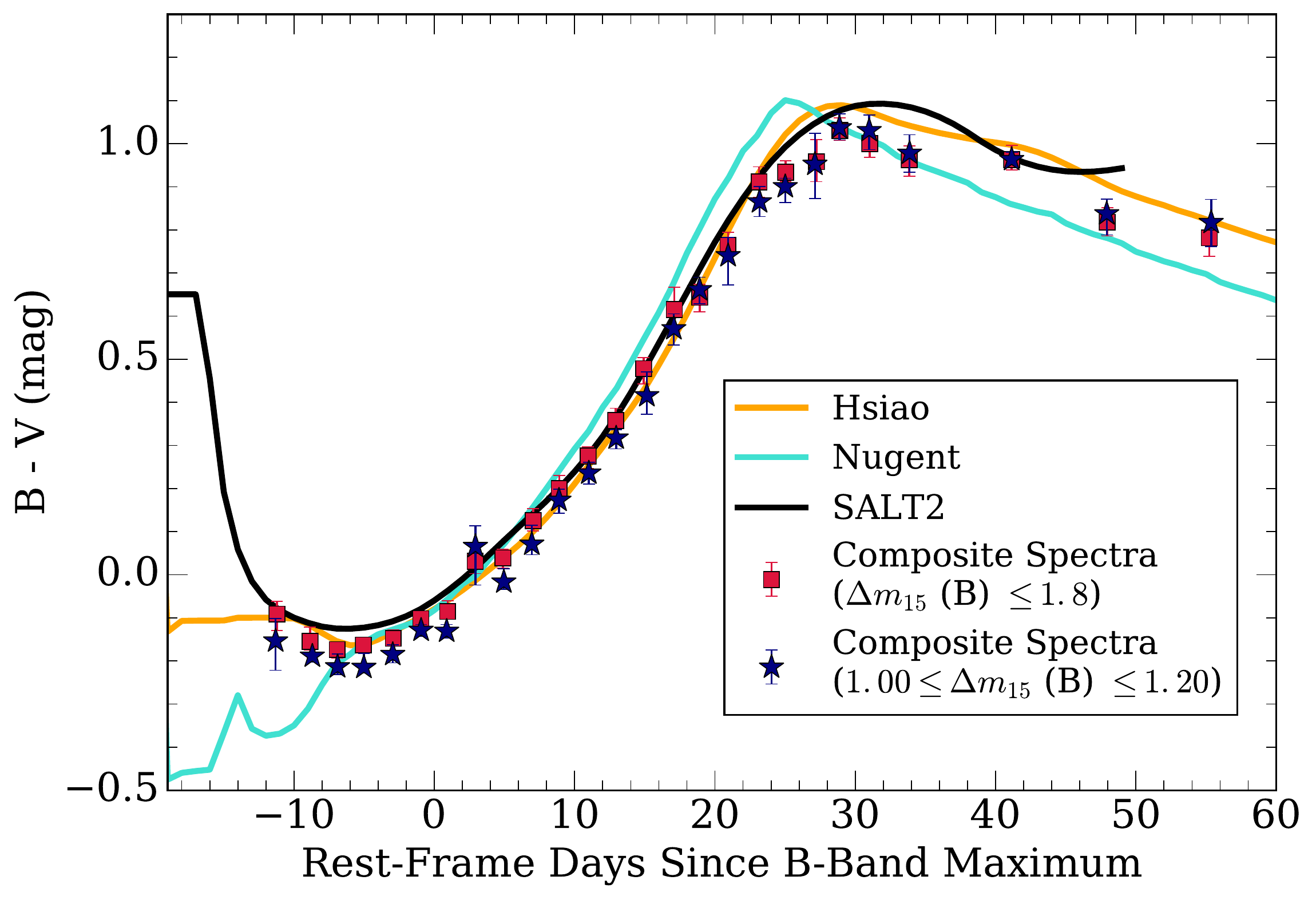}
  \caption{Comparison of the $B - V$ color evolution of various SN Ia template spectra and our composite spectra created from the nominal sample. The orange, light-blue, and black curves are the Hsiao, Nugent, and SALT2 template spectra respectively. The red squares are color measurements from composite spectra generated from the nominal sample with \dm\ $<1.8$ mag. The blue stars are color measurements from composite spectra generated from the nominal sample using only SNe where $1.00 \leq$ \dm\ $\leq 1.20$ mag. Bin sizes of 2 days are used for phase $\leq$+30 days and bin sizes of 4 days are used for phase $>$+30 days.}\label{fig:color}
\end{figure}

In Figure \ref{fig:color}, we show the $B-V$ color evolution of our composite spectra with \dm\ $<1.8$ mag (red squares) and compare to the Nugent (cyan), Hsiao (yellow), and SALT2 template spectra (black) for phases ranging from $-12$ to $+56$ days. We also investigate the effects of binning our composite spectra with $1.05 \leq$ \dm\ $\leq 1.25$ mag (dark blue stars). We chose this bin size for \dm\ since the Hsiao and Nugent template spectra were created to represent $s=1.0$ (\dm\ $\approx 1.1$ mag) SNe, and the SALT2 base template spectrum has $x_1=0$ (\dm\ $\approx 1.1$ mag). Thus our sets of composite spectra with large and small \dm\ windows have average values of \dm\ that range from $1.03 - 1.17$ mag and $1.07 - 1.11$ mag respectively. This means that on average these two sets of composite spectra have very similar light curve shapes. 

Overall, we see good agreement with the the three template spectra. We also find that our two sets of composite spectra with different \dm\ windows are consistent at all phases. This further reinforces the fact our composite spectra with larger sample sizes (due to having larger ranges of light curve shape) are representative of SNe~Ia with the average properties of our underlying samples. On average, our composite spectra are most consistent with Hsiao and slightly bluer than the SALT2 and Nugent color curves. The slope of the Nugent template spectrum between the epochs of $-6$ and +24 days is significantly larger than the other color curves. This difference may be due to the influence of atypical SNe on the Nugent template spectrum. Both the Hsiao and Nugent spectral templates are warped to match the photometry of template light curves in the literature (Hsiao; $B$-band \citealt{goldhaber01}, $UVR$-band \citealt{knop03}, Nugent; $B$-band \citealt{riess99}, $RI$-band \citealt{knop03}). Our composite spectra reproduce a normal SN Ia $B-V$ color curve without the need for spectral warping and thus preserve absorption features, lines ratios, and velocities that are representative of the underlying spectra. 

\begin{figure}
\includegraphics[angle=0,width=3.2in]{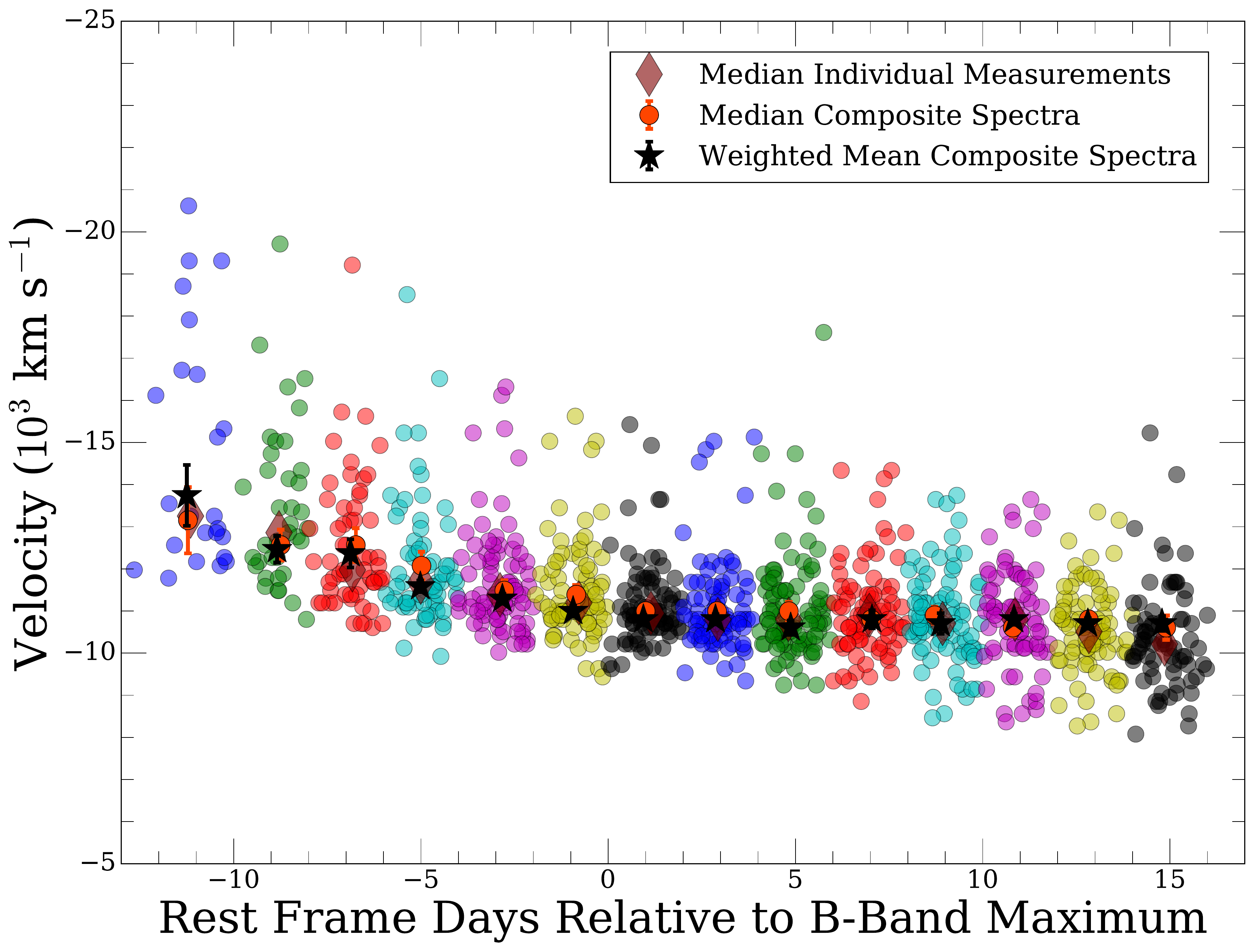}
\caption{Velocity evolution of our composite spectra created from the nominal sample. Black stars are measurements from the weighted mean composite spectra and orange circles are measurements from median composite spectra. The smaller circle points are measurements from the individual spectra. These are color coded to distinguish the borders of the phase bins defined for the composite spectra. The brown diamonds correspond to the median of the individual measurements within each bin.}\label{fig:vel}
\end{figure}

Si II $\lambda 6355$ line velocities measured from these composite spectra are also representative of their respective subsets. In Figure \ref{fig:vel}, we compare the velocity evolution of our composite spectra (black) to the distributions of velocities of the individual SNe in each phase bin (distinguished by color) covering phases between $-13$ and $+16$ days. We also compare to the velocity evolution of the median velocity measurement in each phase range (brown), and median composite spectra generated from the same subsets (orange). 

The velocities of our composite spectra are consistent with these measurements across all phases and range from a blueshift of $-13{,}700$ km s$^{-1}$ at -11 days to $-10{,}700$ km s$^{-1}$ at +15 days. The velocities of our weighted mean and median composite spectra steadily decline with phase until around maximum light where they level off at $\sim$\ $-10{,}800$ km s$^{-1}$. Due to the lack of data at very early epochs (${<}-10$ days), it is likely that the velocity of our composite spectrum is slightly underestimated at this phase. 

\subsection{Maximum-Light \dm -binned spectra} \label{sec:dm15}
\begin{figure}
\includegraphics[angle=0,width=3.2in]{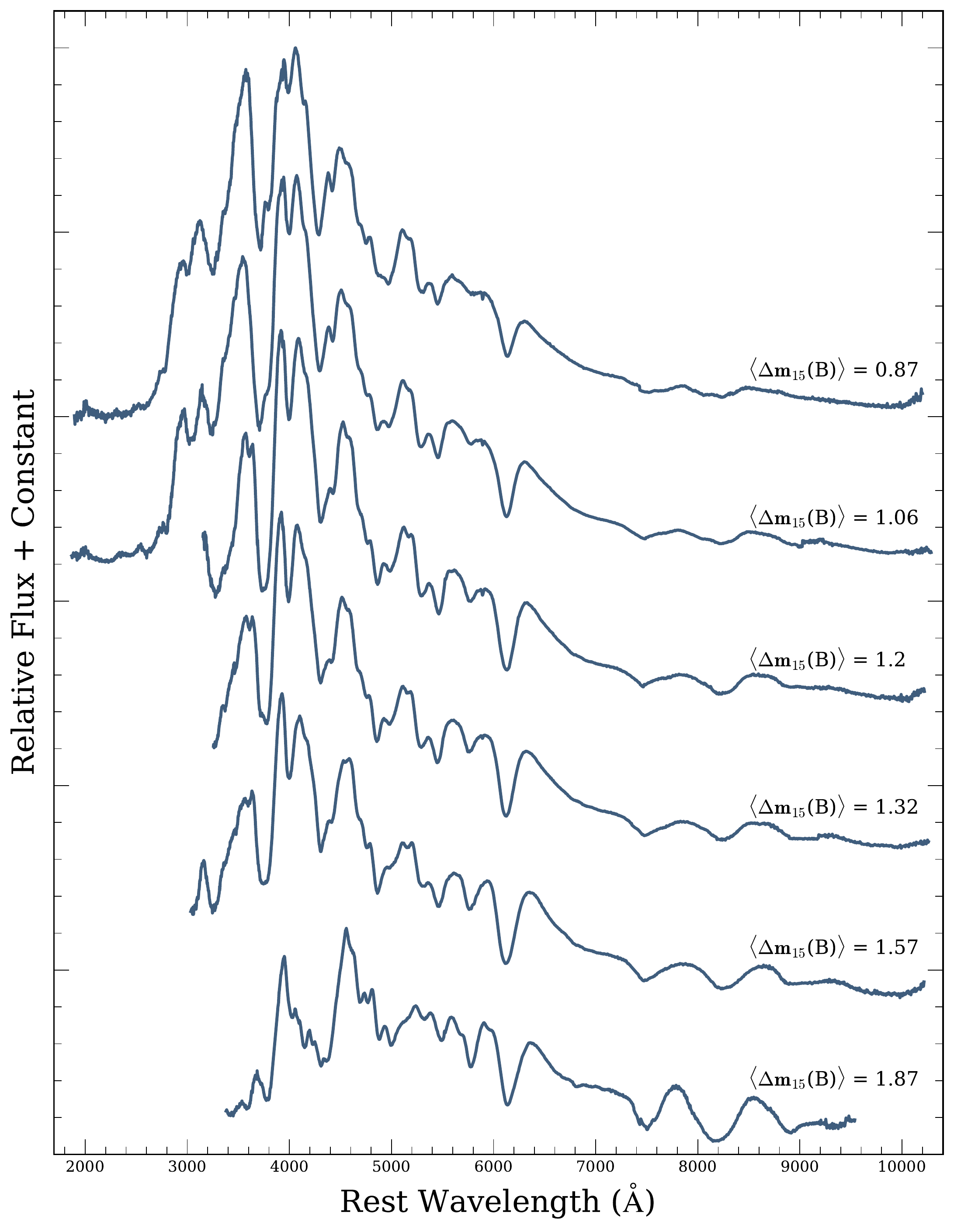}
\caption{\dm -binned maximum-light composite spectra. From top to bottom the spectra has \dm\ bins of \dm\ $< 0.95$ mag,  $0.95<$ \dm\ $< 1.15$ mag, $1.15<$ \dm\ $< 1.25$ mag, $1.25<$ \dm\ $< 1.42$ mag, $1.42<$ \dm\ $< 1.7$ mag, and \dm\ $> 1.7$ mag. Phase bins of $-3 <\tau < +3$ days were used for each composite spectrum. The composite spectra consist of 30, 61, 16, 23, 21, and 13 individual SNe respectively.} The average \dm\ for each composite spectrum is shown above each spectrum. \label{fig:dm15_max}
\end{figure}

We also examine luminosity-dependent spectral differences through the light-curve decline rate parameter \dm\ for events near maximum light. In Figure \ref{fig:dm15_max} we present 6 maximum light composite spectra where we have controlled for light-curve shape. Here we expand our phase range to include spectra with phase between $-$3 and +3 days. We require at least 5 spectra per wavelength bin which results in varying wavelength ranges among the composite spectra. Since there are more UV spectra for slow-declining events that tend to have higher luminosities, we are able to extend two of these composite spectra into the UV. 

This subdivision by \dm\ clearly demonstrates several known spectral sequences with light-curve shape. In particular the continua become redder with narrower light-curves (larger \dm).  We also see clear evidence of Ti II absorption at ${\sim} 4300$~\AA\ in the lowest luminosity composite spectrum (\dm\ $ = 1.87$ mag). This is a characteristic feature of the SN 1991bg-like subclass of SNe~Ia \citep{filippenko92a, leibundgut93} and is expected for these faster declining events. This composite spectrum also has stronger O I $\lambda 7773$ and Ca near-IR triplet absorption than the other composite spectra. 

\begin{figure}
\includegraphics[width=3.2in]{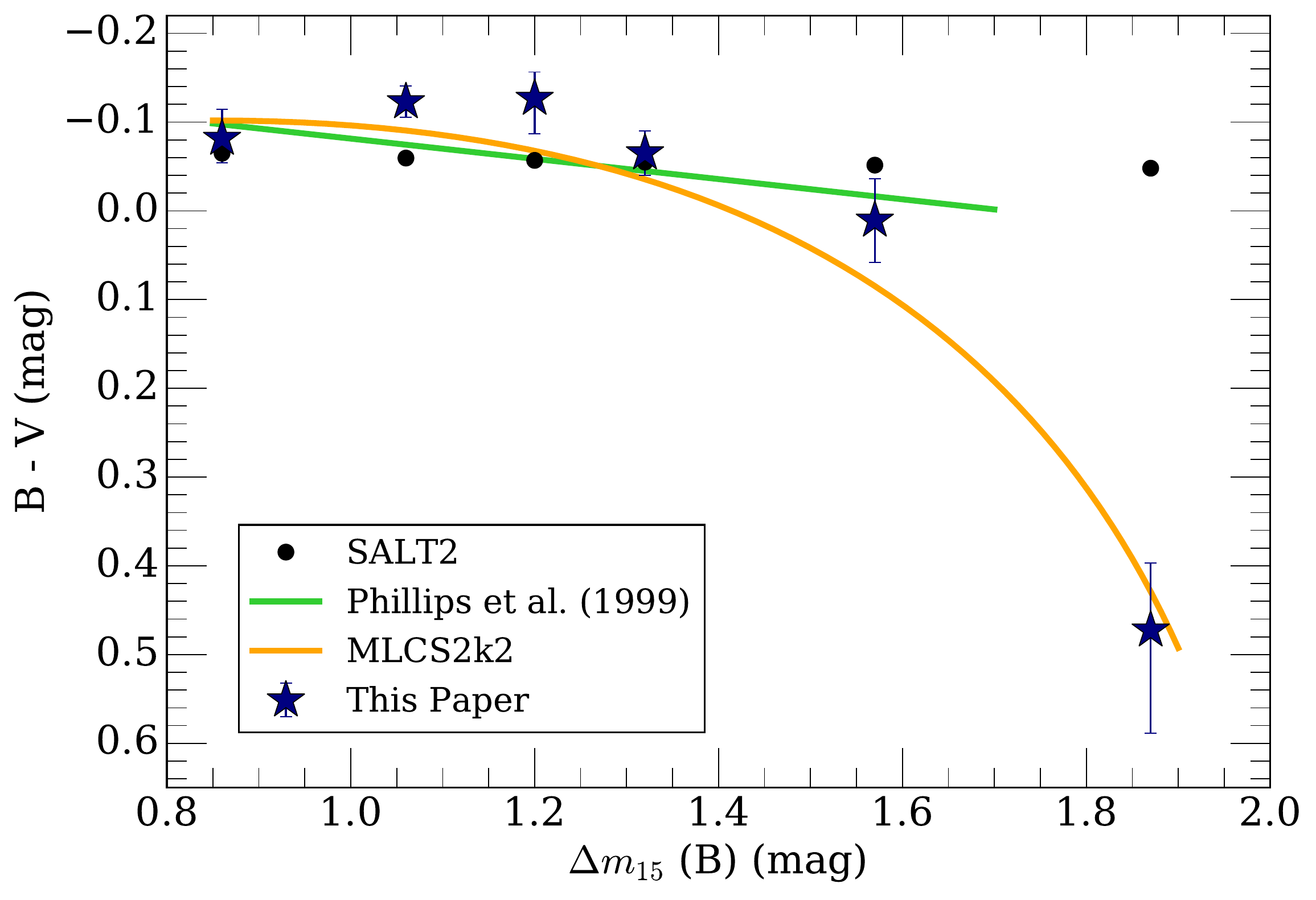}
\caption{The relationship between \dm\ and $B-V$ color at maximum-light. The blue stars are measurement made from our composite spectra in Figure \ref{fig:dm15_max}. The black circles are measurements made from maximum-light SALT2 template spectra with effective \dm\ corresponding to the mean values of \dm\ from our composite spectra. The green and orange curves are fits from \citet{phillips99} and \citet{phillips93} respectively. \label{fig:color_dm15}}
\end{figure}

We measure the $B-V$ colors of our composite spectra from Figure \ref{fig:dm15_max} and present them in Figure \ref{fig:color_dm15} (blue stars). For comparison, we show measurements made from maximum-light SALT2 template spectra with effective \dm\ corresponding to the mean values of \dm\ from our composite spectra (black circles), a linear fit of this relationship from \citet{phillips99} (green curve), and the relationship that is produced by the MLCS2k2 model (\citealt{jha07}; orange curve).

While the SALT2 spectral feature strengths differ greatly, the $B-V$ colors of the SALT2 maximum-light spectra, by design, do not vary much with light-curve shape. The $B-V$ colors of our maximum-light composite spectra slowly become redder with \dm\ and then change dramatically for the fastest-declining SNe. Our composite spectra are consistent with with \citet{phillips99} for the slower-declining SNe~Ia and produce the same general trend seen in MLCS2k2. This is not surprising because we rely on an estimate of $A_V$ from MLCS light-curve fits in order to correct individual spectra for host-galaxy extinction.  

\begin{figure}
\includegraphics[angle=0,width=3.2in]{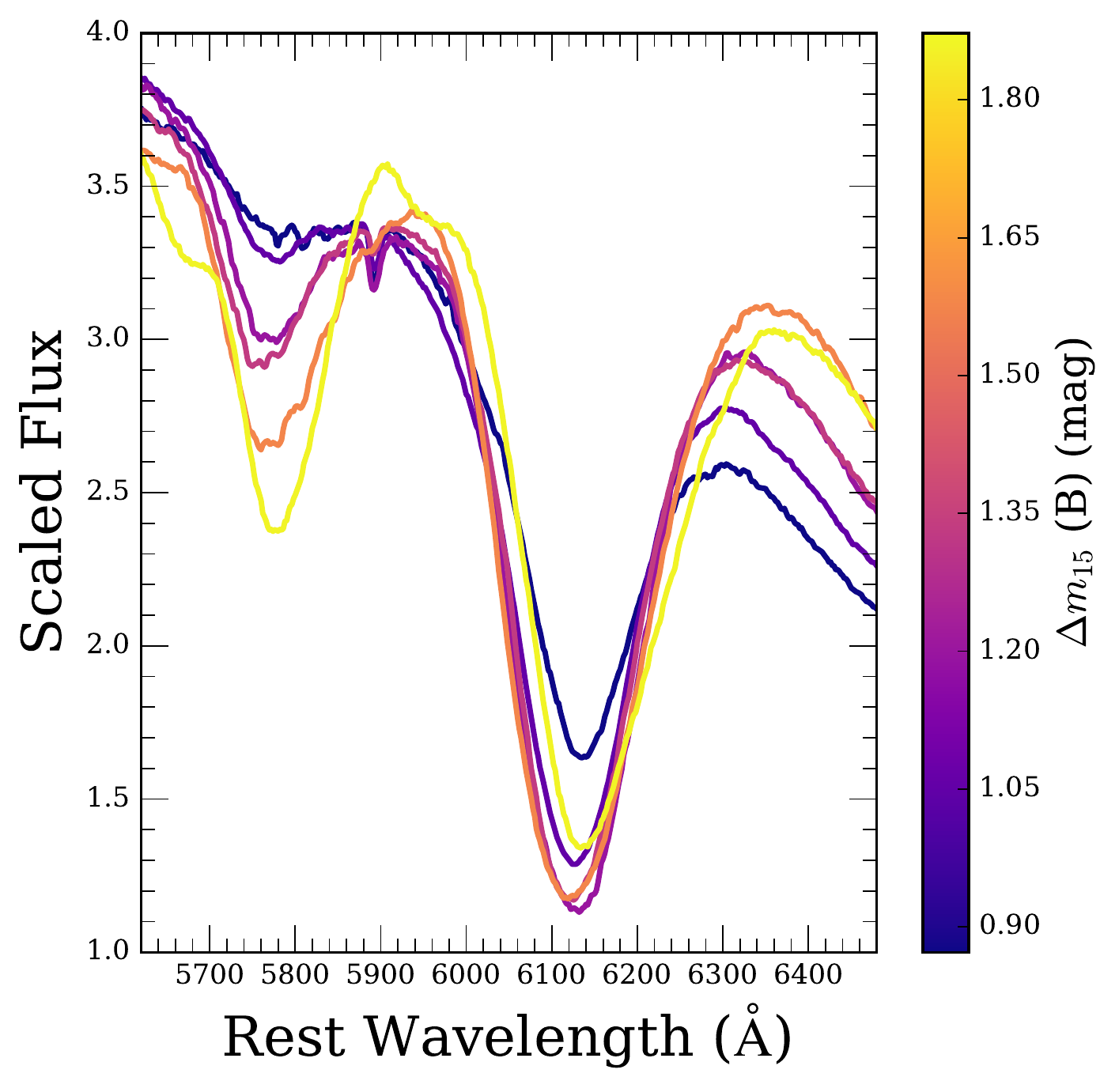}
\caption{The Si II $\lambda 5972$ and $\lambda 6355$ absorption features of our maximum-light composite spectra from Figure \ref{fig:dm15_max}. The color bar ranges from slow-declining light-curve shapes (\dm\ = 0.86 mag; dark purple) to fast-declining light-curve shapes (\dm\ = 1.87 mag; yellow). \label{fig:si_dm15}}
\end{figure}

Figure \ref{fig:si_dm15} highlights the subtle differences between the Si II $\lambda 5972$ and $\lambda 6355$ absorption features of \dm-binned composite spectra. Here the continuum difference between these composite spectra is obvious. Since intrinsically bright SNe tend to have bluer colors, our composite spectra progress towards having redder continua as the average light-curve shape becomes narrower (\dm\ increases). Additionally, we see evidence of Na I D absorption in all composite spectra except the two with the largest effective \dm\ values. This trend is likely because fast-declining SNe~Ia tend to occur in early-type environments devoid of gas \citep{hamuy00,howell01}. 

From the slowest-declining to fastest-declining composite spectra (top to bottom in Figure \ref{fig:dm15_max}) the Si II $\lambda 6355$ line velocities of these composite spectra are $-10{,}500$, $-11{,}000$, $-10{,}800$, $-11{,}200$, $-11{,}200$, and $-10{,}500$ km s$^{-1}$. These measurements are consistent with the lack of an observed correlation between light-curve shape and velocity for the bulk of SNe~Ia and the tendency for events with extremely high or low values of \dm\ to have slightly lower velocities \citep{wang09}.

\begin{figure}
\includegraphics[width=3.2in]{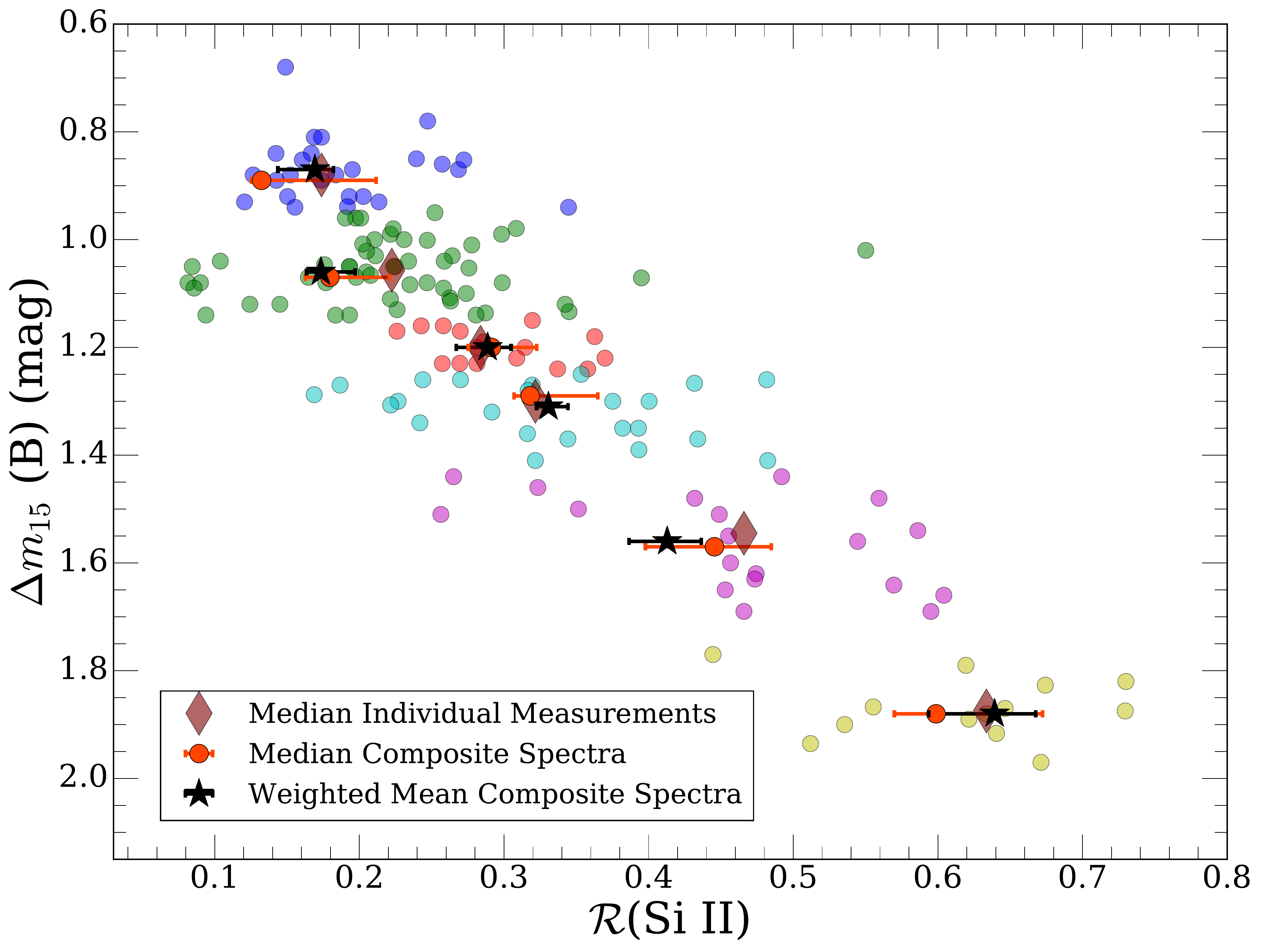}
\caption{$\mathcal{R}$(Si II) of our \dm -binned composite spectra created from the nominal sample. Black stars are measurements from the weighted mean composite spectra and orange circles are measurements from median composite spectra. The smaller circle points are measurements from the individual spectra. These are color coded to distinguish the borders of the phase bins defined for the composite spectra. The brown diamonds correspond to the median of the individual measurements within each bin.\label{fig:si_ratio}}
\end{figure}

In addition, the ratio of the depth of the two main Silicon absorption features ($\mathcal{R}$(Si II)) changes with light-curve shape. This sequence, originally discovered by \citet{nugent}, results from the changing abundance of the ionization species of Silicon \citep{hachinger08}. Using a similar approach to our analysis of velocity near maximum light, we compare $\mathcal{R}$(Si II) from our composite spectra (black) to those measured from the individual spectra in each \dm\ bin (distinguished by color; Figure \ref{fig:si_ratio}). We also compare the median $\mathcal{R}$(Si II) measurement in each \dm\ range (brown), and $\mathcal{R}$(Si II) measured from median composite spectra generated from the same subsets (orange). Here we clearly show that $\mathcal{R}$(Si II) of a composite spectrum is representative of its effective decline rate. The measurements of $\mathcal{R}$(Si II) reproduce the properties of their underlying subsets and are consistent with the median properties of these individual spectra.

\newsavebox{\myimage}
\begin{figure*}
  \centering
  \savebox{\myimage}{\rule{700pt}{0pt}}
  \rotatebox{90}{
    \begin{minipage}{\wd\myimage}
      \usebox{\myimage}
      \includegraphics[width=9.4in]{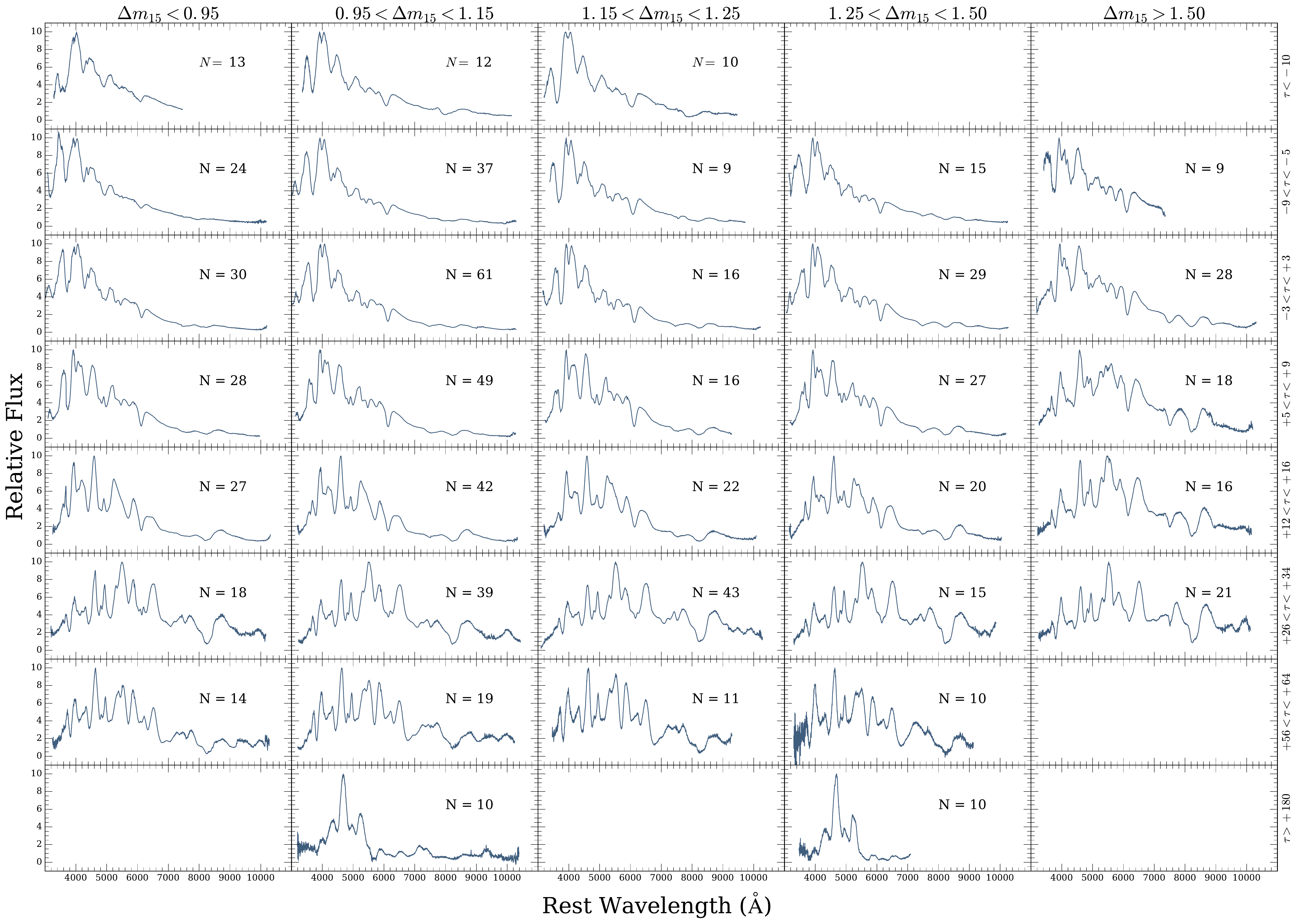}
      \caption{{Phase and \dm -binned composite spectra. Columns from left to right correspond to \dm\ bins of \dm\ $< 0.95$ mag,  $0.95<$ \dm\ $< 1.15$ mag, $1.15<$ \dm\ $< 1.25$ mag, $1.25<$ \dm\ $< 1.5$ mag, and \dm\ $> 1.5$ mag.}. Rows from top to bottom corresponds to phase bins of $\tau < -10$ days, $-9 <\tau < -5$ days, $-3 <\tau < +3$ days, $+5 <\tau < +9$ days, $+12 <\tau < +16$ days, $+26 <\tau < +34$ days, $+56 <\tau < +64$ days, and $\tau > +180$ days. The number of individual spectra included in each composite spectrum is stated in the corresponding panel. Composite spectra have varying wavelength ranges because we require at least 5 individual spectra at any given wavelength. Blank panels correspond to regions of this parameter space where we do not currently have enough data to generate a representative composite spectrum.}\label{fig:grid}
    \end{minipage}}
\end{figure*}

The SNe in our nominal sample span a large range of light-curve shapes and their spectra cover a large range of phases. These are the primary parameters that control the spectral features of SNe~Ia and we have demonstrated that our composite spectra reproduce evolution and variation with these properties. In Figure \ref{fig:grid}, we present a grid of composite spectra generated by controlling for both phase and light-curve shape. From top to bottom composite spectra are increasing in their effective phase, and from left to right composite spectra are increasing in their effective \dm. So the maximum-light composite spectra from Figure \ref{fig:dm15_max} are represented by the third row in this grid. For clarity we have used slightly different \dm\ ranges and the rightmost column contains all SNe with \dm\ $> 1.50$ mag (including SNe Ia spectroscopically similar to SN 1991bg). The amount of data contributing to each composite spectrum is related to observing strategies, intrinsic brightness, and rates. For example, the composite spectrum for bins with $-3 <\tau < +3$ days and $0.95<$ \dm\ $< 1.15$ mag contains the most individual spectra (61) likely because most SNe~Ia are discovered near maximum brightness and normal SNe~Ia (\dm\ $\approx 1.1$ mag) are relatively bright and common. Despite slower declining SNe~Ia (\dm\ $<0.95$ mag) being intrinsically brighter, fewer spectra contribute to these composite spectra because the rates of these events are lower \citep{li11}. Similarly, faster declining SNe~Ia (\dm\ $>1.50$ mag) are both intrinsically fainter and have lower rates than normal SNe~Ia \citep{li11}. For these fast-declining events, we again see Ti II absorption at ${\sim} 4300$~\AA\ at all phases. Naturally, we also see a decrease in the number of spectra with phase as the SNe dim and get more difficult to observe. 

This figure also highlights some spectral correlations that are slightly degenerate with both phase and light-curve shape. As expected, the continua of our composite spectra are bluer for earlier phases and smaller values of \dm\ and redder for later phases and larger values of \dm. This continuum relationship is consistent with Figure \ref{fig:color} which shows that the $B-V$ color of our composite spectra become redder with time for phases between $-6$ and $+28$ days, and Figure \ref{fig:si_dm15} which also shows the tendency for maximum-light composite spectra with larger effective \dm\ to have redder continua. Additionally, light-curve shape and phase produce some similar effects on spectral feature strengths. For example, the strength of the Si II $\lambda 4130$ absorption feature is weaker with earlier epochs and broader light-curve shapes. The flux of the peak to the right the $\lambda 4130$ absorption feature has the tendency to decrease with phase relative to the peak to the right of the Ca H\&K absorption feature. A similar correlation between these peaks exists with increasing \dm\ near maximum light. We also observe a slight positive correlation between $\mathcal{R}$  (Si II) and phase between $-11$ and $+9$ days. For these reasons, some composite spectra look very similar despite having very different average properties. For example, our composite spectra constructed using $-9 <\tau < -5$ and $1.25<$ \dm\ $< 1.5$ mag (row 2, column 4) and $+5 <\tau < +9$ and \dm\ $< 0.95$ mag (row~4, column~1) have similar continua and spectral feature strengths.

\subsection{Template Spectrum Comparisons}
In this section, we compare our composite spectra to the Nugent, Hsiao, SALT2 and Foley template spectra using various bins of phase and \dm.

\begin{figure*}
\begin{center}
\includegraphics[width=6.4in]{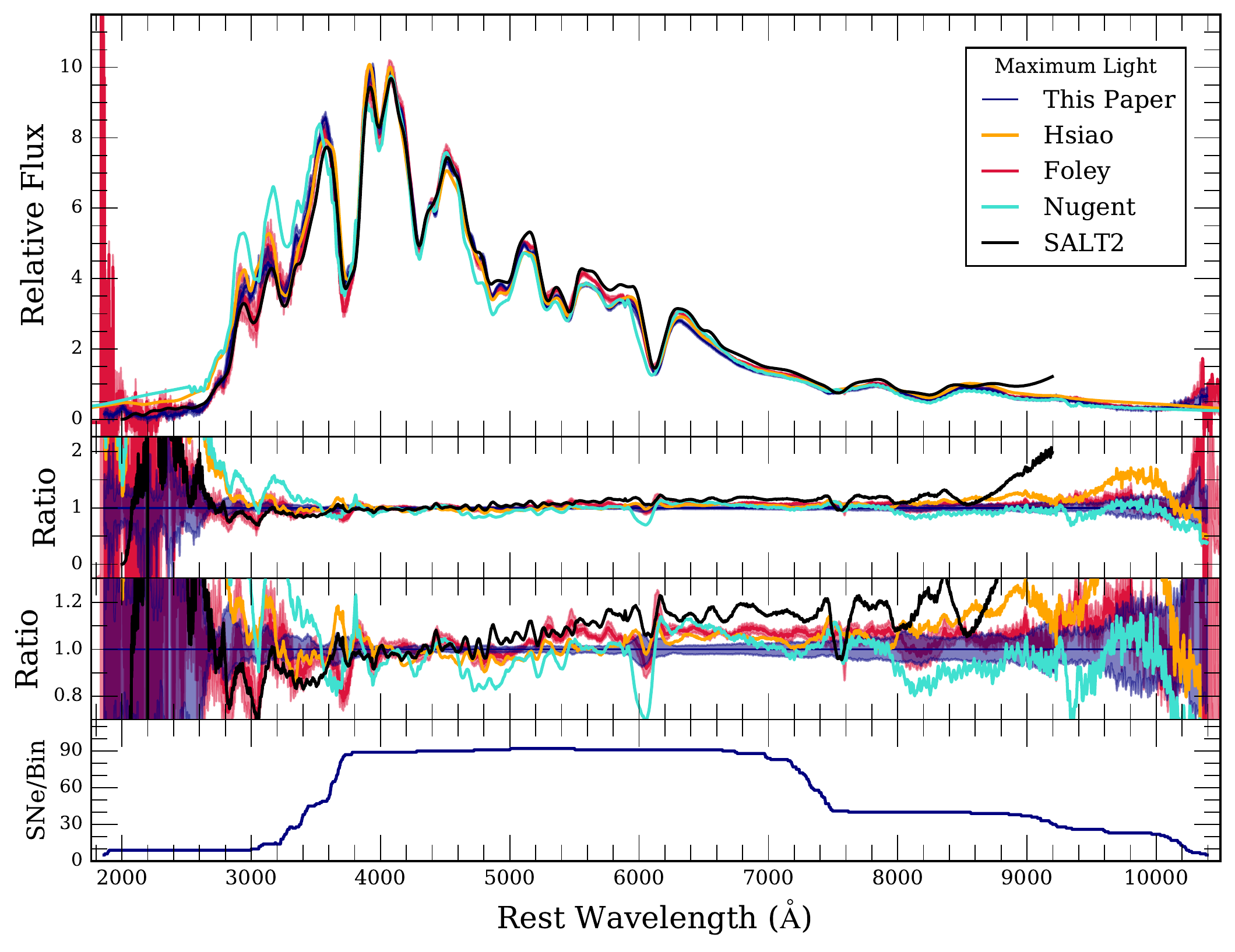}	
\caption{(\textit{first panel}): Comparison of our maximum-light composite spectrum (blue curve) from Figure \ref{fig:max} to maximum-light template spectra from Hsiao (orange curve), Foley (red curve), Nugent (light-blue curve), and SALT2 (black curve). The blue- and red-shaded regions are the $1\sigma$ bootstrapping uncertainty on the composite spectra from this work and Foley respectively. (\textit{second panel}): The ratio as a function of wavelength of a given template spectrum to our maximum-light composite spectrum. (\textit{third panel}): Zoomed in version of the second panel to emphasize smaller differences between template spectra. The light-blue shaded region is the $1\sigma$ bootstrap variation of the total maximum-light composite spectrum. (\textit{fourth panel}): The number of individual spectra contributing to each wavelength bin of the total maximum-light composite spectrum. \label{fig:temp_max}}.
\end{center}
\end{figure*}

In Figure \ref{fig:temp_max} we investigate the normal SN Ia template spectra near maximum light. We use our total maximum-light composite ($-1 < \tau < +1$ days and \dm\ $<1.8$ mag) spectrum from Figure \ref{fig:max} as a comparison. Overall, our composite spectrum is most similar to the Hsiao and Foley template spectra. The majority of the Hsiao anf Foley template spectra fall within the $1 \sigma$ bootstrap uncertainty of our composite spectrum and the near-UV peaks blueward of $4000$~\AA\ are very well matched. Hsiao has slightly higher flux in some regions ${>}8000$~\AA. The Nugent template spectrum is slightly bluer than the composite spectrum especially in the UV and near-UV. The SALT2 maximum-light ($x_1=0$) template spectrum has a slightly redder continuum than the composite spectrum and only extends from $2000-9200$ ~\AA.

Despite the slight color differences, the relative strengths of most absorption features are very similar for our composite spectrum and the three template spectra. The Nugent template spectrum has a larger Si II $\lambda 6355$ blueshift of $-12{,}300$ km/s in comparison to our composite spectrum ($-10{,}900$ km s$^{-1}$), Hsiao ($-11{,}000$ km s$^{-1}$), Foley ($-11{,}500$ km s$^{-1}$), and SALT2 ($-11{,}300$ km s$^{-1}$).  

\begin{figure}
\includegraphics[width=3.2in]{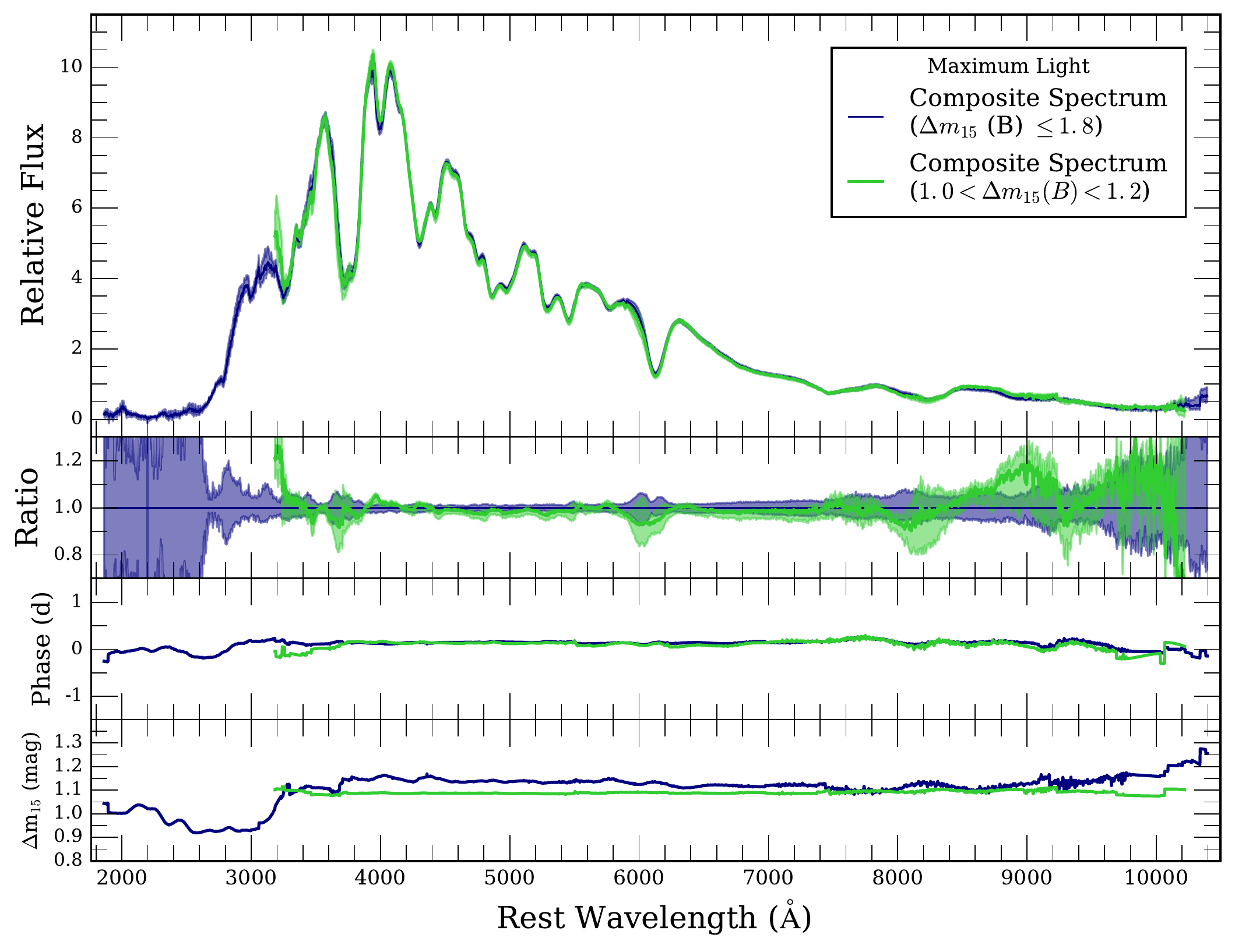}	
\caption{Same format as Figure \ref{fig:samp_size} except we compare our maximum-light composite spectrum (blue curve) from Figure \ref{fig:temp_max} and our maximum-light composite spectrum with $1.0<$ \dm\ $<1.2$ mag (green curve).}\label{fig:temp_comp}
\end{figure}

In section 5.1, we investigated the effect our \dm\ bin size on the color evolution of our composite spectra. Here we perform a similar analysis to see if spectral feature strengths differ for composite spectra with the same average properties but different sample sizes. In Figure \ref{fig:temp_comp} we compare our maximum-light composite spectrum with \dm\ $<1.8$ (blue curve) to our maximum-light composite spectrum with $1.0<$ \dm\ $<1.2$ mag (green curve). The latter composite spectrum was constructed from 51 spectra of 34 SNe. The differences between this composite spectrum and our maximum-light composite spectrum with  \dm\ $<1.8$ (blue curve) are very small. There does seem to be larger variation near the stronger spectral features, but these differences are all captured by the bootstrapping uncertainty. We also see that the average phase and \dm\ are very similar in the wavelength range where these composite spectra overlap. For these reasons, we elect to use our composite spectra constructed from larger sample sizes when comparing to other template spectra of normal SNe~Ia. 

\begin{figure}
\includegraphics[width=3.2in]{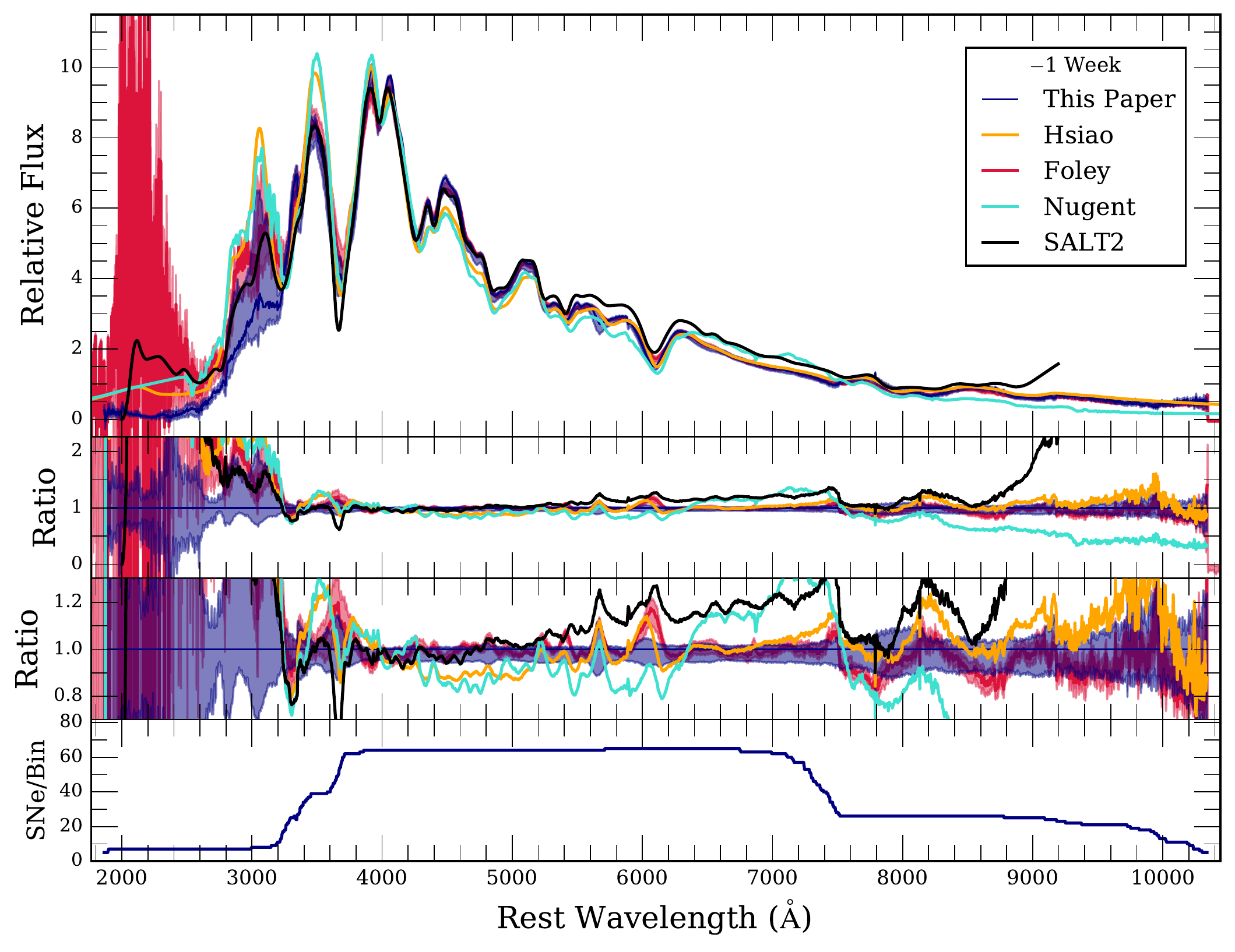}	
\caption{Same as Figure \ref{fig:temp_max} except the composite spectrum (blue curve) is constructed using a phase bin of $-8 < \tau < -6$ days and \dm\ $<1.8$ mag.\label{fig:temp_m7}}
\end{figure}

\begin{figure}
\includegraphics[width=3.2in]{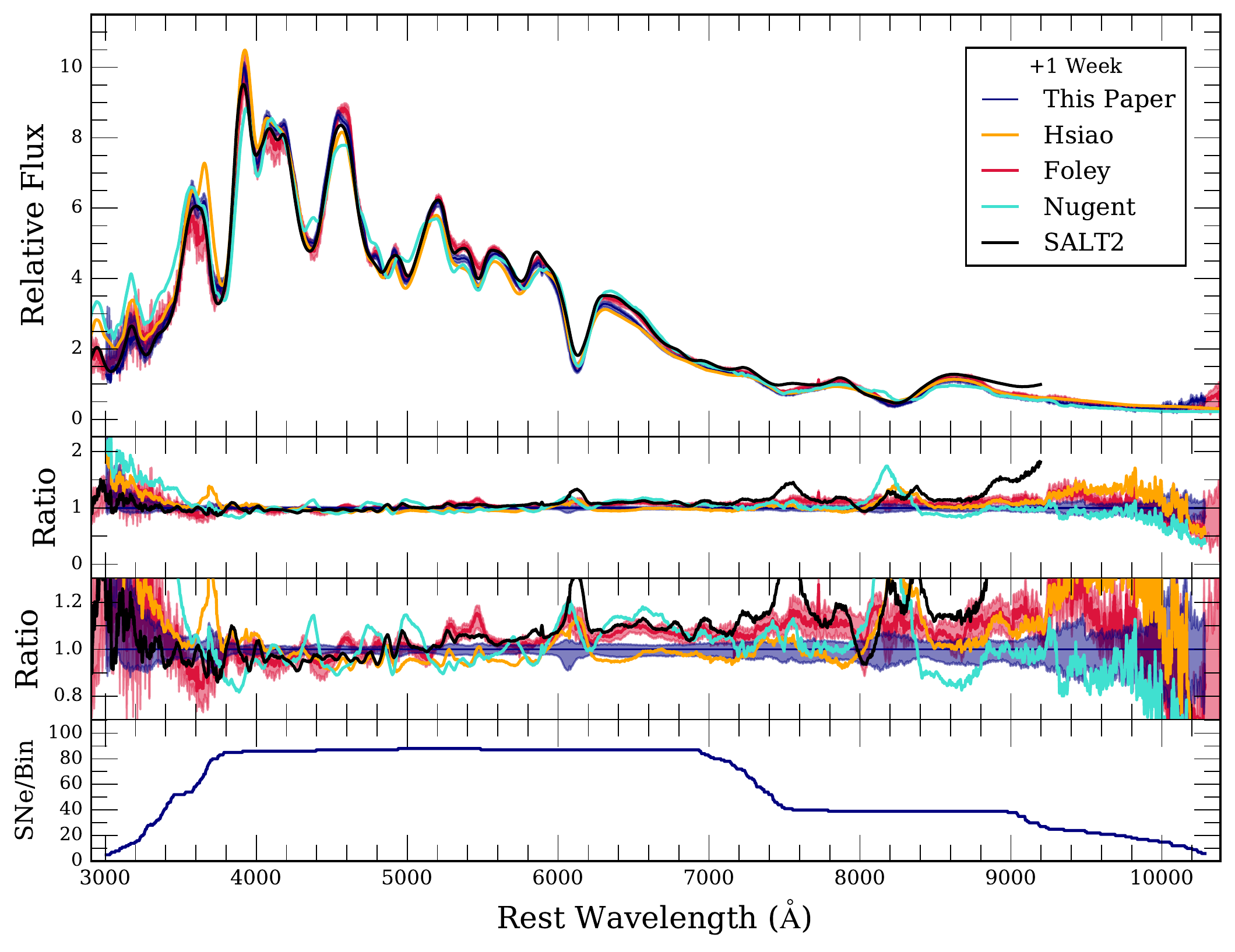}	
\caption{Same as Figure \ref{fig:temp_max} except the composite spectrum (blue curve) is constructed using a phase bin of $+6 < \tau < +8$ days and \dm\ $<1.8$ mag.\label{fig:temp_p7}}
\end{figure}

\begin{figure}
\includegraphics[width=3.2in]{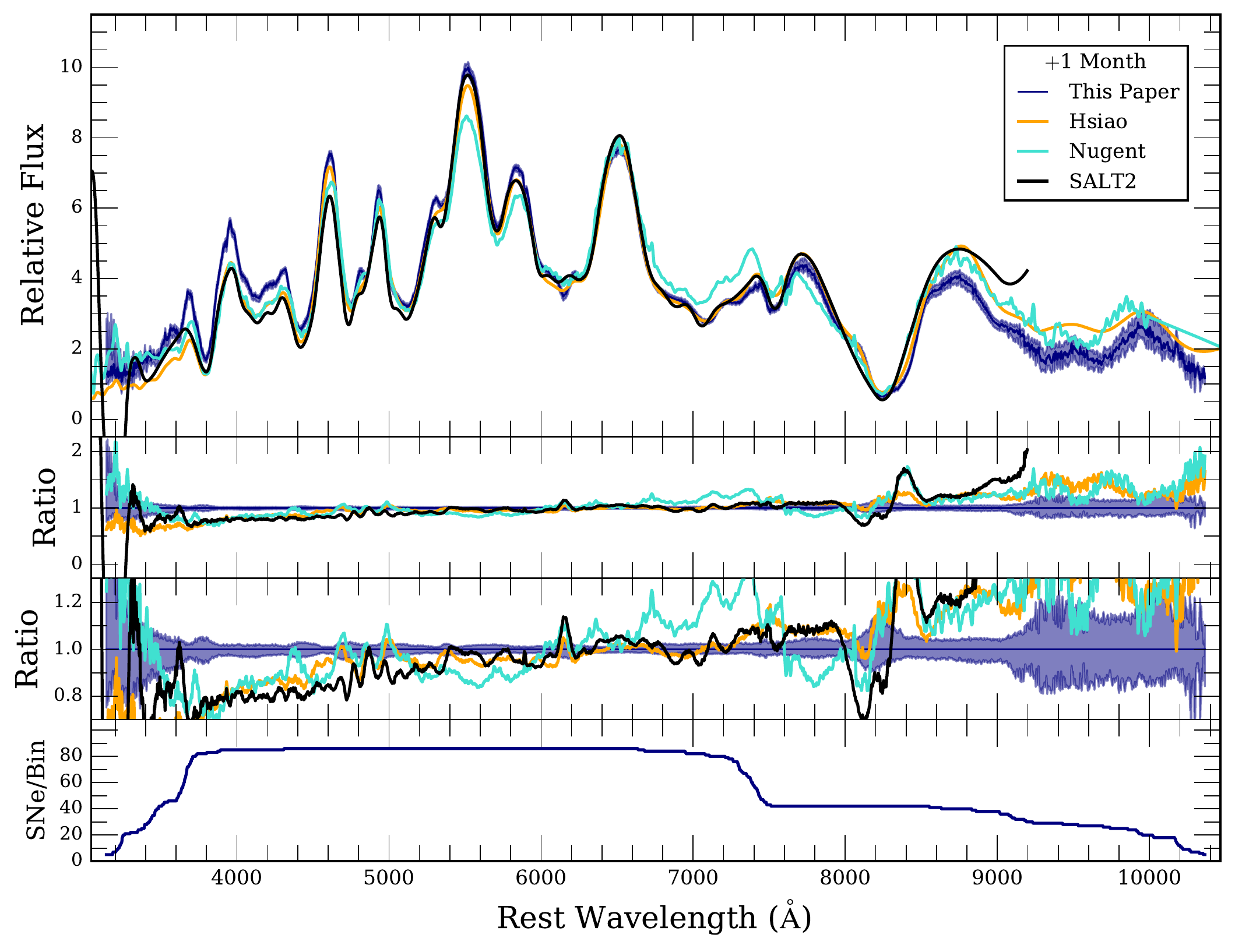}	
\caption{Same as Figure \ref{fig:temp_max} except the composite spectrum (blue curve) is constructed using a phase bin of $+28 < \tau < +32$ days and \dm\ $<1.8$ mag.\label{fig:temp_p30}}
\end{figure}

Figures \ref{fig:temp_m7}, \ref{fig:temp_p7} and \ref{fig:temp_p30} compare our composite spectra to the Nugent, Hsiao, SALT2, and Foley template spectra at different epochs ($\sim -1$ week, $\sim +1$ week, and $\sim +1$ month respectively). The line velocities of these composite spectra are well matched by the SALT2 template spectra. We measure a Si II $\lambda 6355$ blueshift of $-12{,}600$ km s$^{-1}$ for our -1 week composite spectrum ($-11{,}700$ km s$^{-1}$, $-12{,}100$ km s$^{-1}$, $-11{,}500$ km s$^{-1}$, and $-12{,}700$ km s$^{-1}$ for the -1 week Hsiao, Foley, Nugent, and SALT2 template spectra respectively). Similarly, we measure a Si II $\lambda 6355$ blueshift of $-10{,}800$ km s$^{-1}$ for our +1 week composite spectrum ($-10{,}900$ km s$^{-1}$, $-11{,}000$ km s$^{-1}$, $-10{,}600$ km s$^{-1}$, and $-10{,}800$ km s$^{-1}$ for -1 week Hsiao, Foley, Nugent, and SALT2 template spectra respectively). At every epoch, our composite spectra best match the spectral features and continua of the Hsiao and Foley template spectra. Additionally, the SALT2 template spectra have redder continua than our composite spectra at every epoch. The largest differences tend to occur near the stronger spectral features such as Si II $\lambda 6355$ and the Ca II NIR triplet. Finally, our +1 month composite spectrum is significantly bluer than the other template spectra.

\begin{figure}
\includegraphics[width=3.2in]{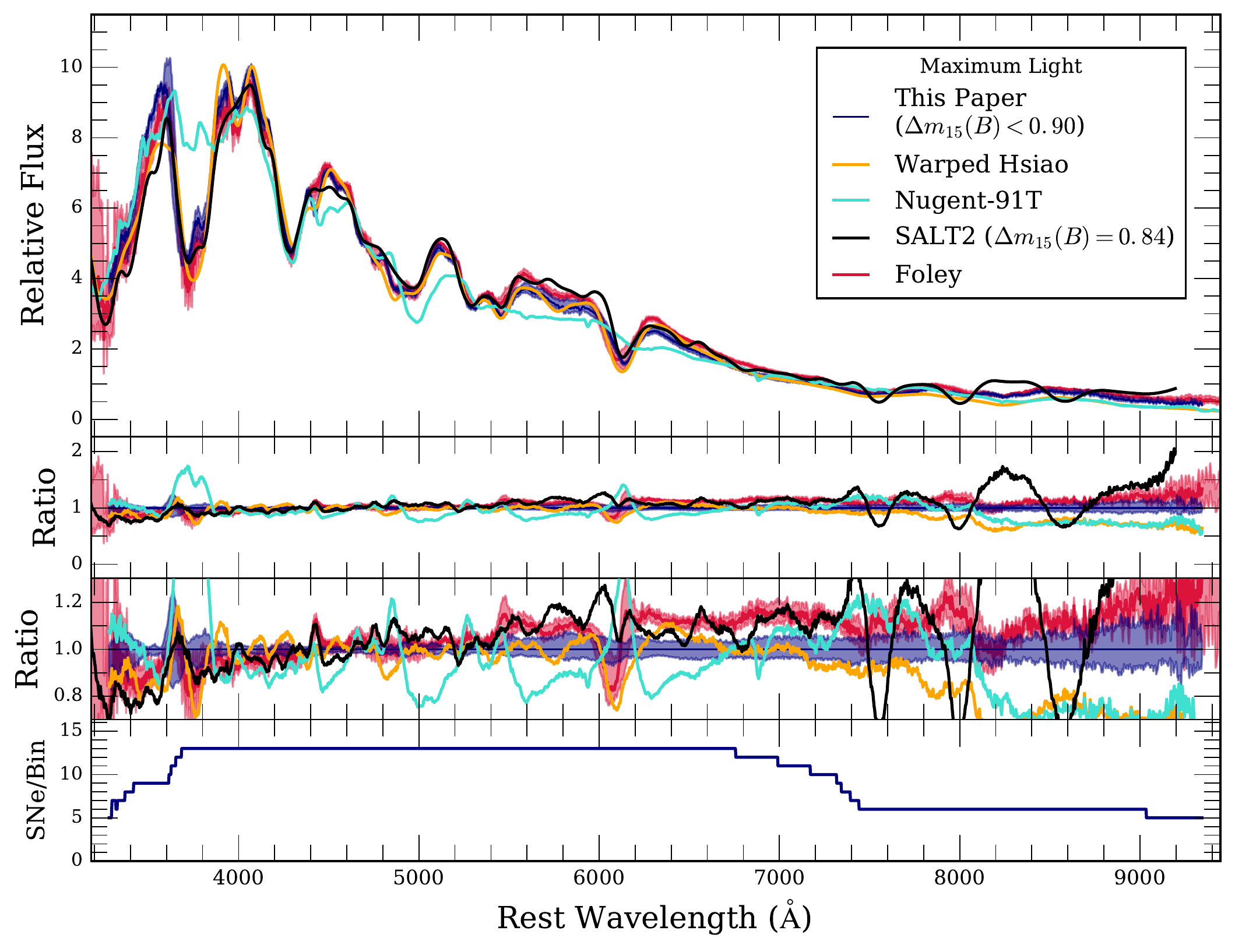}	
\caption{Same as Figure \ref{fig:temp_max} except the composite spectrum (blue curve) is constructed using a phase bin of $-1 < \tau < +1$ days and a \dm\ bin of \dm\ $<0.90$ mag. The maximum-light Hsiao template spectrum has been warped to match the $B$, $V$, and $R$-band photometry measured from the composite spectrum. The red curve is the Foley maximum-light, $\Delta = -0.27$ template spectrum. The light-blue is the Nugent-91T template spectrum, and the black curve is the maximum-light SALT2 model spectrum corrected for light-curve shape such that it has an effective \dm\ $=0.84$ mag.\label{fig:temp_low_dm}}
\end{figure}

\begin{figure}
\includegraphics[width=3.2in]{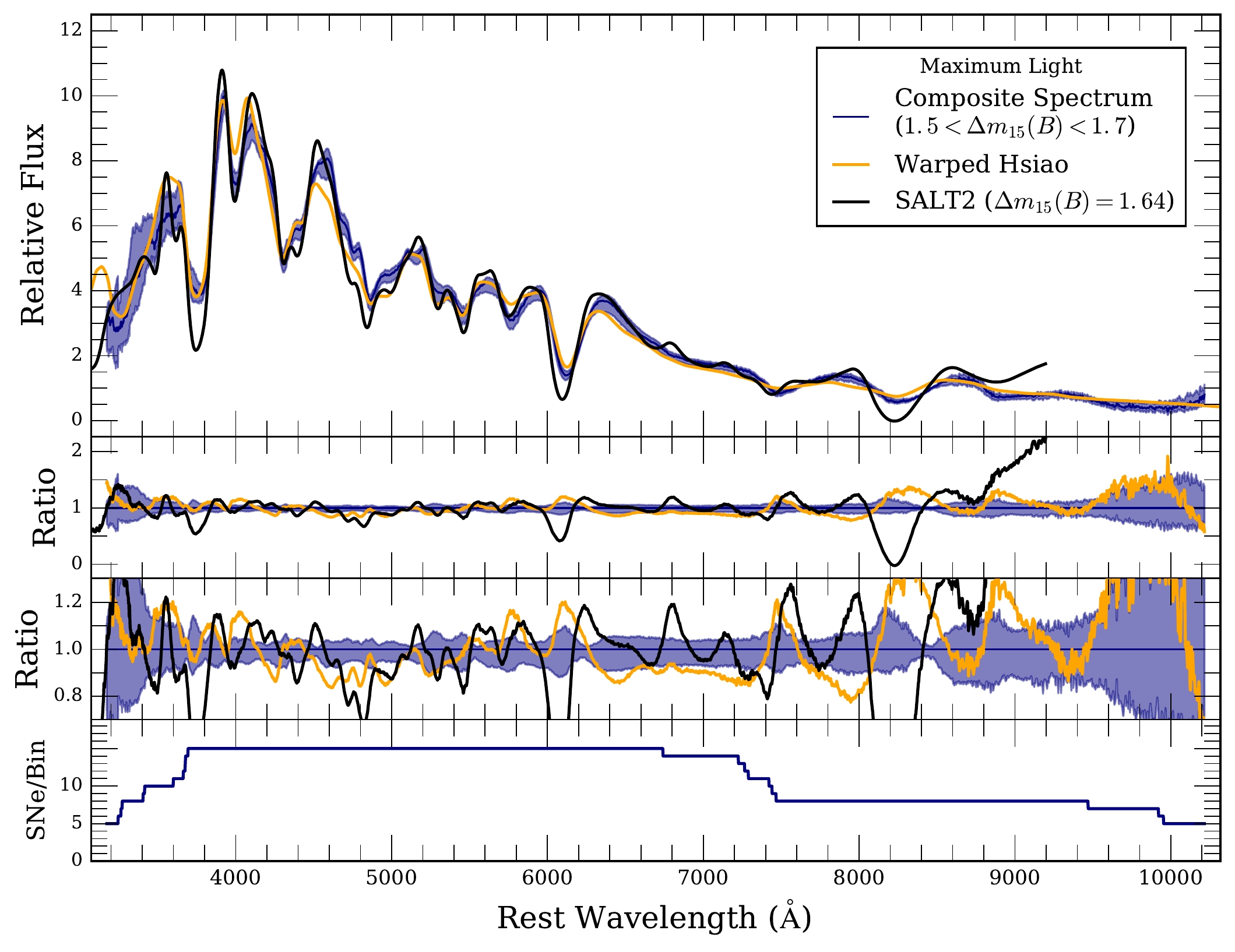}	
\caption{Same as Figure \ref{fig:temp_low_dm} except the composite spectrum (blue curve) is constructed using a phase bin of $-3 < \tau < +3$ days and a \dm\ bin of $1.50<$ \dm\ $<1.70$ mag. The maximum-light SALT2 model spectrum is corrected for light-curve shape such that it has an effective \dm\ $=1.64$ mag. \label{fig:temp_mid_dm}}
\end{figure}

\begin{figure}
\includegraphics[width=3.2in]{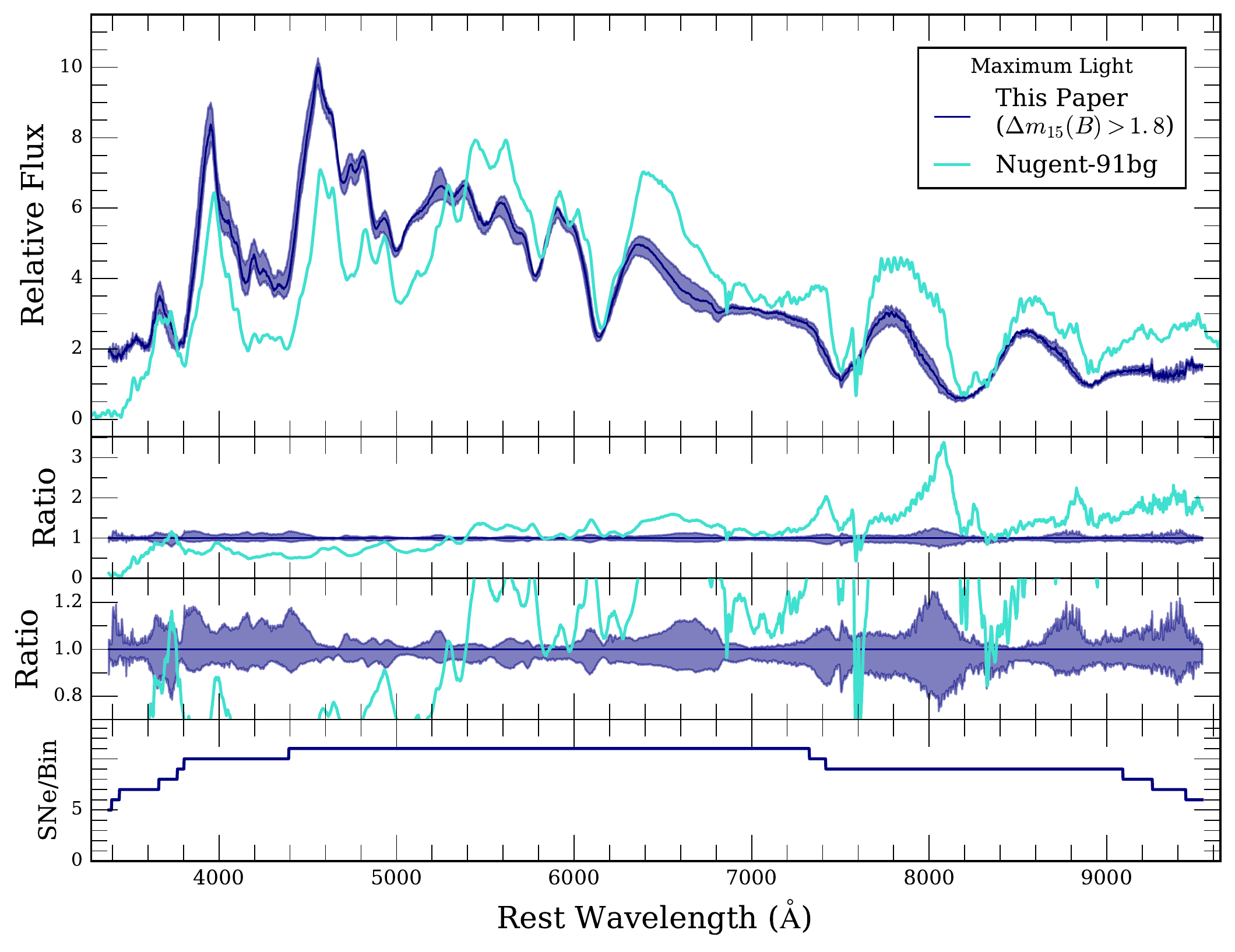}	
\caption{Comparison between a composite spectrum constructed from the fastest declining SNe~Ia and the Nugent-91bg template spectrum. The composite spectrum (light-blue curve) is constructed using a phase bin of $-3 < \tau < +3$ days and a \dm\ bin of \dm\ $>1.80$ mag.\label{fig:temp_91bg}}
\end{figure}

We also compare SN Ia template spectra with maximum-light composite spectra for both slow and fast declining events. In Figure \ref{fig:temp_low_dm} we present our composite spectrum constructed using only slow-declining events (\dm\ $<0.90$ mag). For this comparison, we warp the Hsiao maximum-light (\dm\ $=1.1$ mag) template spectrum to match the $B$, $V$, and $R$-band photometry of the composite spectrum. We also compare to the Foley $\Delta = -0.27$ (\dm\ $=0.90$) template spectrum, the Nugent-91T template spectrum \citep{stern04}, and the SALT2 model spectrum for an event with \dm\ $=0.84$ mag. This value of \dm\ is the average \dm\ of our composite spectrum. The value of $x_1$ was chosen using the polynomial relationship between \dm\ and $x_1$ shown in Figure \ref{fig:dm15_stretch}. The Si II $\lambda 6355$ feature is blueshifted $10{,}300$ km s$^{-1}$, $11{,}000$ km s$^{-1}$, $12{,}000$ km s$^{-1}$, and $10{,}700$ km s$^{-1}$ for the composite spectrum, warped Hsiao template spectrum, Foley $\Delta = -0.27$ template spectrum, and SALT2 model spectrum respectively. Overall, the Foley $\Delta = -0.27$ template spectrum and SALT2 model reproduce the Si II $\lambda 6355$ and $\lambda 5972$ absorption features the best; however, both of these template spectra have slightly redder continua than the other spectra. The Nugent-91T template spectrum is almost exclusively constructed from SN 1991T which has \dm\ $=0.90$ mag. The Nugent-91T template spectrum reproduces the color of the composite spectrum, but does not do well reproducing individual absorption features. This difference is because our composite spectrum was constructing using all slow-declining events, and not just those from the SN 1991T-like subclass. The Foley $\Delta = -0.27$ template spectrum and our composite spectrum are well matched towards bluer wavelengths.

Similarly, we also compare template SN Ia spectra to a composite spectrum constructed from faster-declining events (Figure \ref{fig:temp_mid_dm}). In this case our composite spectrum created from a \dm\ bin of $1.50<$ \dm\ $<1.70$ mag. In this comparison we also warp the maximum-light (\dm\ $=1.1$ mag) template spectrum to the $B$, $V$, and $R$-band photometry of the composite spectrum. The SALT2 spectrum was created to represent a SN Ia with \dm\ $=1.64$ mag (the average \dm\ of the composite spectrum). It is important to note that the SALT2 template spectrum has $x_1=-3.06$. Since SALT2 training sample only included SNe~Ia with \dm\ $<1.6$ ($x_1>-3$)\citep{guy07}, we caution that the SALT2 template may not be adequately representative of these lower-luminosity SNe~Ia. This SALT2 template spectrum does not match the composite spectrum as well as the slow-declining template spectrum matches the corresponding composite spectrum in Figure \ref{fig:temp_low_dm}. Most of the absorption features of the faster-declining SALT2 template spectra are stronger than the respective features in the composite spectrum. However, the overall change in these features going from slow to faster declining events is similar between the composite spectra and SALT2 template spectra. Overall, after warping, the Hsiao template spectrum matches up well with our composite spectra for different values of \dm . As expected, the largest differences between the Hsiao template spectra and our composite spectra occur in the features that strongly correlate with \dm\ (e.g., Si II and Ca II).

Finally, in Figure \ref{fig:temp_91bg} we compare composite spectra for the fastest declining events \dm\ $> 1.80$ mag to the Nugent-91bg template spectrum \citep{nugent02}. The Nugent-91bg template spectrum is constructed from SN 1991bg and SN 1999by and has telluric absorption included. Our composite spectrum has a bluer continuum and slightly larger line velocities. The composite spectrum also shows the Ti II absorption feature at ${\sim} 4300$~\AA\ characteristic of the SN 1991bg-like subclass.

\subsection{Host-galaxy Morphology}
In this section we investigate SN Ia spectral variations related to their host-galaxy environments. As seen above, SN Ia spectra strongly depend on phase and light-curve shape. We have shown that our composite spectra reproduce known spectral trends with these properties. Since low-luminosity SNe~Ia tend to occur in early-type environments while high-luminosity SNe~Ia tend to occur in late-type environments \citep{hamuy00,howell01}, composite spectra (for a specific epoch) generated by solely splitting the sample by host galaxy morphology will reflect this correlation. Therefore, we must control for light-curve shape in order to uncover the differences between similar SNe~Ia that occur in varying environments.

We make a single cut on host galaxy morphology via its Hubble classification. Our composite spectra represent SNe~Ia originating from late-type galaxies classified as Sa to Sd/Irr and early-type galaxies classified as E to S0a. If we split our nominal sample into these two subgroups, we see the expected trend that SNe residing in late-type galaxies have an average \dm\ $=1.10$ mag and the SNe residing in early-type galaxies have an average \dm\ $=1.47$ mag. Since the spectral features of our composite spectra change significantly with \dm\ (section \ref{sec:dm15}), we need to choose ranges of \dm\ that produce composite spectra with similar average properties. Since the two populations have differing average light-curve shapes we create our composite spectra to have an effective \dm\ $\approx 1.3$ mag in order to incorporate the largest number of spectra. We match \dm\ between samples by making hard cuts on the allowed \dm\ ranges for each composite spectrum. 

\begin{figure}
\includegraphics[width=3.2in]{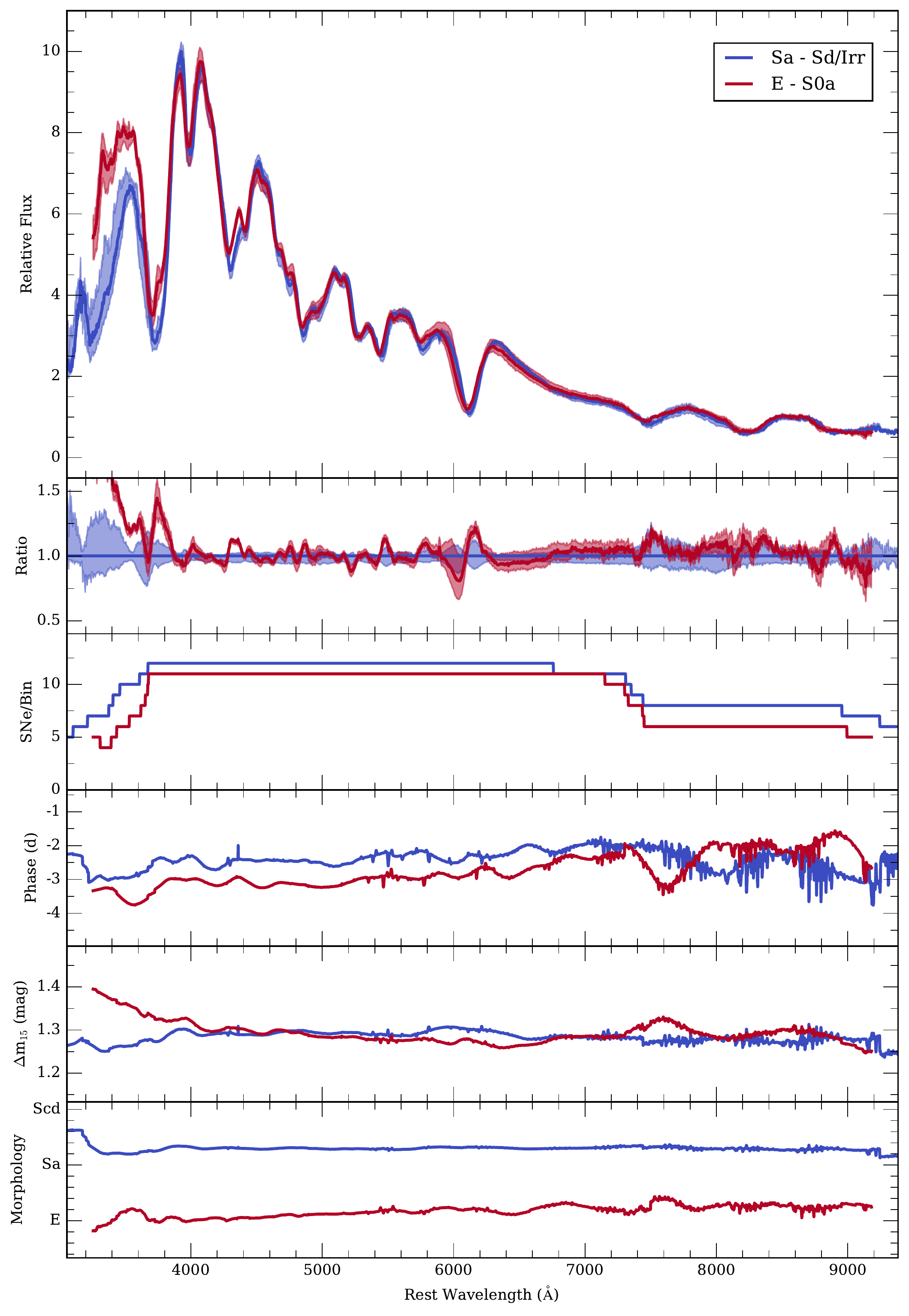}
\caption{Same format as Figure \ref{fig:max} except the bottom panel now displays the average host-galaxy morphology of a given composite spectrum as a function of wavelength. The blue curves show the properties of our late-type composite spectrum and the red curves show the properties of our early-type composite spectrum. Both of these composite spectra were constructed using a phase bin of  $-6 < \tau < 0$. The late-type composite spectrum was constructed using a light-curve shape bin of $1.20 <$ \dm\ $< 1.45$ mag. The early-type composite spectrum was constructed using a light-curve shape bin of $1.15 <$ \dm\ $< 1.50$ mag. \label{fig:host1}}
\end{figure}

\begin{figure}
\includegraphics[width=3.2in]{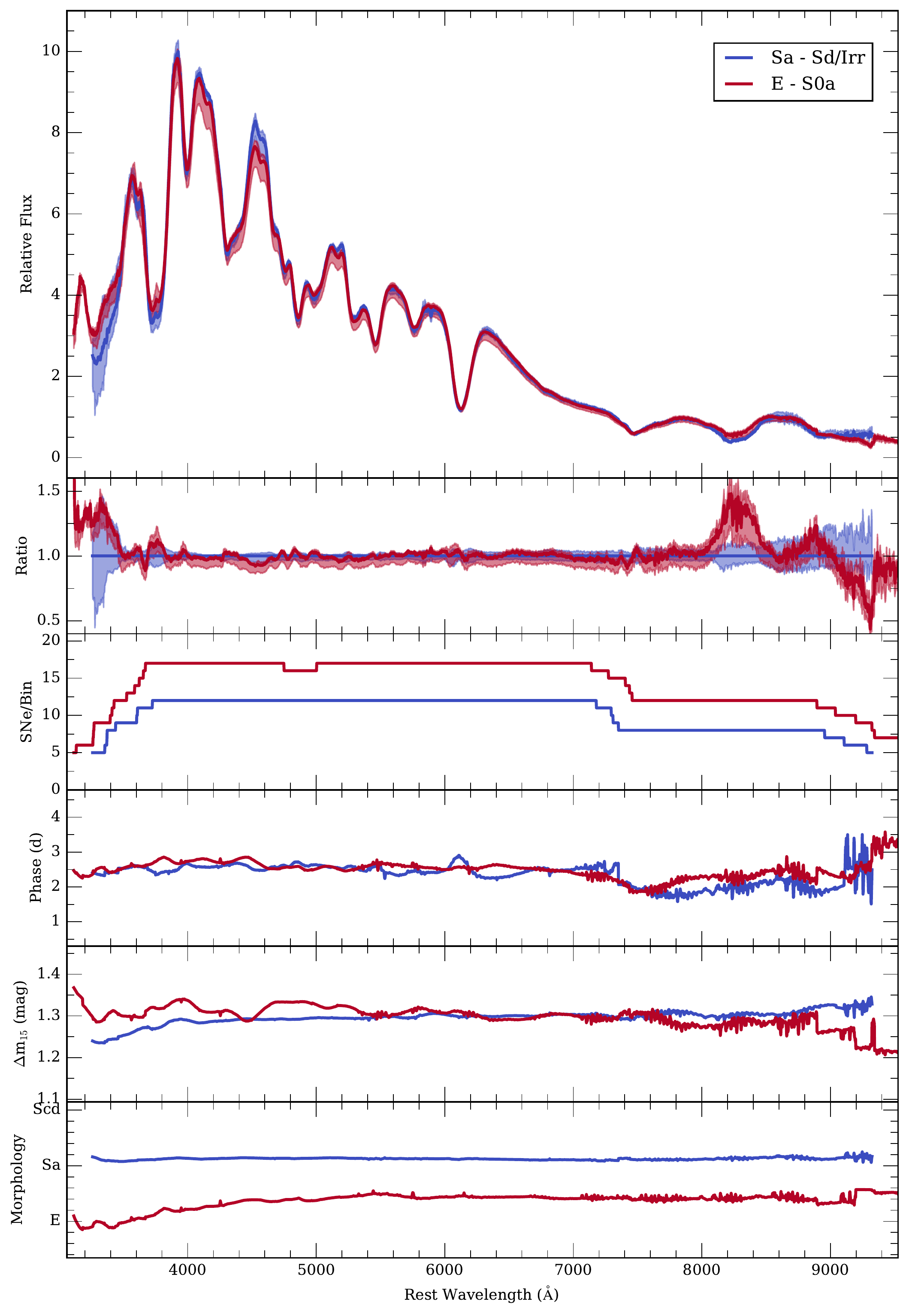}
\caption{Same format as Figure \ref{fig:max}. The blue curves show the properties of our late-type composite spectrum and the red curves show the properties of our early-type composite spectrum. Both of these composite spectra were constructed using a phase bin of  $0 < \tau < +6$. The late-type composite spectrum was constructed using a light-curve shape bin of $1.20 <$ \dm\ $< 1.45$ mag. The early-type composite spectrum was constructed using a light-curve shape bin of $1.10 <$ \dm\ $< 1.50$ mag.\label{fig:host2}}
\end{figure}

\begin{figure}
\includegraphics[width=3.2in]{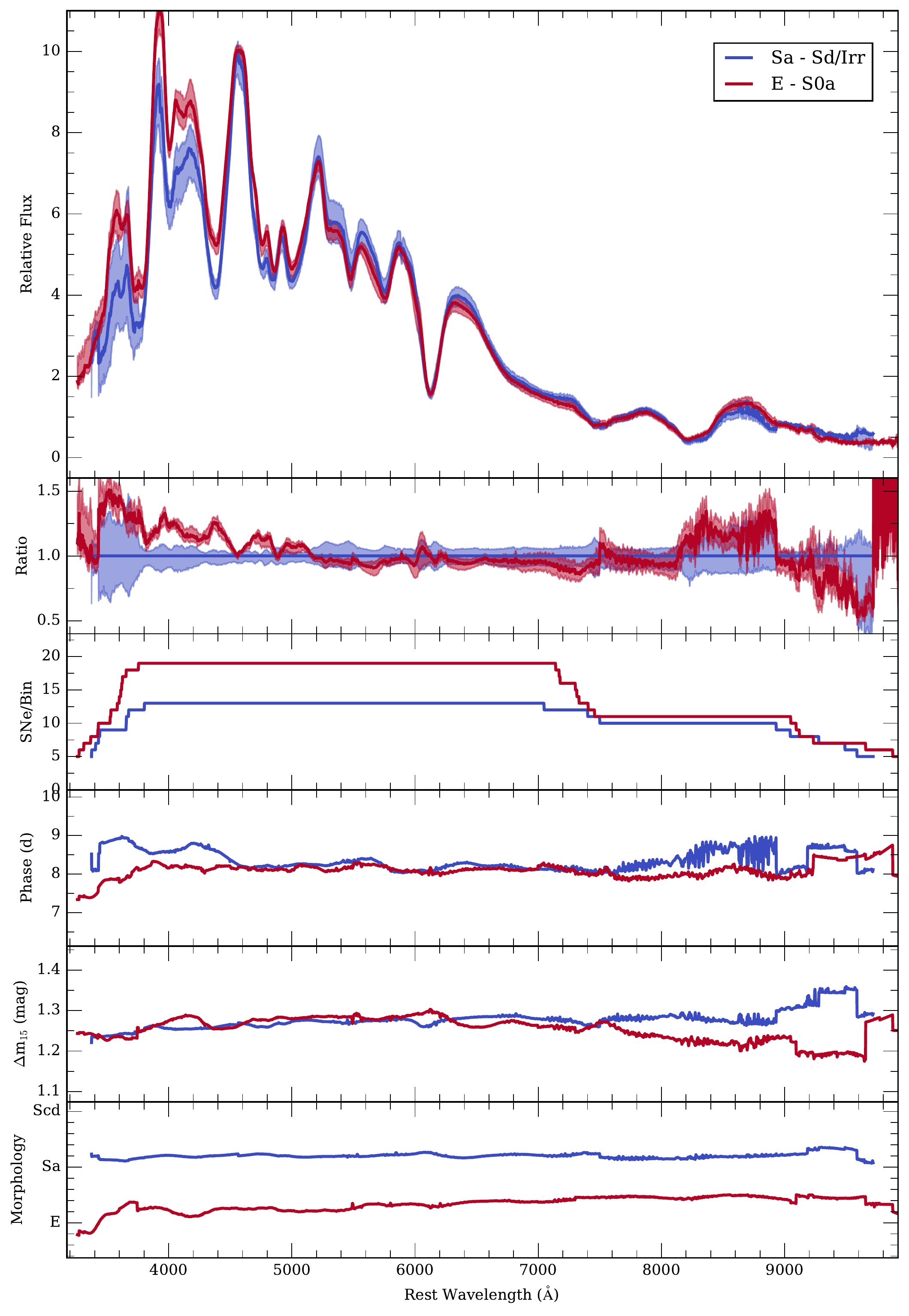}
\caption{Same format as Figure \ref{fig:max}. The blue curves show the properties of our late-type composite spectrum and the red curves show the properties of our early-type composite spectrum. Both of these composite spectra were constructed using a phase bin of  $+6 < \tau < +11$. The late-type composite spectrum was constructed using a light-curve shape bin of $1.17 <$ \dm\ $< 1.48$ mag. The early-type composite spectrum was constructed using a light-curve shape bin of $1.15 <$ \dm\ $< 1.45$ mag.\label{fig:host3}}
\end{figure}

We have generated composite spectra for each host morphology subgroup at three separate epochs, $-6 < \tau < 0$ days (Figure \ref{fig:host1}), $0 < \tau < +6$ days (Figure \ref{fig:host2}), and $+6 < \tau < +11$ days (Figure \ref{fig:host3}). In each of these Figures the blue curve and shaded region are constructed from spectra of SNe~Ia occuring in late-type galaxies and the red curve and shaded region are constructed from SNe~Ia occuring in early-type galaxies. These Figures have similar format to Figure \ref{fig:max} where we presented our total maximum-light composite spectrum. The second panels show the ratio of the $1 \sigma$ bootstrapping uncertainty regions relative to the late-type composite spectrum. The third panel shows the number of spectra contributing to each wavelength bin (we again require at least 5 spectra per wavelength bin). The other panels show the average values of phase, \dm, and host-galaxy morphology as a function of wavelength for each composite spectrum. At all three epochs, we see evidence for stronger Na I D absorption in composite spectra representative of SNe residing in late-type galaxies which indicates more gas in the surrounding environment.

Overall we see very little difference between composite spectra constructed from SNe with differing host galaxy environments. In Figure \ref{fig:host1} the late- and early-type composite spectra have effective phases of $-2.4$ and $-2.7$ days and an effective \dm\ of 1.28 and 1.29 mag respectively. The 1$\sigma$ bootstrapping uncertainty region of the early-type composite spectrum is within the 1$\sigma$ bootstrapping uncertainty region of the late-type composite spectrum at all wavelengths except for the red side of the Ca H\&K absorption feature, and at the bluest wavelengths (${<}3500$~\AA). This Ca H\&K feature is also double troughed in the early-type composite spectrum unlike the late-type composite spectrum. Since the average values of \dm\ are different by ${\sim}0.1$ mag at the bluest wavelengths, we may be able to attribute the differences in composite spectra at these wavelengths to a difference in average light-curve shape. 

Post maximum light the host morphology-controlled composite spectra are also very similar with some consistent differences. The composite spectra in Figure \ref{fig:host2} represent the late- and early-type subsamples and have effective phases of $+2.35$ and $+2.52$ days and an effective \dm\ of 1.30 and 1.29 mag respectively. There is one region near ${\sim}\ 4600$~\AA\ where the late-type composite spectrum has a slightly larger flux. There are also some slight differences in the Calcium absorption features. The early-type composite spectrum has slightly larger flux in the Ca II near-infrared (NIR) triplet absorption region. 

Our final set of composite spectra for late and early-type host morphologies have effective phases of $+8.4$ and $+8.1$ days and an effective \dm\ of 1.28 and 1.25 mag respectively. The 1$\sigma$ bootstrapping uncertainty regions are consistent at all wavelengths except in two regions. We see an excess flux in the early-type composite spectrum near the peak redward of the Si II $\lambda 4130$ feature. Additionally, we see slightly weaker Ca near-infrared triplet absorption in the early-type composite spectrum. We see slightly weaker Ca H\&K absorption in the early-type composite spectrum at +3 and +8 days, but these differences are consistent with the 1$\sigma$ bootstrapping uncertainty regions of the late-type composite spectra at those epochs. It is important to note that while the spectra between each of these subsamples are different, some of the same SNe may contribute to the composite spectra at different epochs.

\begin{figure}
\includegraphics[width=3.2in]{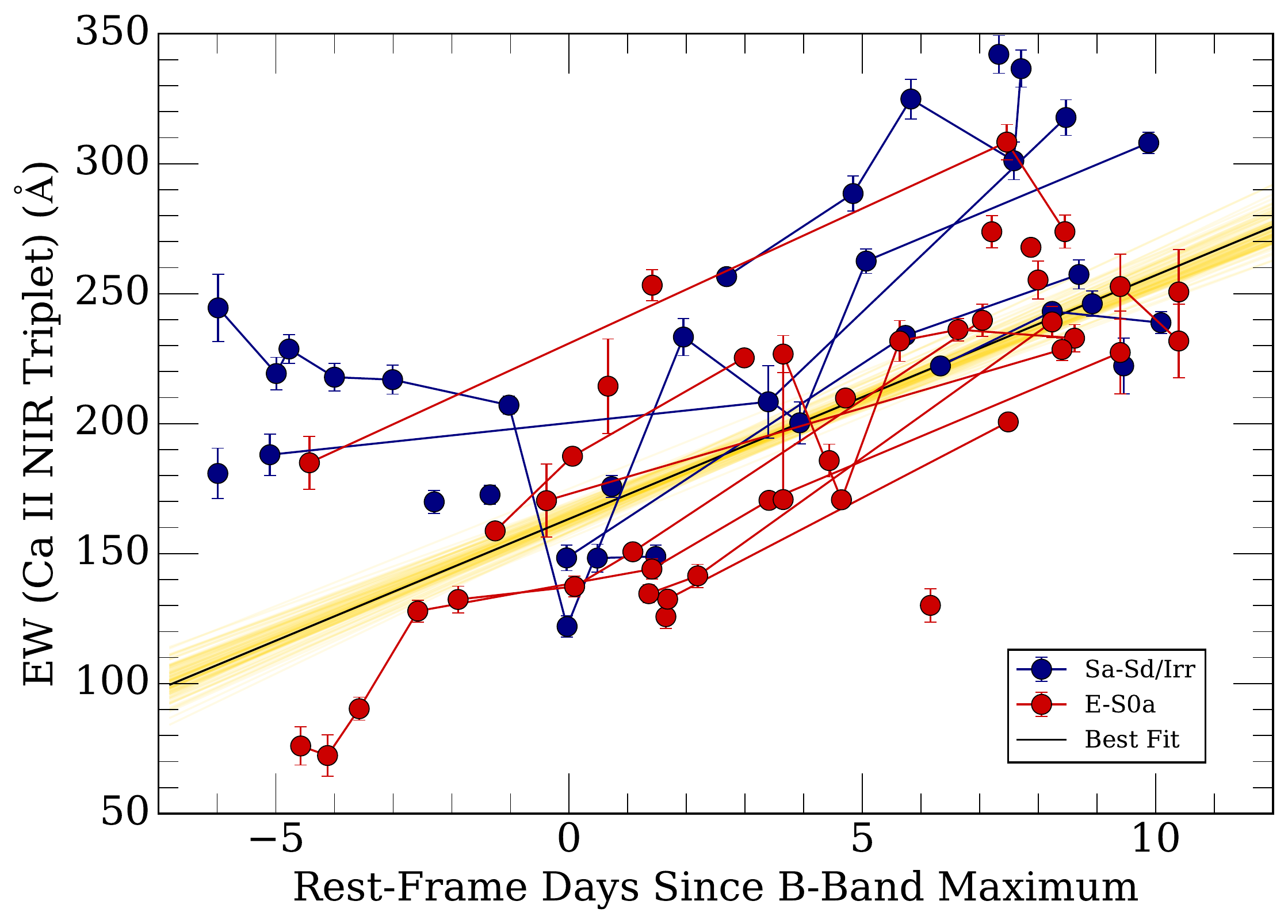}
\caption{Ca II NIR Triplet EW evolution for SNe residing in late (blue points) and early (red points) type galaxies. Measurements coming from the same SNe are connected by lines. The black curve is our linear best fit to the evolution of the whole sample ($y = (9.4\pm 0.6)x + 163 \pm 4$)). The yellow curves are best fit lines generated from bootstrap resampling of the data. \label{fig:ca_ew}}
\end{figure}

\begin{figure}
\includegraphics[width=3.2in]{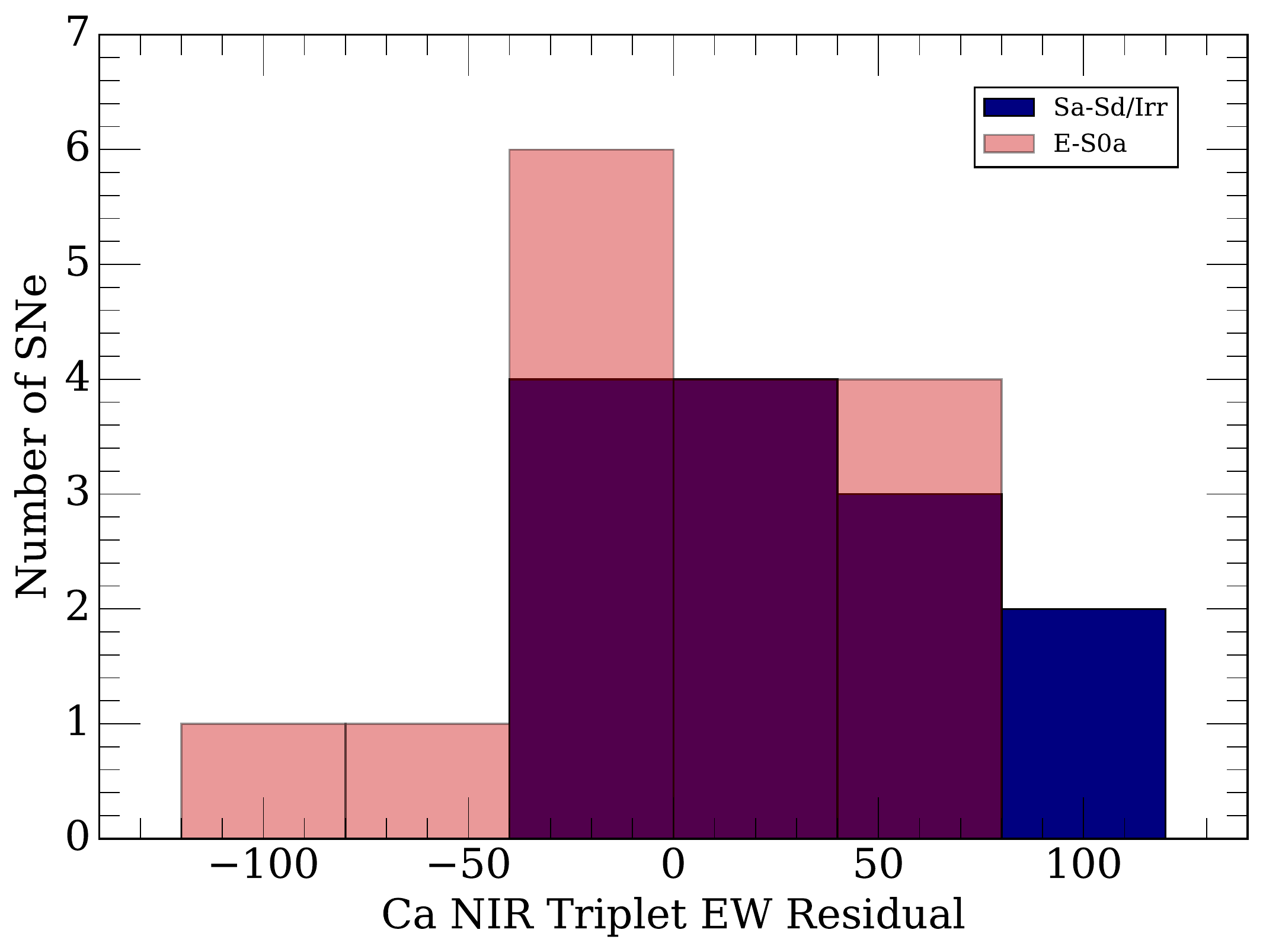}
\caption{Histograms of the late (blue) and early (red) type mean Ca II NIR Triplet EW residuals per SN from Figure \ref{fig:ca_ew}. \label{fig:ew_res}}
\end{figure}

Across multiple epochs there is some evidence for weaker Ca II absorption in the early-type composite spectra. Since this difference is more evident in the Ca II NIR triplet absorption feature, we investigate the evolution of the equivalent width (EW) of this feature from the individual spectra contributing to these composite spectra. These measurements are displayed in Figure \ref{fig:ca_ew}. Here, measurements coming from the same SNe are connected by lines and the black curve is the best fit linear evolution for the whole sample. We see a very slight tendency for the evolution of the Ca II NIR triplet EW for SNe residing in late-type galaxies (blue points) to lie above the mean evolution, and a very slight tendency for the evolution of the Ca II NIR triplet EW for SNe residing in early-type galaxies (red points) to lie below the mean evolution. To quantify this difference, we measure the mean residual per SN relative to our best fit evolution of the EW of the Ca II NIR triplet. In Figure \ref{fig:ew_res}, we show the distribution of these residuals for two bins of host galaxy morphology. The mean Ca II NIR triplet EW residual for late- and early-type host galaxies are $29\pm 2$~\AA, and $4\pm 2$~\AA\ respectively. We perform a Kolmogorov-Smirnov test with these data and find a p-value of 0.60, suggesting that we cannot reject the hypothesis that these residual distributions are drawn from the same population. Nonetheless, the tentative evidence for a difference in the strength of the Ca feature is intriguing.

It is possible that these weaker Ca features can be explained by differences in progenitor metallicity. The simulations presented in \citet{de14} and \citet{miles16} show decreasing $^{40}$Ca yields with increasing metallicity while $^{28}$Si yields stay  constant. At each phase, our early-type composite spectra show weaker Ca II features than their late-type analogs and Si II features remain roughly constant (especially near maximum light). However, \citet{miles16} also predict that the the strongest effects of metallicity on spectra of SNe~Ia should occur in the regions near ${\sim}4200$\AA\ and ${\sim}5200$\AA. Our early- and late-type composite spectra tend to be consistent and within the bootstrapping uncertainty regions in these wavelength regions. 

The subdivision of our sample by host-galaxy morphology also shows our need for more data in these fine-grained groupings. Since there is a strong correlation between host-galaxy morphology and light-curve shape, disentangling effects of \dm\ from effects caused by the SN environment is difficult with our limited sample sizes. Although not significant, the differences that we see in Ca II absorption features between our late- and early-type composite spectra is intriguing and could be further investigated with more spectra. This analysis also shows the potential for our composite spectra to uncover minor trends in the spectral features of SNe~Ia.

\section{Discussion and Conclusions}
We have compiled an open-source SQL relational database (\code{kaepora}) for observations of SNe~Ia. We currently include all spectroscopic data from the largest low-z spectroscopic samples as well as UV data from HST and Swift. This amounts to 4975 spectra of 777 SNe. We have inspected these data for quality and enhanced the data by removing galactic emission lines and cosmic rays, generating variance spectra, and correcting for the reddening caused by both MW and host-galaxy dust. The resulting product is a fully homogenized and accessible sample of SN~Ia spectra.

We construct Gini-weighted composite spectra from subsets of these data with sufficient metadata to estimate phase and correct for host-galaxy extinction. We generate 62 independent, phase-binned composite spectra with at least 20 contributing spectra covering phases $-11$ - $+271$ days. We also generate \dm -binned composite spectra for a variety of effective light-curve shapes and phases. The maximum-light \dm -binned composite spectra range from \dm\ $=0.86-1.87$ mag. Finally, we investigate the differences between composite spectra generated from SNe~Ia residing and late-type and early-type galaxies. Wavelength ranges vary between composite spectra because we require at least 5 SNe to contribute at any given wavelength. Using these composite spectra we are able to reproduce and verify the following trends:

\begin{itemize}
  \item General evolution of spectral features with phase for normal SNe~Ia
  \item $B-V$ color evolution of normal SNe~Ia
  \item Si II $\lambda 6355$ velocity evolution of normal SNe~Ia
  \item General spectral trends with \dm\ at maximum light
  \item Correlation of continuum with \dm\
  \item Correlation of $\mathcal{R}$ (Si II) with \dm
  \item Lower Si II $\lambda 6355$ velocities in SNe with extreme values of \dm\
  \item Na I D absorption in slower-declining SNe
  \item Na I D absorption in SNe residing in late-type galaxies
\end{itemize}
These numerous verifications assure us that the data in our database are of very high quality, which is encouraging for the large number of studies that we include. The additional metadata in our database (see Table \ref{tab:2}) will be useful for future studies.

Our phase and \dm -binned composite spectra are in good agreement with other template SN Ia spectra. In general the continua of composite spectra for normal SNe~Ia are best matched by the Hsiao and Foley template spectra. This was achieved without any generic spectral warping. We have also demonstrated that larger bin sizes in phase and \dm\ can be used to produce composite spectra that are representative of SNe with the same average properties.

Finally, we examine the subtle spectral differences of composite spectra generated from samples with differing host-galaxy morphologies. Since we can control for both phase and light-curve shape, the remaining significant spectral differences should be related to the progenitor environments of these SNe~Ia. Overall the composite spectra representing SNe occurring in late- and early-type environments are very similar at multiple epochs. We see a slight trend for weaker Ca II absorption in composite spectra representing SNe~Ia occurring in early-type environments. However, by investigating the Ca NIR triplet EW evolution of these SNe, we do not find significant evidence for the association of these differences with two distinct populations.

Our primary goal for this work is to present \code{kaepora} to the community along with the tools that we developed for its validation. We encourage members of the community to use these tools where they see fit. We also strongly encourage the analysis of individual spectra which \code{kaepora} makes very easy.

We hope that our database will become a valuable tool for investigating the spectral properties of SNe~Ia. We have demonstrated its utility for making template spectra with arbitrary average properties. While we have investigated some parameters here to
demonstrate the power of this new database and technique, there are
many potential future investigations using other parameters. Also, the large amount of photometric metadata present in the database should be useful for those not solely interested in spectra. Finally, we would like to emphasize that a particular strength this tool provides is fast testing of hypotheses. Since we have homogenized these observations, subsets of spectra with specific properties can be acquired and investigated via a few lines python code.

\section*{Acknowledgements}
This work relies on data obtained over several decades by many researchers.  Critically, these scientists made their data publicly available for additional studies such as this.  We thank all of the telescope operators, observers, data reducers, and PIs who's whose hard work and attention to detail were crucial for this project's success. We thank G. Dimitriadis for helpful comments and insights, and K. Siebert for a variety of helpful discussions.

This project was initiated as part of an undergraduate and graduate seminar at the University of Illinois by R.J.F.  We thank A. Beaudoin, Y. Cao, R. Chue, B. Fry, Y. Lu, S. Rubin, and A. Snyder for their contributions in the project's infancy.

M.R.S. is supported by the National Science Foundation Graduate Research Fellowship Program Under Grant No. 1842400. D.O.J. is supported by a Gordon \& Betty Moore Foundation postdoctoral fellowship at the University of California, Santa Cruz. The UCSC team is supported in part by NASA grant NNG17PX03C; NSF grants AST--1518052 and AST--1815935; the Gordon \& Betty Moore Foundation; the Heising-Simons Foundation; and by a fellowship from the David and Lucile Packard Foundation to R.J.F.

Based on observations made with the NASA/ESA Hubble Space Telescope, obtained from the Data Archive at the Space Telescope Science Institute, which is operated by the Association of Universities for Research in Astronomy, Inc., under NASA contract NAS 5--26555. These observations are associated with programs GO--4016, GO--12298, GO--12582, GO--12592, GO--13286, and GO--13646.  Swift spectroscopic observations were performed under programs GI--04047, GI--5080130, GI--6090689, GI--8110089, GI--1013136, and GI--1215205; we are very grateful to N.\ Gehrels, S.B.\ Cenko, and the Swift team for executing the observations quickly.

\bibliographystyle{mnras}
\bibliography{ref} 

\bsp	
\label{lastpage}
\end{document}